\documentclass[aps,prl,twocolumn,superscriptaddress,10pt]{revtex4}
\usepackage{amsmath}
\usepackage{amssymb}
\usepackage{graphicx}
\usepackage{subcaption}
\usepackage{epsfig}
\usepackage{dcolumn}
\usepackage{bm}
\usepackage{blindtext}
\usepackage{natbib}
\usepackage{multirow}
\usepackage{bbm}
\usepackage{placeins}
\usepackage{caption}
\usepackage{verbatim}
\usepackage{subcaption}
\setcitestyle{numbers,square}
\makeatletter
\def\@biblabel#1{[#1]}
\renewcommand\NAT@bibsetup%
   {\setlength{\leftmargin}{\z@}%
    \setlength{\itemindent}{\parindent}%
    \setlength{\itemsep}{\z@}%
    \setlength{\parsep}{\z@}}
\makeatother
\usepackage{color}
\usepackage{comment}
\usepackage{xr-hyper}
\usepackage{hyperref}

\usepackage{mathtools}

\def\be{\begin{equation}} \def\ee{\end{equation}}
\def\bea{\begin{eqnarray}} \def\eea{\end{eqnarray}}

\def\nn{\nonumber}

\begin{document}

\title{Topological multicomponent superconductivity with sizable $s$-wave admixture in twisted bilayer cuprates} 

\author{Yu-Hang Li}
\affiliation{School of Physics, Nankai University, Tianjin 300071, China}

\author{Congjun Wu}
%\email{wucongjun@westlake.edu.cn}
\affiliation{New Cornerstone Science Laboratory, Department of Physics, School of Science, Westlake University, Hangzhou 310024, Zhejiang, China}
\affiliation{Institute for Theoretical Sciences, Westlake University, Hangzhou 310024, Zhejiang, China}
\affiliation{Key Laboratory for Quantum Materials of Zhejiang Province, School of Science, Westlake University, Hangzhou 310024, Zhejiang, China}
\affiliation{Institute of Natural Sciences, Westlake Institute for Advanced Study, Hangzhou 310024, Zhejiang, China}

\author{Wang Yang}
\email{wyang@nankai.edu.cn}
\affiliation{School of Physics, Nankai University, Tianjin 300071, China}

\begin{abstract}

We investigate multicomponent superconductivity in twisted bilayer cuprates with order parameter $s+d_1 e^{i\phi_1}+d_2 e^{i\phi_2}$, where $s=s_1+s_2$ is the layer-symmetric $s$-wave pairing and $d_i$ is the $d$-wave pairing in layer $i$.
When $\phi_1-\phi_2\neq 0,\pi$, this three-component state spontaneously breaks time-reversal and $C_4$ rotational symmetries and can realize a topologically nontrivial phase.
Ginzburg-Landau analysis and self-consistent mean-field calculations show that this state is stabilized over a broad parameter regime.
We further identify nematic Kerr anisotropy as a key signature distinguishing it from $s+id$ and $d_1+e^{i\phi}d_2$,
and find that its Josephson critical current remains sizable near the $45^{\circ}$ twist.
Our results show that $s$-wave pairing does not preclude chiral topological superconductivity, pointing to twisted cuprates as a more robust platform than previously appreciated.

\end{abstract}

\maketitle

%%%%%%%%%%%%%%%%%%%%%%%%%%%%%%%%%%%%%%%%%%%%%%%%%%%%

\textit{Introduction.} --
Twist engineering, initially developed to generate correlated phases in van der Waals moir\'e materials \cite{Yankowitz2019,Cao2021,Oh2021,Park2021,Tarnopolsky2019,Cao2020,Lisi2021,Kerelsky2019,Samajdar2020,Kim2022,Chou2019,Xu2018,Saito2020,Törmä2022,Arora2020,Xia2025,Guo2025,Kim2025,Valagiannopoulos2022},
has been extended to cuprate superconductors to manipulate pairing symmetry \cite{Can2021,Lu2022,Bélanger2024,Fidrysiak2023,Tummuru2022,Volkov2025,Yuan2023}.
A prime target is chiral topological superconductivity hosting Majorana zero modes \cite{He2021,Zhang2016,Cheng2010,Garaud2011,Garaud2013,Yerin2021,Flensberg2021,Lutchyn2018,Sun2016,Sarma2015,RRoy2010,Brosco2024}. % for topological quantum computing
Among various candidate platforms \cite{Morf1998,Fu2016,Nelson2004,Maeno2024,Luke1998,Maeno2011,Kallin2009,Riseman1998,Brydon2019,Wu2013,Black-Schaffer2014,Ying2020,BRoy2010,Lin2018,Su2009,Xu2016,Balents2020,Uri2023,Andrei2021,Kezilebieke2022,Chen2019,cao2021,Park2026,Axe1989},
twisted bilayer cuprates are uniquely promising as they inherit high transition temperatures from parent compounds, providing a large energy scale for the topological $d+id$ phase \cite{Zhang2008,Martini2023,can2021,Yu2019,Lichtenstein2000,Tsuei2000,Kamimura1996,Song1995,Newns1995,Aji2010}. % $d+id$
However, their exact pairing symmetry remains debated.
While Ref.~\onlinecite{Zhao2023} observed suppressed interlayer Josephson coupling indicating predominantly $d$-wave pairing that supports $d+id$,
Ref.~\onlinecite{Zhu2021,HWang2023,Zhu2025,Zhu2023} found no such strong suppression, implying an appreciable $s$-wave component.
Given that existing theories suggest that the $s$-wave channel drives the system into a topologically trivial $s+id$ phase \cite{Lucht2025,Pixley2026,Zheng2025,Panda2026}, reexamining its exact role on a firmer theoretical basis is essential.

In this work, we %revisit the superconducting pairing structure of twisted bilayer cuprates,
demonstrate that twisted bilayer cuprates can remain topologically nontrivial even with an admixed $s$-wave component.
Solving the linearized gap equation near the superconducting transition % for a bilayer hopping model, we find 
reveals that interlayer tunneling significantly enhances the $s$-wave transition temperature, %by up to one order of magnitude,
while mildly suppressing the $d$-wave channel,
establishing an extended regime of competing $s$- and $d$-wave pairings.
Motivated by this degeneracy, our Ginzburg-Landau (GL) free energy analysis stabilizes a frustrated three-component pairing state $s + e^{i\phi_1} d_1 + e^{i\phi_2} d_2$, where $s=s_1+s_2$, and $s_{1,2}$ and $d_{1,2}$ are the layer-resolved $s$- and $d$-wave pairings. % across different twist angles
By simultaneously breaking time-reversal and $C_4$ lattice rotational symmetries for $\phi_1-\phi_2 \neq 0,\pi$,
this state transcends the $s+id$ paradigm, realizing a chiral superconducting state with nematic and potentially topological characteristics. % conventional intertwined

\begin{figure*}[htb]
    \centering
    \begin{subfigure}[b]{0.28\textwidth}
        \hspace*{-0.36cm}
        \includegraphics[width=\textwidth]{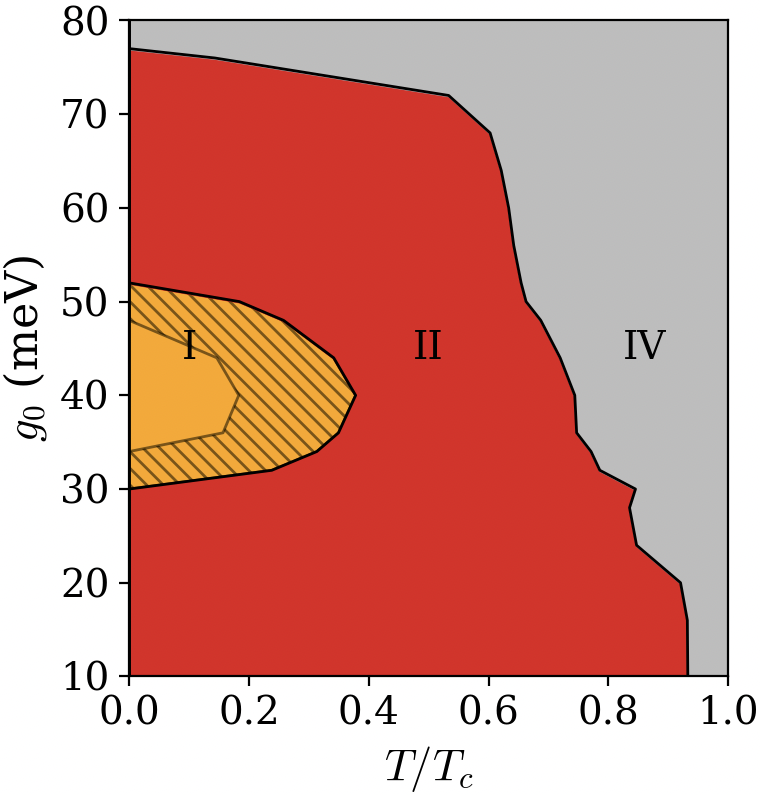}
        \caption{$\theta=43.6^\circ$, $\mu=-0.9t$.}
        \label{fig:v25mu-1.05}
    \end{subfigure}
    \hspace*{1.1cm}
    \begin{subfigure}[b]{0.28\textwidth}
       \hspace*{-0.36cm}
        \includegraphics[width=\textwidth]{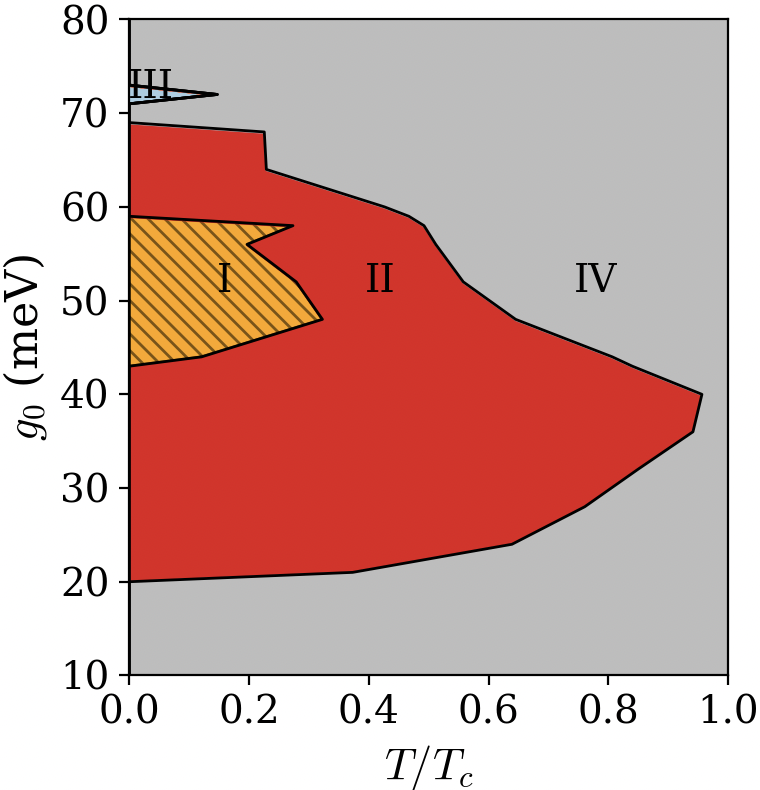}
        \caption{$\theta=53.13^\circ$, $\mu=-0.5t$.}
        \label{fig:v12mu-0.5}
    \end{subfigure}
    \hspace*{1.1cm}
    \begin{subfigure}[b]{0.2825\textwidth}
         \hspace*{-0.6cm}
        \includegraphics[width=\textwidth]{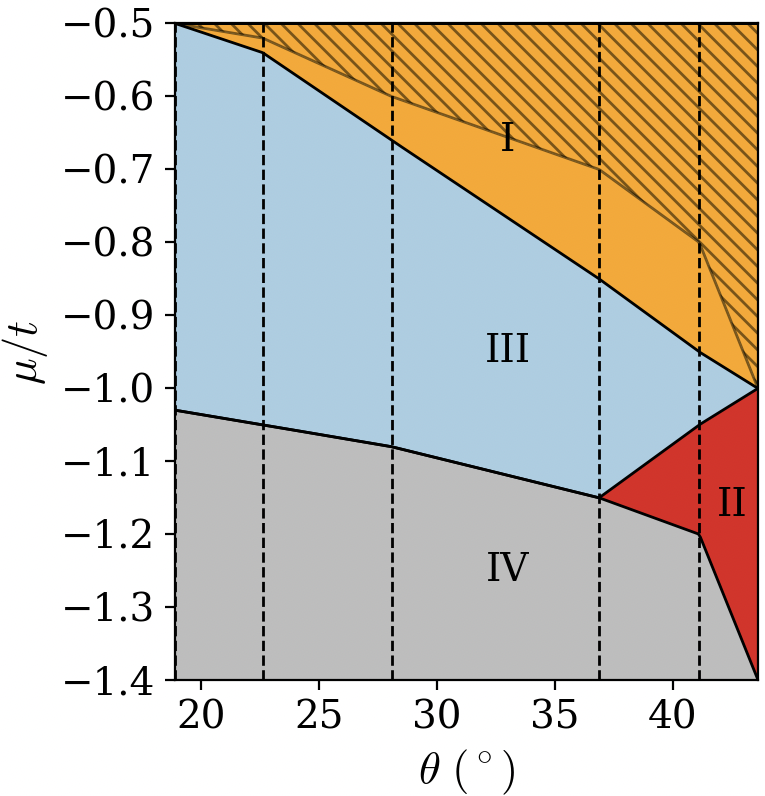} % 765*796
        \caption{$g_0=50$meV, $T=0$K.}
        \label{fig:mu_theta}
    \end{subfigure}
    \captionsetup{justification=raggedright}
    \caption{Evolution of distinct pairing states obtained from self-consistent calculations for the bond order parameters on the lattice.
    Regions I, II, III, and IV correspond to the pairing states $s+d_1e^{i\phi_1}+d_2e^{i\phi_2}$, $d_1e^{i\phi_1}+d_2e^{i\phi_2}$, $s+i(d_1\pm d_2)$ and $d_1\pm d_2$.
    Within Region I, the shaded and unshaded portions are topologically nontrivial and trivial, respectively, separated by bulk-gap closing lines.
    Region II is topologically nontrivial, while Regions III and IV are topologically trivial.
    In panel (c), dashed lines denote the discrete twist angles (including the two endpoints) used in exact lattice calculations, with continuous phase boundaries obtained via interpolation.}
\label{fig:PhaseDiagram}
\end{figure*}

Microscopically, our self-consistent mean-field calculations reveal a stable frustrated state, $s+e^{i\phi_1}d_1+e^{i\phi_2}d_2$, across a broad parameter space (Fig.~\ref{fig:PhaseDiagram}). % three-component , as shown in
The topologically nontrivial phase persists over wide ranges of temperature $T$ and interlayer tunneling strength $g_0$. % is accessible
% By tracking the evolution of distinct pairing states across the $g_0$-$T$ and $\mu$-$\theta$ planes, we find a stable three-component frustrated pairing state of the form $s+e^{i\phi_1}d_1+e^{i\phi_2}d_2$ in a broad parameter regime, where $g_0$, $T$, $\mu$, and $\theta$ denote the interlayer tunneling strength, temperature, chemical potential, and twist angle, respectively.
% Notably, the topologically nontrivial phase can be accessed through the temperature evolution over a wide range of $g_0$.
Furthermore, even in regimes prone to $s+id$, tuning the chemical potential $\mu$ and twist angle $\theta$ between two layers can rescue the nontrivial $s+e^{i\phi_1}d_1+e^{i\phi_2}d_2$ topology (Fig.~\ref{fig:PhaseDiagram}~(c)). % the topologically trivial state as can be seen from
Therefore, an $s$-wave component does not necessarily destroy the topological superconductivity, % Consequently; the emergence of; emergent
%Instead, twisted bilayer cuprates can remain chiral and topologically nontrivial even with a substantial $s$-wave pairing,
establishing twisted bilayer cuprates as a more robust platform for topological superconductivity than previously appreciated. % chiral
Nematic Kerr responses exhibiting $C_4$-symmetry breaking provide experimental tests for this frustrated three-component pairing, distinguishing it from both $d+id$ and $s+id$ pairings. % $s + e^{i\phi_1}d_1 + e^{i\phi_2}d_2$; decisive; conventional chiral; topologically trivial
In addition, its Josephson critical current remains finite near $\theta=45^\circ$, consistent with the robust experimental signals  \cite{Zhu2021,HWang2023,Zhu2025,Zhu2023}. % experimentally observed robust response.

%-----------------------------------------------------------------------------------
%---------------------------------------------------------------------------------

 %%%%%%%%%%%%%%%%%%%%%%%%%%%%%%%%%%%%%%%%%%%%%%%%%%%%%%
 
\textit{Superconducting transition temperatures.} -- % for $d$- and $s$-wave channels
 % \label{sec:pairing_kernel} 
 %---------------------------------------------------------------------------------------------------------------------
 % \subsection{Linearized gap equations}
 % \label{subsec:gap_equation} 
%As the system approaches the superconducting transition, the gap amplitude in channel $\alpha$ becomes infinitesimally small but nonzero ($\Delta_\alpha \neq 0$). 
The critical temperature $T_c^\alpha$ for channel $\alpha$ ($\alpha=s,d_1,d_2$) can be determined by the linearized gap equation
%\begin{eqnarray}
$V_\alpha \Pi_{\alpha}(T_c^\alpha)=1$,
%\label{eq:criterion}
%\end{eqnarray}
where $V_\alpha$ is attractive interaction strength in channel $\alpha$, and % $\Pi_{\alpha}(T)$ is the corresponding pairing susceptibility, given by
% \begin{equation}
$\Pi_{\alpha}(T)=(T/L^d) \sum_k \mathrm{Tr} \left[\hat{\phi}_\alpha^{\dagger}(\boldsymbol{k}) \hat{\mathcal{G}}^0_{k} \hat{\phi}_\alpha(\boldsymbol{k}) \hat{\mathcal{G}}^{0,T}_{-k} \right]$
% \label{eq:PiT}
% \end{equation}
is the corresponding pairing susceptibility.
Here, $\hat{\phi}_{\alpha}(\boldsymbol{k})$ specifies the pairing form factor % of $\alpha$-channel % in the combined layer and moir\'e unitcell site space,
and $\hat{\mathcal{G}}^0_{k}=(i\omega_n\hat{I}-h_{\boldsymbol{k}}^0)^{-1}$ is the normal-state Green's function, % matrix
% $k=(\boldsymbol{k},i\omega_n)$, $\omega_n=(2n+1)\pi/\beta$ ($n \in \mathbb{Z}$, $\beta=1/T$) is the fermionic Matsubara frequency.
in which $h_{\boldsymbol{k}}^0$ originates from a real-space Hamiltonian that incorporates intralayer nearest- and next-nearest-neighbor hoppings $t$ and $t^\prime$, chemical potential $\mu$, and interlayer tunneling modeled as % $g_{ij}$  originates from a real-space
%\begin{equation}
$g_{ij}=g_0\exp{\left[-\left(\sqrt{r_{ij}^2+d^2}-d \right)/\rho \right]}$ \cite{Can2021}.
%\label{eq:gij}
%\end{equation}
% where $g_0$ is the interlayer tunneling strength.
%Other geometric parameters in Eq.~\eqref{eq:gij}, together with the 
The explicit Hamiltonian and derivation of the pairing susceptibility are detailed in Sec.~I~A in Supplementary Materials (SM) \cite{SM}. % Explicit normal-state; Eq.~\eqref{eq:PiT}

%-------------------------------------------------------------------------------------------------------- 
 % \subsection{Critical temperatures of competing pairing symmetries}
 % \label{subsec:Tc}
 %---------------------------------------------------------------------------
\begin{figure}[h]
\hspace*{-0.6cm}
\includegraphics[width=0.265\textwidth]{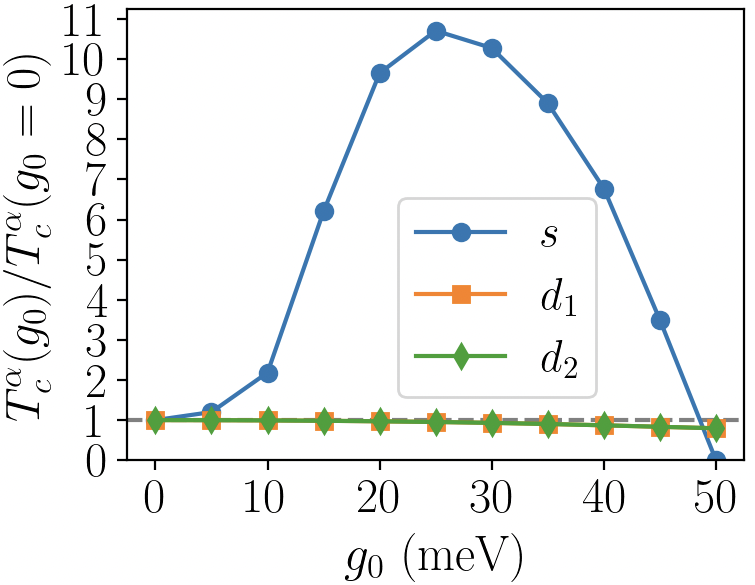}
\captionsetup{justification=raggedright}
\caption{
$T_c^\alpha(g_0)/T_c^\alpha(g_0=0)$ as functions of $g_0$, where $\alpha=s,d_1,d_2$. % $T_c^\alpha/T_c$
% and $T_c^\alpha$ is the superconducting transition temperature in the $\alpha$-channel.
% and $T_c^\alpha(g_0)$ denotes the superconducting transition temperature in the $\alpha$ channel at interlayer tunneling strength $g_0$. and $T_c$ is the critical temperature for monolayer superconducting cuprate. in the Hamiltonian chosen as; of the square lattice
 The parameters are $t=0.153$~eV, $t^\prime=-0.45t$, $\mu=-1.3t$, $d=2.22a$, and $\rho=0.39a$, where $a$ is the lattice constant, with twist angle $\theta=53.13^\circ$.}
\label{fig:Tc_g0}
%\vspace{-10pt}
\end{figure}
%---------------------------------------------------------------------------

By numerically solving the linearized gap equation, we evaluate the superconducting instabilities for $s$-, $d_1$-, and $d_2$-wave pairing channels,
where $d_{1,2}$ are the layer-resolved $d_{x^2-y^2}$-wave pairings and $s=s_1+s_2$ denotes the layer-symmetric combination of the intralayer $s$-wave components \cite{s12d12}.
Fig.~\ref{fig:Tc_g0} shows $T_c^\alpha(g_0)/T_c^\alpha(0)$ as functions of $g_0$ at $\theta=53.13^\circ$, revealing distinct responses among the pairing channels. % ($\alpha=s,d_1,d_2$) normalized critical temperatures
% The $s$- and degenerate $d_{1,2}$-wave channels exhibit distinct responses upon introducing the interlayer coupling.
While the degenerate $d$-wave are slightly suppressed by interlayer coupling,
$T_c^s$ is strongly enhanced, reaching a peak value eleven times larger than the $g_0=0$ case. % before decreasing at larger $g_0$.
% Specifically, $T_c^d$ is only lightly suppressed, while $T_c^s$ is rapidly enhanced and reaches a peak value eleven times larger than the $g_0=0$ case before decreasing at larger $g_0$.
% This pronounced enhancement of the $s$-wave instability, together with the robustness of the $d$-wave channels, 
This contrasting behavior suggests the possible coexistence of the $s$- and $d_{1,2}$-wave components.

The enhancement of the $s$-wave instability originates from the reconstruction of the normal-state Fermi surface. % induced by interlayer tunneling. % interlayer-tunneling-induced reconstruction
As $g_0$ increases, an additional annular Fermi surface emerges around the $\Gamma$ point.
% strongly enhancing the $s$-wave pairing susceptibility while the $d$-wave contribution is suppressed. % from this Fermi surface; by symmetry.
Since the little group at $\Gamma$ coincides with the lattice point group $D_4$, the Bloch states at $\Gamma$ can be chosen as eigenstates of $C_4$. % of the $\Gamma$ point
The $d$-wave form factor changes sign under a $C_4$ rotation, % forcing its intraband pairing amplitude to vanish at $\Gamma$,
forcing its projection onto the degenerate normal-state Bloch eigenspace at the $\Gamma$ point to vanish,
with the leading symmetry-allowed contribution scaling quadratically with the momentum deviation from $\Gamma$. % Because; must; growing only
In contrast, the $s$-wave form factor is invariant under $C_4$, % allowing its contribution to remain finite at $\Gamma$ with a constant leading contribution.
allowing its projection onto the Bloch eigenspace at $\Gamma$ to remain finite.
% Thus, the emergent Fermi surface couples efficiently to the $s$-wave channel but only weakly to the $d$-wave channel.
As a result, the newly emerged Fermi surface near the $\Gamma$ point strongly enhances the $s$-wave pairing, while having only a weak effect on the $d$-wave channel. % not touching the $d$-wave component much.
Details of the analysis are provided in Sec.~I~C in SM \cite{SM}.

% where the $s$-wave pairing susceptibility is enhanced while the $d$-wave response is suppressed because of the $C_4$ covariance of the normal-state Hamiltonian in the Bloch basis. At $\Gamma$ point, the normal-state band eigenstates can be chosen as $C_4$ eigenstates, whereas the $d$-wave form factor changes sign under $C_4$. Therefore, the intraband $d$-wave pairing vertex vanishes at $\Gamma$ and grows only quadratically in momentum nearby, strongly suppressing its contribution from this Fermi surface while leaving the $s$-wave channel unsuppressed. With further increasing $g_0$, the annular Fermi surface expands and eventually disappears, accounting for the subsequent decrease of $T_c^s$ at larger $g_0$. 

%%%%%%%%%%%%%%%%%%%%%%%%%%%%%%%%%%%%%%%%%%%%%%%%%%%%

\textit{GL free energy analysis.} --
% \label{sec:GL}
At $\theta=45^\circ$, the system respects the $D_{4d}$ point group symmetry (see Sec.~II~A in SM \cite{SM}). % twist between two layers; for details
% \subsection{$45^\circ$ twist between layers}
% \label{subsec:45}
% \subsubsection{Symmetries}
% \subsubsection{Ginzburg-Landau free energy}
% Including the $s$-, $d_1$-, and $d_2$-wave pairing components,  
The most general GL free energy for the $s$-, $d_1$-, and $d_2$-wave order parameters that respects $U(1)$ gauge, time-reversal ($\mathcal{T}$), and $D_{4d}$ symmetries is given by % three-component
% \begin{equation}
$F=F_{s}^{(0)} + F_{d}^{(0)}+F_{dd}+F_{sd}$,
%\label{eq:F}
%\end{equation}
where $F_s^{(0)}=\alpha_{s}|\psi_{s}|^2+\beta_{s}|\psi_{s}|^4/2$, $F_d^{(0)}=\alpha_d \left( |\psi_1|^2+|\psi_2|^2 \right) + \beta_d \left( |\psi_1|^4+|\psi_2|^4 \right)/2$, and
\begin{eqnarray}
F_{dd}&=&\gamma_{dd}|\psi_1|^2|\psi_2|^2+g_{dd} \left(\psi_1^2 \psi_2^{*2}+\psi_1^{*2} \psi_2^2 \right), \nn\\
F_{sd}&=&\gamma_{sd}|\psi_s|^2 \left(|\psi_1|^2+|\psi_2|^2 \right) \nn\\
&&+ g_{sd} \left[\psi_s^2 \left(\psi_1^{*2}+\psi_2^{*2} \right)+\psi_s^{*2} \left(\psi_1^2+\psi_2^2 \right)\right].
\label{eq:F4}
\end{eqnarray}
Here, $\psi_1$, $\psi_2$ and $\psi_s$ denote the complex order parameters of $d_1$-, $d_2$-, and $s$-wave order parameters, respectively.
%For completeness, the explicit $D_{4d}$ symmetry operations and the full four-component GL free energy involving $s_{1,2}$ and $d_{1,2}$, are presented in Appendix~\ref{app:GL45}.
Fixing the overall $U(1)$ gauge by choosing $\psi_s$ to be real, we have $\psi_s=|\psi_s|$, $\psi_1=|\psi_1|e^{i\phi_1}$, and $\psi_2=|\psi_2|e^{i\phi_2}$, where $\phi_{1,2}$ denote the phases of $\psi_{1,2}$ relative to $\psi_s$.

While the phase-independent coefficients are detailed in Ref.~\cite{GLc},
the positive quadratic Josephson couplings, % between $d_1$-, $d_2$-, and between $s$-, $d$-wave components, 
$g_{dd}$ and $g_{sd}$, favor $\phi_1-\phi_2=\pm \pi/2$ and $\phi_{1,2}=\pm \pi/2$, respectively.
%Since these conditions cannot  be satisfied simultaneously,
This mutual incompatibility leads to a frustrated pattern where not all pairwise relative phases are exactly $\pm \pi/2$,
giving rise to a time-reversal-symmetry-breaking (TRSB) superconducting state with nematic order and potentially nontrivial topological character.
%Indeed, if both $\phi_1$ and $\phi_2$ were equal to $\pm \pi/2$, then their difference $\phi_1-\phi_2$ could only be $0$ or $\pi$, rather than $\pm \pi/2$. Conversely, imposing $\phi_1-\phi_2=\pm \pi/2$ together with one of $\phi_1=\pm \pi/2$ or $\phi_2=\pm \pi/2$ necessarily drives the remaining phase away from its individually preferred value.

%--------------------------------------------------------------------------------------
% \subsubsection{Topological frustrated three-component pairing}
% \begingroup
% \clubpenalty=0
Minimizing the free energy resolves the frustrated phase configurations.
% As shown in Fig.~\ref{fig:TD4d_dds} (a), 
The equal amplitudes of order parameters ($|\psi_{s,1,2}|=1\,k_B T_c$) and Josephson couplings ($g_{sd}=g_{dd}$) yield a symmetric compromise with $\phi_1=\pi/3$ and $\phi_2=2\pi/3$.
% balancing the three competing phase lockings.
Breaking this balance shifts $\phi_{1,2}$ closer to $\pi/2$ for $g_{sd}>g_{dd}$,
% the couplings between $\psi_s$ and $\psi_{1,2}$ dominate, driving $\phi_{1,2}$ closer to $\pi/2$.
or $\phi_1-\phi_2$ closer to $\pi/2$ for $g_{sd}<g_{dd}$.
% the coupling between $\psi_1$ and $\psi_2$ dominates, moving $\phi_1-\phi_2$ toward $\pi/2$, while $\phi_1$ and $\phi_2$ deviate further from $\pi/2$, reflecting the stronger frustration imposed on the $s$--$d_1$ and $s$--$d_2$ phase lockings.
%Crucially, the frustrated configurations with $\phi_1-\phi_2\neq 0,\pi$ spontaneously breaks time-reversal symmetry, %combining the two layer-resolved $d$-wave components into a chiral superconducting state.
Crucially, when $\phi_1-\phi_2\neq 0,\pi$, $d_{1,2}$-wave components can be combined into a topologically nontrivial superconducting state \cite{Li2026_tri}.
The inclusion of an $s$-wave component does not necessarily destroy the topological nature,
allowing the $s+e^{i\phi_1}d_1+e^{i\phi_2}d_2$ state to remain topologically nontrivial over a finite parameter regime.
% \par
% \endgroup

% \subsubsection{Symmetry breaking pattern}
Notably, the invariance of the pairing state under the combined anti-unitary operations $\mathcal{T} l_{x\pm y}$ imposes the phase constraint $\phi_1+\phi_2=\pi$, where $l_{x\pm y}$ are twofold interlayer rotations.
The three-component pairing gap function thus takes the compact form of $s+i d_1 e^{i\phi/2}+i d_2 e^{-i\phi/2}$,
where $\phi=\phi_1-\phi_2$.
Including the invariance under the $C_2$ rotation about the $c$-axis,
the unbroken symmetry group is $\{ E, C_2, \mathcal{T} l_{x+y}, \mathcal{T} l_{x-y} \} \cong D_2$,
and the symmetry-breaking pattern is % corresponding; can be determined as; for the configuration in Fig.~\ref{fig:TD4d_dds}~(a)
$D_{4d}\times \mathbb{Z}_2^{\mathcal{T}}\to D_2$,
where $\mathbb{Z}_2^{\mathcal{T}}$ denotes the two-element group generated by $\mathcal{T}$.
Since $|D_{4d}\times \mathbb{Z}_2^{\mathcal{T}}|/|D_2|=8$,
there are eight degenerate ground-state configurations, detailed in Sec.~II~B in SM \cite{SM}.
% pairing; as shown in Fig. \ref{fig:TD4d_dds}~(a)-(h). Symmetry operations that can transform the configuration in Fig. \ref{fig:TD4d_dds}~(a) to (b)-(h) are indicated on top of each subfigure.
%with the the symmetry operations defined in Appendix~\ref{app:GL45}.

This three-component pairing state spontaneously breaks not only $\mathcal{T}$ but also the in-plane $C_4$ rotational symmetry down to $C_2$, realizing a TRSB and nematic superconducting phase.
%As a result, physical observables are expected to develop an intrinsic twofold anisotropy within the plane.
Such nematicity may be detected through angle-resolved measurements, which should exhibit a characteristic $C_2$ periodicity rather than the $C_4$ behavior of the underlying lattice.

% \textit{Topological transition.} --
We further discuss the possibility of a topological phase transition.
%now specify the conditions \textcolor{red}{where $\phi_1-\phi_2=0, \pi$} and 
% rendering the pairing state topologically trivial.
Setting $|\psi_1|=|\psi_2|=|\psi_s|=|\psi|$, the phase-dependent part of the free energy can be simplified to
\begin{equation}
F_{\phi}=C_{sd}[\cos(2\phi_1)+\cos(2\phi_2)]+C_{dd}\cos[2(\phi_1-\phi_2)],
\label{eq:F_phase_simple}
\end{equation}
in which $C_{sd}=2g_{sd}|\psi|^4$ and $C_{dd}=2g_{dd}|\psi|^4$ are both positive.
Minimizing Eq. (\ref{eq:F_phase_simple}) reveals a topological phase transition at $C_{sd}/C_{dd}=2$.
For $C_{sd}/C_{dd}>2$, we find $\phi_{1,2}=\pm\pi/2$, % and $\phi_1-\phi_2=0,\pi$,
corresponding to a topologically trivial pairing state $s+i(d_1\pm d_2)$ (``$d+is$'' in Ref. \onlinecite{Can2021}). % configuration denoted as 
Conversely, the regime $C_{sd}/C_{dd}<2$ maintains $\phi_1-\phi_2 \neq 0, \pi$, preserving the topologically nontrivial pairing.
% Therefore, topological phase transition occurs at $C_{sd}/C_{dd}=2$. 

For general twist angles, the four-component and reduced three-component GL free energy analysis % governed by $D_4$,
% $\theta\neq 45^\circ$ complete involving $s_1$, $s_2$, $d_1$, and $d_2$ is provided in Appendix~\ref{app:GLgeneral}, point-group symmetry
% along with the reduced three-component ($s$, $d_{1,2}$) GL free energy analysis, %
are provided in  Secs.~II~C and II~D in SM \cite{SM}.

%Appendix~\ref{app:GLgeneral} and Appendix~\ref{app:general}, respectively.
% while detailed pairing configurations and symmetry-breaking pattern for the reduced three-component model with mixed $s$-, $d_1$-, and $d_2$-wave symmetries are presented in Appendix~\ref{app:general}.

%%%%%%%%%%%%%%%%%%%%%%%%%%%%%%%%%%%%%%%%%%%%%%%%%%%%%%%%%

% \enlargethispage{\baselineskip}

\textit{Self-consistent mean field solution.} --
% \label{sec:selfconsistent}
% In this section, we self-consistently calculate the pairing order parameters on twisted bilayer lattice from a microscopic model, and identify the topologically nontrivial three-component-pairing order parameter, establishing the corresponding temperature-dependent phase diagrams for two different twisting angles.
% \subsection{Self-consistent solution on the lattice}
% \label{subsec:selfconsistent}
Including the electron-electron interactions described by an extended Hubbard model to $h_{\boldsymbol{k}}^0$, the full Hamiltonian is given by % the normal-state noninteracting Hamiltonian \cite{Can2021}; with onsite repulsion and nearest-neighbor attraction
\begin{align}
H=&-\sum_{ij,\sigma l}t_{ij}\left(c_{i\sigma l}^\dagger c_{j\sigma l}+h.c. \right) -\mu \sum_{i\sigma l}n_{i\sigma l} \notag\\
&-\sum_{ij\sigma}\left(g_{ij}c_{i\sigma 1}^\dagger c_{j\sigma 2}+h.c. \right) + \sum_{ij,l}V_{ij}n_{il} n_{jl},
\label{eq:Hfull}
\end{align}
where $l=1,2$ are layer indices, $t_{ij}$ includes $t$, $t^\prime$, % the intralayer hoppings $t$, $t^\prime$; $g_{ij}$ is the interlayer tunneling
and $V_{ij}$ is the density-density interaction with onsite repulsion $U$ and nearest-neighbor attraction $V$.

Focusing on a nearest-neighbor spin-singlet pairing interaction, we introduce the Hubbard--Stratonovich pairing fields $\Delta_{ij,l}$,
and the self-consistent mean field equation determines the bond order parameters as \cite{Can2021}
\begin{equation}
 \Delta_{ij,l}=-V \sum_{\boldsymbol{k}} \text{Tr} \left[ \frac{\partial h_{\boldsymbol{k}}}{\partial \Delta_{ij,l}^*} U_{\boldsymbol{k}} n_F\left(E_{\boldsymbol{k}} \right) U_{\boldsymbol{k}}^\dagger \right],
 \label{eq:Delta_ijl}
 \end{equation}
where $h_{\boldsymbol{k}}$ is the Bogoliubov-de Gennes (BdG) matrix corresponding to Eq.~\eqref{eq:Hfull} % transformed
with $U_{\boldsymbol{k}}^\dagger h_{\boldsymbol{k}} U_{\boldsymbol{k}} = E_{\boldsymbol{k}}$, % unitary matrix $U_{\boldsymbol{k}}$ diagonalizes $h_{\boldsymbol{k}}$ as
and $n_F$ is the Fermi distribution. % \left(E_{\boldsymbol{k}} \right) matrix
Solving Eq.~\eqref{eq:Delta_ijl} self-consistently and performing a Fourier transform within each layer,
we obtain the momentum-space pairing functions. %(see Sec.~III in SM \cite{SM}).
%A path-integral derivation of Eq.~\eqref{eq:Delta_ijl} and the corresponding plots of the momentum-space pairing functions are provided in Sec. III in SM \cite{SM}.

Projecting the obtained pairing functions onto the $D_4$ irreducible representations (see Sec.~IV~A in SM \cite{SM} for details), % momentum-space
we find that the two-dimensional $E$ sector vanishes numerically.
The surviving one-dimensional sectors can be reconstructed into the layer-resolved $s$- and $d$-wave components via
% $s_{1,2}(\boldsymbol{k_{1,2}})=[f^{A_1}(\boldsymbol{k}) \pm f^{A_2}(\boldsymbol{k})]$ and $d_{1,2}(\boldsymbol{k_{1,2}})=[f^{B_1}(\boldsymbol{k}) \pm f^{B_2}(\boldsymbol{k})]$, where $f^{\Gamma}(\boldsymbol{k})$ is the projection onto channel $\Gamma$,
$A_{1,2}=s_1\pm s_2$ and $B_{1,2}=d_1\pm d_2$,
from which the relative phases between different pairing components can be extracted. % uniform; across the Brillouin zone (BZ)
% $\phi_{s_{1,2}}$ and $\phi_{d_{1,2}}$,
On the other hand, the pairing magnitudes are inhomogeneous in the Brillouin zone,
hindering a direct comparison between different pairing channels.
% To quantify the effective pairing strengths, we use the Cooper-pair condensate density of each surviving channel as a proxy (see Sec.~IV~B in SM \cite{SM}).
To quantify the effective pairing strengths, we calculate the Cooper-pair condensate density for each pairing component as a measure of the pairing strengths (see Sec.~IV~B in SM \cite{SM}). % , using it
%Consequently, we can apply the aforementioned projection and quantification procedures to the self-consistent order parameters obtained across different twist angles and parameter regimes.

%---------------------------------------------------------------------------
\begin{figure}[h]
\centering
\begin{subfigure}{0.16\textwidth}
    \hspace{-0.1cm}
    \includegraphics[width=\textwidth]{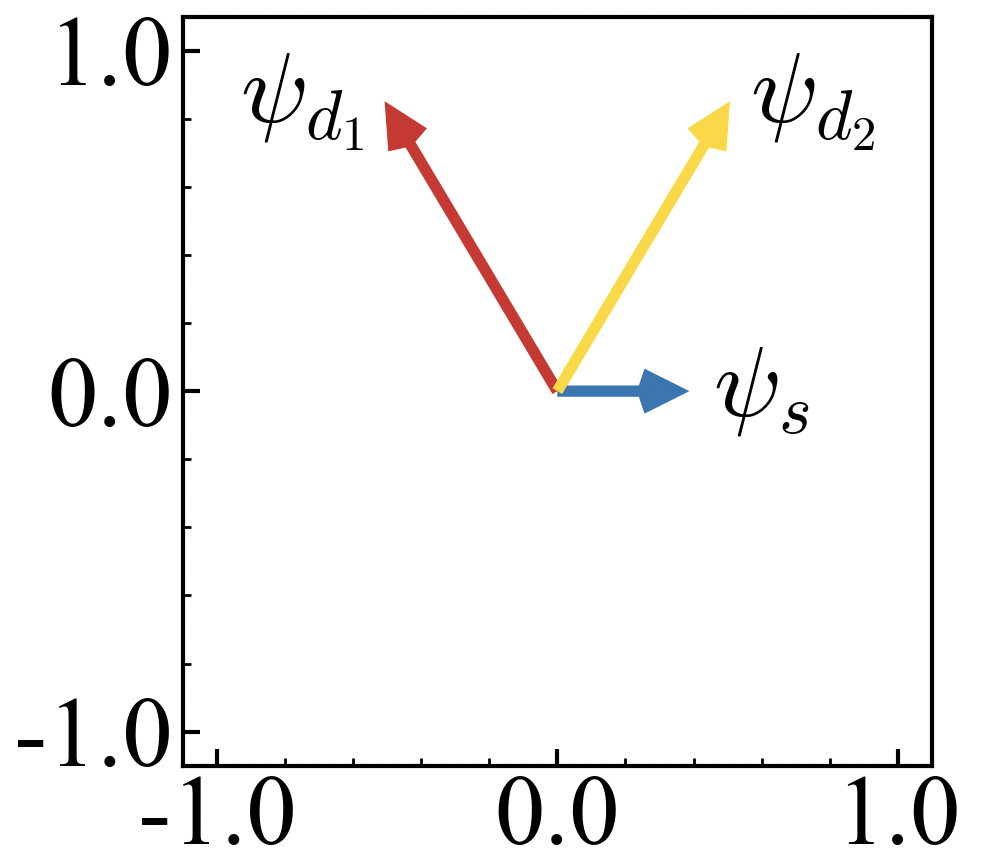}
    \caption{$\theta=43.6^\circ$}
    \label{fig:sd1d2_v25g036}
\end{subfigure}%
\hspace{0.6cm}
\begin{subfigure}{0.16\textwidth}
    \hspace{-0.36cm}
    \includegraphics[width=\textwidth]{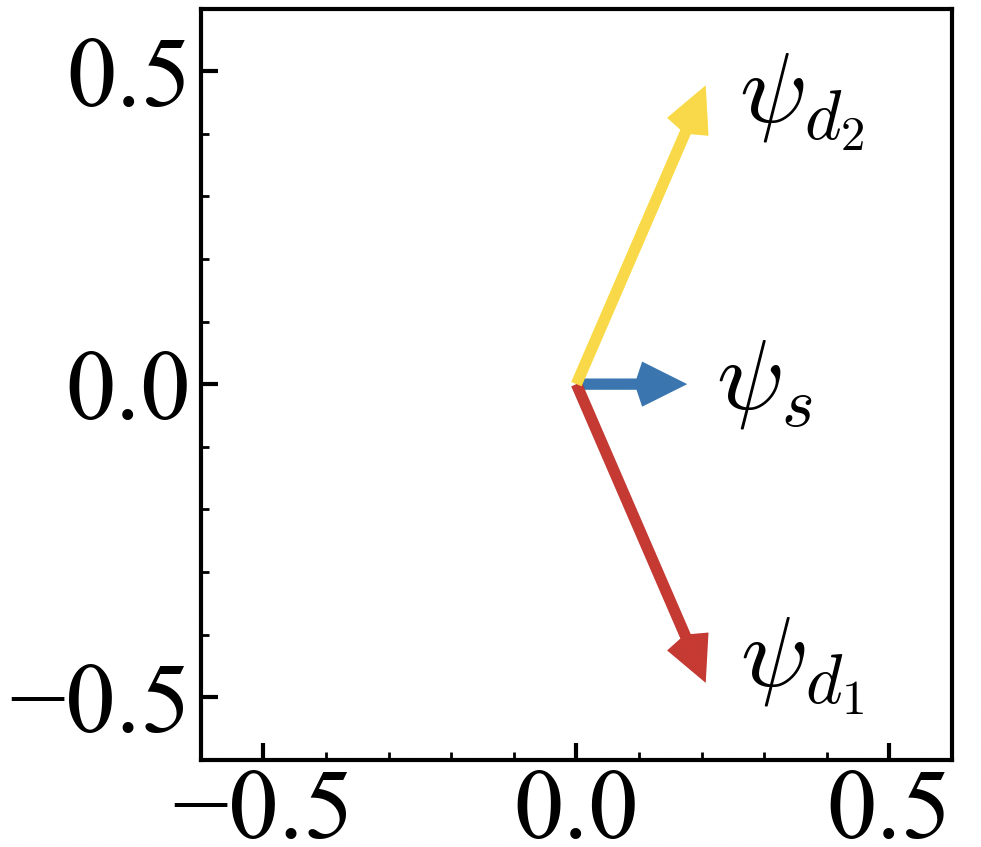}
    \caption{$\theta=53.13^\circ$}
    \label{fig:sd1d2_v12g032}
\end{subfigure}
\captionsetup{justification=raggedright}
\caption{Configurations of the three-component pairing $s+d_1e^{i\phi_1}+d_2e^{i\phi_2}$ with fixing $\phi_s=0$. % at different twist angles.
In panel (a), $|\psi_s|=0.388k_BT_c$, $|\psi_{1,2}|=$ $0.991k_BT_c$, $\phi_1=0.671\pi$, and $\phi_2=0.329\pi$,
while in panel (b), $|\psi_s|=0.177k_BT_c$, $|\psi_{1,2}|=$ $0.522k_BT_c$, $\phi_1=-0.372\pi$, and $\phi_2=0.372\pi$.}
\label{fig:sd1d2}
% \vspace{-10pt}
\end{figure}
%---------------------------------------------------------------------------

The resulting pairing strengths reveal that $A_2$ is negligibly small compared with the other pairing channels, % the component
being about two orders of magnitude smaller than $A_1$. % (see Secs.~IV~A and IV~B in SM \cite{SM}).
Therefore,  the pairing is well described by the three components $(s_1+s_2)$-, $d_1$-, and $d_2$, with the gap function % state
% Neglecting this subdominant component, the pairing state is therefore well described by the three components $(s_1+s_2)$-, $d_1$-, and $d_2$, with the gap function
\begin{equation}
\hat{\Delta}(\boldsymbol{k})=|\psi_s|e^{i\phi_s}\hat{s}+|\psi_1|e^{i\phi_1}\hat{d}_1+|\psi_2|e^{i\phi_2}\hat{d}_2.
\label{eq:Delta3}
\end{equation}
Exploiting the $U(1)$ gauge freedom to set $\phi_s=0$, % global
we obtain the pairing configurations at $\theta=43.6^\circ$ and $53.13^\circ$, as shown in Fig.~\ref{fig:sd1d2}, % two
which are characterized by $\phi_1+\phi_2=\pi$ and $0$ and preserve the unbroken symmetry groups $\{E, C_2, \mathcal{T}l_{x\pm y} \}$ and $\{E, C_2, \mathcal{T}l_{x,y} \}$, respectively.
Here, $l_{x\pm y}$ and $l_{x,y}$ denote the twofold interlayer rotations about their respective in-plane axes,
and $x$-, $y$-axes are defined along the two nearest-neighbor directions of the original untwisted lattice, with layers 1, 2 rotated by $\pm \theta/2$ relative to this reference orientation, respectively.
% where $x$-axis is defined in such a way that the angle between  layer $1$ (layer $2$) and the $x$-axis is $\theta/2$ ($-\theta/2$).
% Notably, these two configurations are characterized by $\phi_{d_1}+\phi_{d_2}=\pi$ and $0$, whose unbroken symmetry groups are $\{ E, C_2, \mathcal{T}l_{x\pm y} \}$ and $\{ E, C_2, \mathcal{T}l_{x,y} \}$, respectively. Here, $l_{x\pm y}$ and $l_{x,y}$ denote the twofold interlayer rotations about their respective in-plane axes.

% $\phi_{d_1}$ and $\phi_{d_2}$ become $0.671\pi$ and $0.329\pi$, respectively, as shown in Fig.~\ref{fig:sd1d2}~(a).
% Through analogous calculations, the pairing configuration for $\theta=53.13^\circ$ is presented in Fig.~\ref{fig:sd1d2} (b). %(with other parameters identical to those in Fig.~\ref{fig:sd1d2_T_vec12} (d) of Appendix~\ref{app:sd1d2T_vec} at $T=0$~K) 

To systematically map the evolution of distinct pairing states, we calculate the phase diagrams in the $g_0$-$T$ and $\mu$-$\theta$ planes, as shown in Fig.~\ref{fig:PhaseDiagram}.
Region I corresponds to a stable frustrated $s+d_1e^{i\phi_1}+d_2e^{i\phi_2}$ state, with topologically nontrivial and trivial sectors indicated by the shaded and unshaded regions, respectively, and separated by bulk-gap-closing lines.
%revealing a stable and topologically nontrivial $s+d_1e^{i\phi_1}+d_2e^{i\phi_2}$ state labeled as Region I.
While this frustrated state is most prominent within $30$~meV$\leq g_0 \leq 60$~meV in Fig.~\ref{fig:PhaseDiagram}~(a, b),
it can persist down to $g_0=20$~meV (see Sec.~V in SM \cite{SM}).
% Across the temperature range, the system hosts several competing pairing states, the frustrated three-component state,
Other competing states include the chiral $d_1e^{i\phi_1}+d_2e^{i\phi_2}$, topologically trivial $s+i(d_1\pm d_2)$, and $d_1\pm d_2$ states (Regions II, III, IV). % ultimately
% Elevating the temperature induces a cascade of phase transitions, including the chiral $d_1e^{i\phi_1}+d_2e^{i\phi_2}$ (Region II), the topologically trivial $s+i(d_1\pm d_2)$ (Region III), and ultimately the pure $d_1\pm d_2$ (Region IV) states.
 % fine-tuning rescue the chemical potential and twist angle
The Chern numbers characterizing the different regions, % in Fig.~\ref{fig:PhaseDiagram},
the topological transition within the $s+d_1e^{i\phi_1}+d_2e^{i\phi_2}$ state driven by mass inversions at four unbroken-symmetry-related Dirac points,
and the corresponding doping ranges are detailed in Secs.~VI~A, VI~B, and VII in SM~\cite{SM}, respectively.

% The Chern numbers characterizing the different regions in Fig.~\ref{fig:PhaseDiagram} are provided in Sec.~VI~A of the SM~\cite{SM}, while the topological transition within the $s+d_1e^{i\phi_1}+d_2e^{i\phi_2}$ state, driven by mass inversions at four unbroken-symmetry-related Dirac points, is discussed in Sec.~VI~B in SM~\cite{SM}. % and further corroborated by a Fermi-surface winding analysis. % distinguished
% The corresponding experimental doping ranges are given in Sec.~VII in SM~\cite{SM}.

% The Chern numbers characterizing the different regions in Fig.~\ref{fig:PhaseDiagram} are provided in Sec.~VI~A in SM \cite{SM},
% and the transition between topologically distinct sectors within the $s+d_1e^{i\phi_1}+d_2e^{i\phi_2}$ state is driven by mass inversions at four symmetry-related Dirac points, with the Chern numbers $C=0$ and $C=-4$ corroborated by the Fermi-surface winding analysis (see Sec.~VI~B of the SM~\cite{SM}).
%The temperature-evolution profiles of the order parameters for several selected parameter sets ($\theta$, $g_0$, $\mu$) are provided in Sec. V in SM \cite{SM}.

Furthermore, even in parameter regimes that typically favor $s+i(d_1\pm d_2)$, %inherently prone to
tuning $\mu$ and $\theta$ can recover the nontrivial chiral topology, as can be seen in Fig.~\ref{fig:PhaseDiagram}~(c).
Tuning the chemical potential provides an efficient means of controlling the $s$-wave pairing strength by modifying the $\Gamma$-centered Fermi-surface loop that favors $s$-wave pairing.
% At fixed $g_0$ and $\theta$, varying $\mu$ shifts the Fermi level relative to the bands near $\Gamma$-point and thereby changes the size of the $\Gamma$-centered Fermi-surface loop that favors $s$-wave pairing.
A larger loop enhances the relative $s$-wave weight, $|\psi_s|/|\psi_d|$, making $|\phi_{1,2}|=\pi/2$ energetically more preferable than $|\phi_1-\phi_2|=\pi/2$, and thereby driving the transition from the $s+d_1e^{i\phi_1}+d_2e^{i\phi_2}$ to the $s+id$ state. % system toward the $s+id$ state.
For example, at $\theta=36.87^\circ$ in Fig.~\ref{fig:PhaseDiagram}~(c), decreasing $\mu$ enlarges this Fermi loop, and drives the system from a topologically nontrivial $s+d_1e^{i\phi_1}+d_2e^{i\phi_2}$ state to its trivial counterpart and then to the $s+id$ state.
Upon further decreasing $\mu$, this loop disappears and the $s$-wave component is lost, eventually yielding the $d_1-d_2$ state.
By contrast, at $\theta=43.6^\circ$, the $s$-wave weight increases moderately before decreasing, allowing the system to remain in the nontrivial three-component regime until the $s$-wave component vanishes and the $d_1e^{i\phi_1}+d_2e^{i\phi_2}$ state emerges.
The evolution of the Fermi surface and $|\psi_s|/|\psi_d|$ with $\mu$ is analyzed in Sec.~VI~C in SM \cite{SM}. % the corresponding $s$-wave weight

Temperature provides a complementary control of the $s$-$d$ competition.
Since $T_c^s<T_c^d$, increasing $T$ suppresses the $s$-wave component more strongly than the $d$-wave component, % preferentially suppresses the $s$-wave component
driving the $s+id$ phase or the topologically trivial sector of the frustrated three-component state into the topologically nontrivial $s+d_1e^{i\phi_1}+d_2e^{i\phi_2}$ regime as $|\psi_s|/|\psi_d|$ decreases.
A further increase in temperature eventually suppresses $|\psi_s|$ altogether, leaving the $d_1e^{i\phi_1}+d_2e^{i\phi_2}$ state.
Thus, both $\mu$ and $T$ serve as control parameters for tuning the relative pairing strengths and switching among the distinct superconducting phases. % provide effective means of; in Fig.~\ref{fig:PhaseDiagram}.

%%%%%%%%%%%%%%%%%%%%%%%%%%%%%%%%%%%%%%%%%%%%%%%%%%%%%

\textit{Nematic Kerr signals.} --
The polar Kerr effect offers a characteristic experimental signature to distinguish the nematic $s+d_1e^{i\phi_1}+d_2e^{i\phi_2}$ state from $s+id$ and $d_1+e^{i\phi}d_2$. % the topologically trivial $s+i(d_1\pm d_2)$; $C_4$-preserving; states; smoking-gun
For the $s+id$ state, the macroscopic $\mathcal{T}C_4$ symmetry requires $\sigma_{xy}=\sigma_{yx}=0$, % the off-diagonal optical conductivities to vanish,
while for the $d_1+e^{i\phi}d_2$ state, the $GC_4$ symmetry enforces $\sigma_{xx}=\sigma_{yy}$ and $\sigma_{xy}+\sigma_{yx}=0$,
where $G$ is a global $\pi$-phase gauge transformation.
In contrast, the $s+d_1e^{i\phi_1}+d_2e^{i\phi_2}$ state simultaneously breaks both $\mathcal{T}$ and $C_4$ symmetries, generally lifting these restrictions and manifesting a polarization-resolved Kerr rotation with nematic anisotropy. % dependence on the incident light.

Crucially, the specific non-vanishing optical responses are dictated by residual symmetries of the phase configurations. % specific (Sec. II C in SM \cite{SM})
For $\phi_1+\phi_2=\pi$, the preserved symmetry $\mathcal{T}l_{x\pm y}$ % composite
% (where $l$ denotes interlayer twofold rotations, defined in Appendix~\ref{app:GLgeneral})
enforces $\sigma_{xx}=\sigma_{yy}$ but permits %a symmetric off-diagonal response, 
$\sigma_{xy}+\sigma_{yx}\neq0$.
Conversely, the $\phi_1+\phi_2=0$ mode preserves $\mathcal{T}l_{x,y}$, which demands $\sigma_{xy}+\sigma_{yx}=0$ while allowing $\sigma_{xx} \neq \sigma_{yy}$. % diagonal anisotropy; (see Sec.~VII~A in SM \cite{SM} for the derivation).

\begin{figure}[h]
\centering
%\hspace*{-0.36cm}
\begin{subfigure}{0.23\textwidth}
    \centering
    %\hspace*{-0.36cm}
    \includegraphics[width=\textwidth]{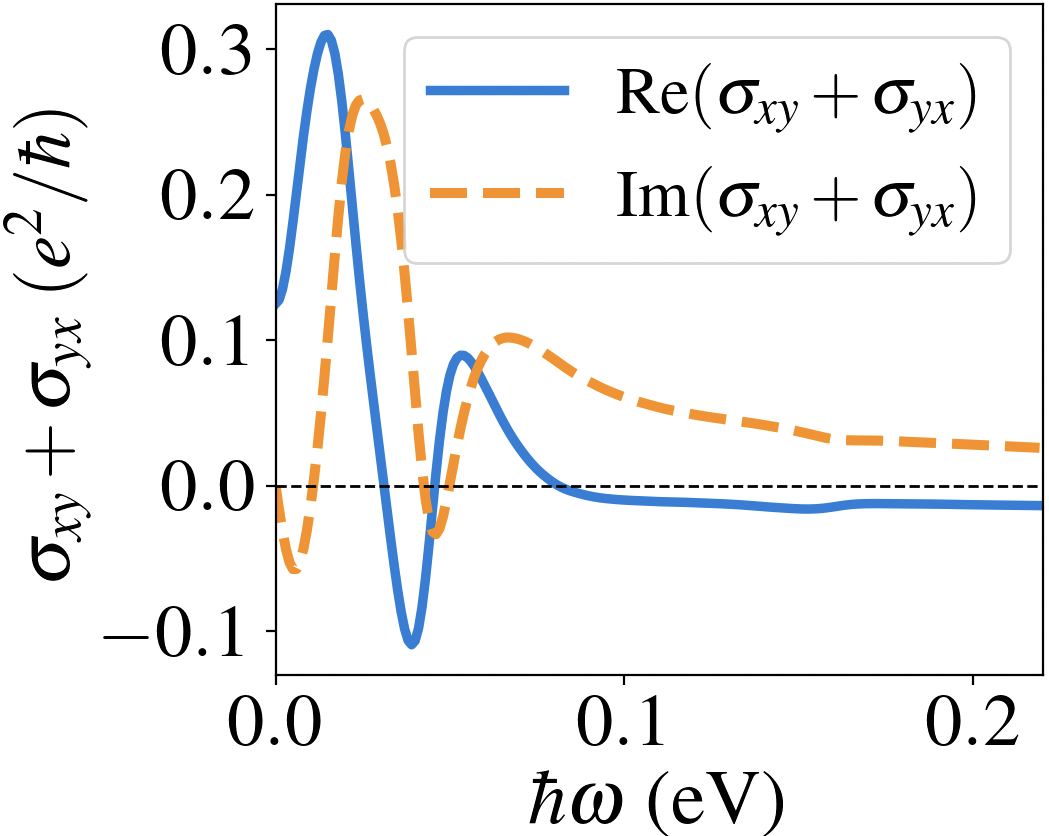} % vec2-5_g0_0.036_mu_-1.05t_M_40_T_1e-06. % 1049*836
    \caption{}
    \label{fig:xy+yx}
\end{subfigure}
\hfill
\begin{subfigure}{0.23\textwidth}
    \centering
    \includegraphics[width=\textwidth]{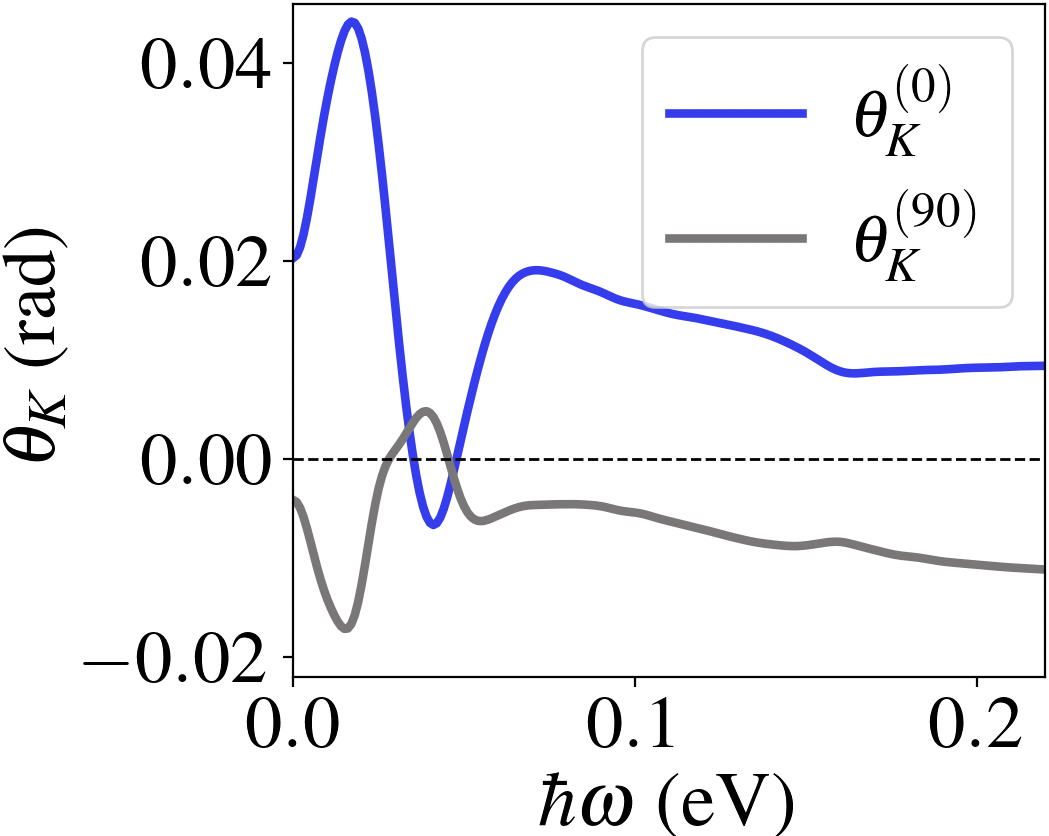} % vec2-5_g0_0.036_mu_-1.05t_M_40_T_1e-06   % 1049*837
    \caption{}
    \label{fig:thetaK}
\end{subfigure}
\captionsetup{justification=raggedright}
\caption{Nematic (a) optical conductivity and (b) Kerr rotation of the $s+d_1e^{i\phi_1}+d_2e^{i\phi_2}$ state at $\theta=43.6^\circ$.
% Solid and dashed lines show the real and imaginary parts, respectively. %for the case $\phi_{d_1}+\phi_{d_2}=0$ at twist angle $\theta=53.13^\circ$.
The parameters used to generate this figure are the same as those in Fig.~\ref{fig:sd1d2}~(a).}
\label{fig:Kerr}
\end{figure}

As concrete demonstrations, Fig.~\ref{fig:Kerr}~(a) plots $\sigma_{xy}+\sigma_{yx}$ for the $\phi_1+\phi_2=\pi$ case at $\theta=43.6^{\circ}$, % the non-vanishing
and Fig.~\ref{fig:Kerr}~(b) shows the resulting polarization-resolved Kerr rotations $\theta_K^{(0)}$, $\theta_K^{(90)}$,
where the superscripts denote two nematic orientations of the sample related by a $90^\circ$ rotation. % corresponding to the original orientation and a $90^\circ$ rotation of the sample, respectively.
Both Kerr rotations are finite and exhibit a sizable nematic splitting, with $\theta_K \sim \mathcal{O}(10)$~mrad and $|\theta_K^{(90)}-\theta_K^{(0)}|$ of the same order.
The corresponding results for the $\phi_1+\phi_2=0$ mode at $\theta=53.13^{\circ}$, including the diagonal anisotropy $\sigma_{xx}-\sigma_{yy}$ and the associated Kerr rotations, are presented in Sec.~VIII of the SM~\cite{SM}.
For the plotted window $\hbar\omega\lesssim0.2$~eV ($f\lesssim48.4$~THz), the optical penetration depth in cuprates is typically $\mathcal{O}(10^2)$~nm, sufficient to reach the twisted interface.
% Light in this range ($\hbar\omega \lesssim 0.2$~eV, $f\lesssim 48.4$~THz) typically penetrates $\mathcal{O}(10^2)$~nm into cuprate superconductors and thus readily reaches the twisted interface.
Taken together, the sizable optical anisotropy establishes a distinct fingerprint of the frustrated three-component superconducting phase.
% nonzero conductivity anisotropies and the polarization-resolved Kerr manifestation establish the nematic optical response as a distinct fingerprint of the frustrated three-component superconducting phase.

% The finite Kerr rotation which changes substantially under the $90^\circ$ rotation directly reflects the nematic optical response.
% The predicted Kerr angle reaches $\theta_K \simeq 10$s~mrad, with the rotation-induced change of the same order, % Quantitatively, our calculation predicts that
% establishing the sizable optical anisotropy as a distinct fingerprint of the frustrated three-component superconducting phase. % These sizable Kerr signals; explicitly nonzero

%-----------------------------------------------------------------------------------
%---------------------------------------------------------------------------------
\begin{comment}
\hspace{0.01cm}
% \hfill
\begin{subfigure}{0.23\textwidth}
    %\centering
    \hspace*{-0.36cm}
    \includegraphics[width=\textwidth]{yy-xx.png} % vec1-2_g0_0.032_mu_-1.0t_M_40_T_1e-06   % 1049*836
    \caption{$\theta=53.13^\circ$}
    \label{fig:yy-xx}
\end{subfigure}
\end{comment}
%----------------------------------------------------------------------------------

%%%%%%%%%%%%%%%%%%%%%%%%%%%%%%%%%%%%%%%%%%%%%

% \enlargethispage{\baselineskip}

\begin{figure}[h]
\hspace*{-0.6cm}
\includegraphics[width=0.36\textwidth]{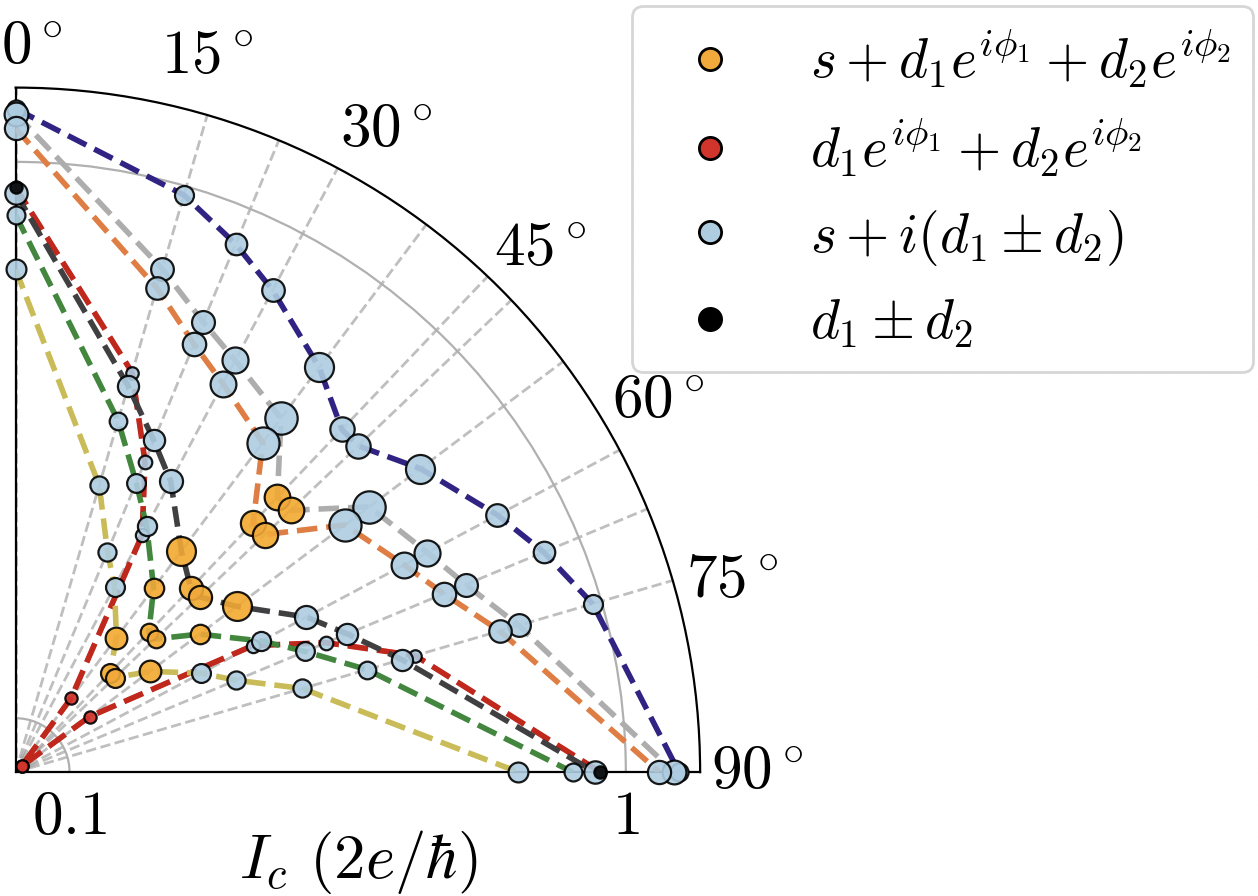} % 1258*896
\captionsetup{justification=raggedright}
\caption{% Josephson critical current at selected twist angles, with dashed radial lines marking those used in the self-consistent calculations. Colored dashed curves connect data obtained with the same model parameters.
Josephson critical current at different twist angles.
Dashed radial lines mark the twist angles used in the self-consistent calculations, while colored dashed curves connect data obtained with the same model parameters.
Marker size scales linearly with $|\psi_s|/|\psi_d|$.} % encodes; Marker size scales linearly with $|\psi_s|/|\psi_d|$. and color;  and the pairing state, respectively
\label{fig:Ic}
\end{figure}

\textit{Josephson critical current.} --
As a complementary diagnostic of the pairing state, we calculate the Josephson critical current $I_c$ at selected twist angles in Fig.~\ref{fig:Ic}.
For the $d_1+e^{i\phi}d_2$ state, symmetry suppresses the Josephson coupling near $\theta=45^\circ$, yielding $I_c(45^\circ)\approx 0$.
In contrast, a finite $s$-wave component lifts this suppression, with $I_c(45^\circ)/I_c(0^\circ)$ generally increasing with $|\psi_s|/|\psi_d|$. % the relative $s$-wave weight.
Importantly, the $s+d_1e^{i\phi_1}+d_2e^{i\phi_2}$ state retains a finite critical current near $45^\circ$, in agreement with the robust experimental response in this angular regime \cite{Zhu2021,HWang2023,Zhu2025,Zhu2023}. % experimentally observed robust response
% whose magnitude tracks its $s$-wave content but remains systematically larger than that of the $d_1e^{i\phi_1}+d_2e^{i\phi_2}$ state.
% Importantly, the $s+d_1e^{i\phi_1}+d_2e^{i\phi_2}$ state can retain a sizable critical current near $45^\circ$, owing to its substantial $s$-wave component.

 %---------------------------------------------------------------------------
%---------------------------------------------------------------------------

%%%%%%%%%%%%%%%%%%%%%%%%%%%%%%%%%%%%%%%%%%%

\textit{Conclusions.} --
In summary, twisted bilayer cuprates can host a stable topologically nontrivial $s+d_1e^{i\phi_1}+d_2e^{i\phi_2}$ state.
This state exhibits frustrated phase locking, spontaneous $\mathcal{T}$ and $C_4$ symmetry breaking, and can be identified through nematic Kerr anisotropy together with a sizable Josephson critical current near $45^\circ$.
Our results show that the system can possibly remain chiral and topological even in the presence of an $s$-wave pairing. % $s$-wave pairing is compatible with chiral topological superconductivity.

%%%%%%%%%%%%%%%%%%%%%%%%%%%%%%%%%%%%%%%%%%%%%%

\textit{Acknowledgments.} --
Y.-H. L. and W. Y. are supported by the Fundamental Research Funds for the Central Universities.
C. W. is supported by the National Natural Science Foundation of China under the Grants No. 12234016 and No.12174317.
This work has been supported by the New Cornerstone Science Foundation.

%%%%%%%%%%%%%%%%%%%%%%%%%%%%%%%%%%%%%%%%%%%%%%

%\makeatletter
%\renewcommand{\bibfont}{\normalsize}
%\renewcommand\NAT@biblabel[1]{[#1]}
%\makeatother

%%%%%%%%%%%%%%%%%%%%%%%%%%%%%%%%%%%%%%%%%%%%%%%%%%%%%%
%%%%%%%%%%%%%%%%%%%%%%%%%%%%%%%%%%%%%%%%%%%%%%%%%%%%%%

\begin{widetext}
\clearpage
\centerline{ {\Large \bf Supplementary Materials} }
\tableofcontents

\section{Analysis of superconducting instabilities}
\label{app:SCinstabilities}

\subsection{Derivation of linearized gap equations}
\label{app:derive_linearized_gap}
%---------------------------------------------------------------------------------------------------------------------
 % \subsection{Normal-state Hamiltonian}
 % \label{subsec:hk_0}
 
In twist bilayer square lattice, the real-space normal-state Hamiltonian incorporating intralayer hoppings, chemical potential, and interlayer tunneling is given by
\begin{equation}
H_0
=-t\sum_{\langle ij \rangle \sigma l}\left(c_{i\sigma l}^\dagger c_{j\sigma l}+h.c.\right)
-t^\prime \sum_{\langle \langle ij \rangle \rangle \sigma l}\left(c_{i\sigma l}^\dagger c_{j\sigma l}+h.c.\right)
-\mu \sum_{i\sigma l}n_{i\sigma l}
-\sum_{ij\sigma}\left(g_{ij}c_{i\sigma 1}^\dagger c_{j\sigma 2}+h.c.\right),
\label{eq:H0}
\end{equation}
where $l=1,2$ are layer indices, $t$, $t^\prime$ denote the nearest-neighbor (NN) and next-nearest-neighbor (NNN) hoppings within each layer, $\mu$ is the chemical potential.
$g_{ij}$ is the interlayer tunneling modeled as
\begin{equation}
g_{ij}=g_0\exp{\left[-\left(\sqrt{r_{ij}^2+d^2}-d \right)/\rho\right]},
\label{eq:gij}
\end{equation}
in which $g_0$ sets the interlayer tunneling strength,
$r_{ij}$ is the in-plane separation between sites $i$ and $j$ on different layers, $d$ is the interlayer spacing, and $\rho$ characterizes the spatial decay length of the interlayer tunneling associated with the extended nature of the Cu $4s$ orbital.

 The twist is incorporated into the bilayer geometry through a twist vector $\boldsymbol{v}=(m,n)$, and the twist angle is
\begin{equation}
\theta=2\arctan\left(m/n\right),
\end{equation}
where $m$, $n$ are integers.
The two layers are rotated by $\pm \theta/2$ around the $c$-axis, respectively,
and the corresponding moir\'e unit cell contains $2q=2(m^2+n^2)$ lattice sites in total, with $q$ sites in each layer \cite{Can2021SM}.

We define the Nambu spinor as
\begin{equation}
\Psi_{\boldsymbol{k}}
=
\left(
c_{\boldsymbol{k}\uparrow 1}^{(0)},
c_{-\boldsymbol{k}\downarrow 1}^{(0)\dagger},
\cdots,
c_{\boldsymbol{k}\uparrow 2}^{(2q-1)},
c_{-\boldsymbol{k}\downarrow 2}^{(2q-1)\dagger}
\right)^T,
\label{eq:Psi_a}
\end{equation}
where the superscripts label the sites within the moir\'e unit cell: $0,\cdots,q-1$ correspond to sites on layer 1, while $q,\cdots,2q-1$ correspond to sites on layer 2. 
Eq.~\eqref{eq:H0} can be rewritten as
\begin{equation}
H_0=\sum_{\boldsymbol{k}}\Psi_{\boldsymbol{k}}^\dagger h_{\boldsymbol{k}}^{\Delta=0}\Psi_{\boldsymbol{k}},
\end{equation}
where $h_{\boldsymbol{k}}^{\Delta=0}$ is the Bogoliubov--de Gennes (BdG) matrix obtained by Fourier transforming Eq. \eqref{eq:H0},% with the pairing terms set to zero. 
which contains both the electron and hole sectors. 
From this matrix, we extract the particle block $h_{\boldsymbol{k}}^0$, which represents the multiband normal-state noninteracting Hamiltonian.
%------------------------------------------------------------------------

To analyze superconducting instabilities in the twisted bilayer square-lattice system, 
we consider the linearized gap equation in matrix form in the combined layer and moir\'e-unit-cell site space, 
which reduces the pairing problem to an eigenvalue equation:
\begin{equation}
\left[\hat{\Delta}(\boldsymbol{k}^\prime) \right]_{ab}
=
\frac{T}{L^d}\sum_{k} \sum_{cd}
V_{abcd}(\boldsymbol{k}^\prime,\boldsymbol{k})
\left[
\hat{\mathcal{G}}^0_{k}\,
\hat{\Delta}(\boldsymbol{k})\,
\hat{\mathcal{G}}^{0,T}_{-k}
\right]_{cd},
\label{eq:Delta_kernel}
\end{equation}
where $k=(\boldsymbol{k},i\omega_n)$, $\omega_n=(2n+1)\pi/\beta$ is the fermionic Matsubara frequency, 
$\beta=1/T$, and $n\in\mathbb{Z}$. 
The $2q\times 2q$ matrix Green's function is given by
\begin{equation}
\hat{\mathcal{G}}^0_k=
\left(i\omega_n \hat{I}-h_{\boldsymbol{k}}^0\right)^{-1},
\label{eq:G0}
\end{equation}
where the matrix indices run over the internal layer and site degrees of freedom within the moir\'e unit cell. 
The effective pairing interaction $V(\boldsymbol{k}^\prime,\boldsymbol{k})$ acts as a kernel in the combined layer and moir\'e-unit-cell site space,
and the product in Eq.~\eqref{eq:Delta_kernel} involves summation over the corresponding internal indices.
To resolve the pairing symmetry, we expand the interaction in different channels $\delta$,
\begin{equation}
V_{abcd}(\boldsymbol{k}^\prime,\boldsymbol{k})
=
\sum_{\delta}
V_{\delta}\,
\left[\hat{\phi}_{\delta}(\boldsymbol{k}^\prime) \right]_{ab}\
\left[\hat{\phi}_{\delta}^{\dagger}(\boldsymbol{k}) \right]_{dc}
\label{eq:Vkkprime}
\end{equation}
where $V_{\delta}>0$ is the effective attraction in channel $\delta$, 
and $\hat{\phi}_{\delta}(\boldsymbol{k})$ specifies the corresponding pairing form factor in the combined layer and moir\'e-unit-cell site space.

To solve Eq.~\eqref{eq:Delta_kernel}, we expand the  gap matrix $\hat{\Delta}(\boldsymbol{k^\prime})$ in the basis of pairing form factors as
\begin{equation}
\left[\hat{\Delta}(\boldsymbol{k^\prime}) \right]_{ab}=\sum_\beta \Delta_\beta \left[\hat{\phi}_\beta(\boldsymbol{k^\prime}) \right]_{ab},
\label{eq:Delta_beta}
\end{equation}
where $\Delta_\beta$ denotes the order-parameter amplitude in pairing channel $\beta$.
Substituting Eqs.~\eqref{eq:Vkkprime} and \eqref{eq:Delta_beta} into Eq.~\eqref{eq:Delta_kernel}, one finds that the contraction over the internal matrix indices can be written as a trace, yielding
\begin{equation}
\sum_\beta \Delta_\beta \left[\hat{\phi}_\beta(\boldsymbol{k}^\prime) \right]_{ab} =\frac{T}{L^d} \sum_{k,\delta,\gamma}V_{\delta} \left[\hat{\phi}_\delta(\boldsymbol{k}^\prime) \right]_{ab}
\mathrm{Tr} \left[ \hat{\phi}_\delta^{\dagger}(\boldsymbol{k}) \hat{\mathcal{G}}^0_{k} \Delta_\gamma \hat{\phi}_\gamma(\boldsymbol{k}) \hat{\mathcal{G}}^{0,T}_{-k} \right].
\label{eq:Delta_Tr}
\end{equation}
To isolate a particular pairing channel $\alpha$, we multiply both sides of Eq.~\eqref{eq:Delta_Tr} by $[\hat{\phi}_\alpha^\dagger(\boldsymbol{k}^\prime)]_{ba}$ and sum over $\boldsymbol{k}^\prime$ and indices $a$, $b$.
Using the orthogonality of basis matrices associated with different pairing channels, the projection removes all contributions except those in the $\alpha$ channel. 
The gap equation decouples into separate equations for different pairing symmetries, and for channel $\alpha$ we obtain
\begin{equation}
\Delta_{\alpha}
=
\frac{T}{L^d} \sum_k
V_\alpha \Delta_\alpha
\mathrm{Tr}
\left[
\hat{\phi}_\alpha^{\dagger}(\boldsymbol{k})
\hat{\mathcal{G}}^0_{k}
\hat{\phi}_\alpha(\boldsymbol{k})
\hat{\mathcal{G}}^{0,T}_{-k}
\right]
\end{equation}
which reduces to 
$1=V_\alpha \Pi_{\alpha}(T_c^\alpha)$
when approaching the superconducting transition temperature.
$\Pi_{\alpha}(T)$ is the pairing susceptibility in channel $\alpha$ given by
\begin{equation}
\Pi_{\alpha}(T)=\frac{T}{L^d} \sum_k \mathrm{Tr} \left[ \hat{\phi}_\alpha^{\dagger}(\boldsymbol{k}) \hat{\mathcal{G}}^0_{k} \hat{\phi}_\alpha(\boldsymbol{k}) \hat{\mathcal{G}}^{0,T}_{-k} \right].
\label{eq:PiT}
\end{equation}
By transforming to the band basis and evaluating the trace in Eq.~\eqref{eq:PiT}, one obtains
\begin{equation}
\Pi_{\alpha}(T)=\frac{T}{L^d} \sum_{k,ij} \left|\hat{\Phi}_\alpha^{ij}(\boldsymbol{k})\right|^2 \frac{1}{i\omega_n-E^i_{\boldsymbol{k}}} \frac{1}{-i\omega_n-E^j_{\boldsymbol{k}}},
\label{eq:PiTij}
\end{equation}
where $\hat{\Phi}_\alpha(\boldsymbol{k})=U_{\boldsymbol{k}}^{0\dagger} \hat{\phi}_\alpha(\boldsymbol{k}) U^0_{\boldsymbol{k}}$,
in which $U^0_{\boldsymbol{k}}$ is the unitary matrix that diagonalizes $h_{\boldsymbol{k}}^0$ as $U_{\boldsymbol{k}}^{0\dagger} h_{\boldsymbol{k}}^0 U^0_{\boldsymbol{k}}=\hat{E}^0_{\boldsymbol{k}}$,
and $i$, $j$ label the normal state bands.
Carrying out the Matsubara-frequency summation in Eq.~\eqref{eq:PiTij}, we finally arrive at
\begin{equation}
\Pi_{\alpha}(T)=\sum_{\boldsymbol{k},ij} \frac{\left|\hat{\Phi}_\alpha^{ij}(\boldsymbol{k})\right|^2}{L^d} \frac{\tanh{\left(\hat{E}^{0i}_{\boldsymbol{k}}/2T\right)}+\tanh{\left(\hat{E}^{0j}_{\boldsymbol{k}}/2T\right)}}{2(\hat{E}^{0i}_{\boldsymbol{k}}+\hat{E}^{0j}_{\boldsymbol{k}})}.
\label{eq:PiTc}
\end{equation}

%%%%%%%%%%%%%%%%%%%%%%%%%%%%%%%%%%%%%%%%%%%%%%%%%%%%%%

\subsection{Critical temperatures of different pairing channels}
\label{app:Tc_alpha}

Our primary focus in the main text is the relative sensitivity of each pairing channel to the interlayer tunneling $g_0$, which can be characterized by $T_c^\alpha(g_0)/T_c^\alpha(0)$ without assuming identical pairing interactions among different channels.
To further compare the absolute critical temperatures across different channels on an equal footing, here, we take a constant effective pairing interaction, $V_\alpha=V$ ($\alpha=s,d_1,d_2$).
This choice serves only as a common reference for direct comparison and does not imply identical microscopic pairing interactions among different channels in real materials.

 %---------------------------------------------------------------------------
\begin{figure}[h]
%\centering
%\begin{subfigure}{0.235\textwidth}
%    \centering
%    \includegraphics[width=\textwidth]{Tc_g0_vec12.png}
%    \caption{}
%    \label{fig:subTc1}
%\end{subfigure}
%\hfill
%\begin{subfigure}{0.235\textwidth}
\hspace*{-0.6cm}
\includegraphics[width=0.3\textwidth]{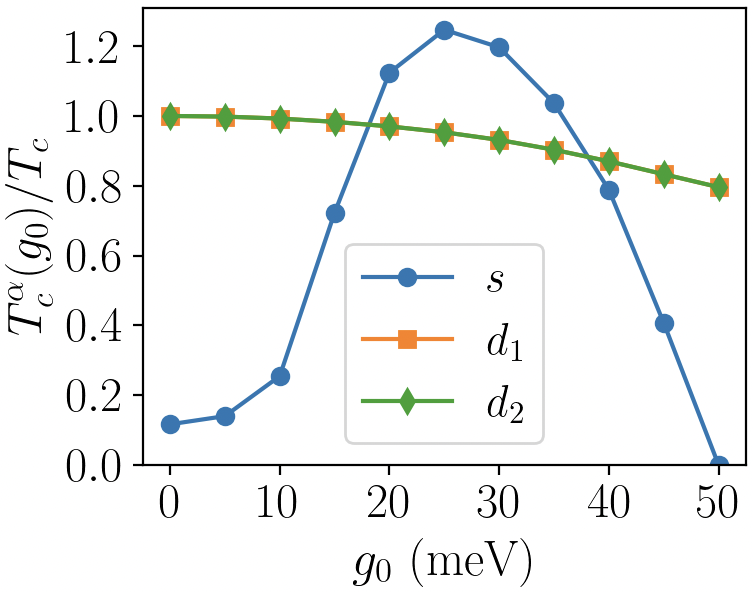}
\captionsetup{justification=raggedright}
\caption{
$T_c^\alpha(g_0)/T_c$ as functions of $g_0$, where $\alpha=s,d_1,d_2$, 
$T_c^\alpha(g_0)$ denotes the superconducting transition temperature in the $\alpha$ channel at interlayer tunneling strength $g_0$, and $T_c$ is the critical temperature for monolayer superconducting cuprate.
The parameters in the Hamiltonian are chosen as $t=0.153$~eV, $t^\prime=-0.45t$, $\mu=-1.3t$, $d=2.22a$, and $\rho=0.39a$, where $a$ is the lattice constant of the square lattice, with $V=0.146$~eV and $\theta=53.13^\circ$.}
\label{fig:Tc_g0_sd}
\end{figure}
%---------------------------------------------------------------------------

Fig.~\ref{fig:Tc_g0_sd} illustrates the resulting critical temperatures of the $s$-, $d_1$-, and $d_2$-wave components as functions of $g_0$.
In the decoupled limit ($g_0=0$), $T_c^s \simeq T_c^d/9$, consistent with the $d_{x^2-y^2}$-wave dominance in isolated monolayer cuprate superconductors.
However, upon introducing the interlayer coupling, $T_c^d$ is lightly suppressed, whereas $T_c^s$ undergoes a rapid enhancement.
Although $T_c^s$  drops for $g_0 > 30$~meV and vanishes around $g_0=50$~meV in the chosen parameter regime, $T_c^s$ and $T_c^d$ become comparable over the range $15$~meV$<g_0<45$~meV.
Furthermore, within the range $16$~meV$<g_0<36$~meV, $T_c^s$ is even larger than $T_c^d$.
% becoming comparable to $T_c^d$ over the range $15$~meV$<g_0<45$~meV$ and even exceeding $T_c^d$ within the range $16$~meV$<g_0<36$~meV$. At larger $g_0$, $T_c^s$ decreases and eventually vanishes near $g_0=50~\mathrm{meV}$.
This regime of comparable critical temperatures suggests the possible emergence of a mixed pairing state in which the $s$-, $d_1$-, and $d_2$-wave symmetries coexist.

%%%%%%%%%%%%%%%%%%%%%%%%%%%%%%%%%%%%%%%%%%%%%%%%%%%%%%

%---------------------------------------------------------------------------
\begin{figure}[h]
\centering
\begin{subfigure}{0.22\textwidth}
    \centering
    \includegraphics[width=\textwidth]{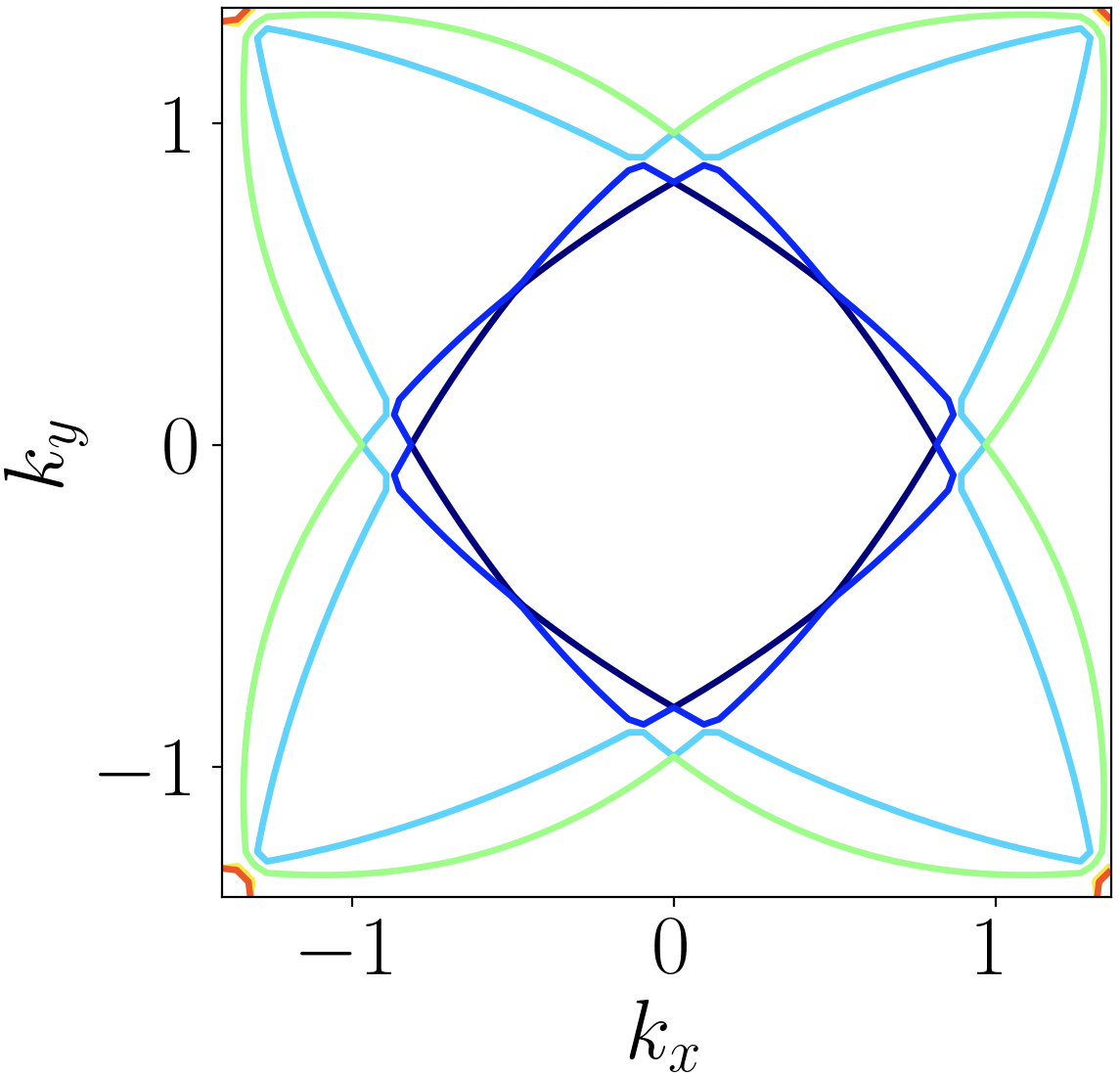} % 1145*1100
    \caption{$g_0=0$.}
    \label{fig:FS_g0}
\end{subfigure}
\hspace{0.06\textwidth}
\begin{subfigure}{0.22\textwidth}
    \centering
    \includegraphics[width=\textwidth]{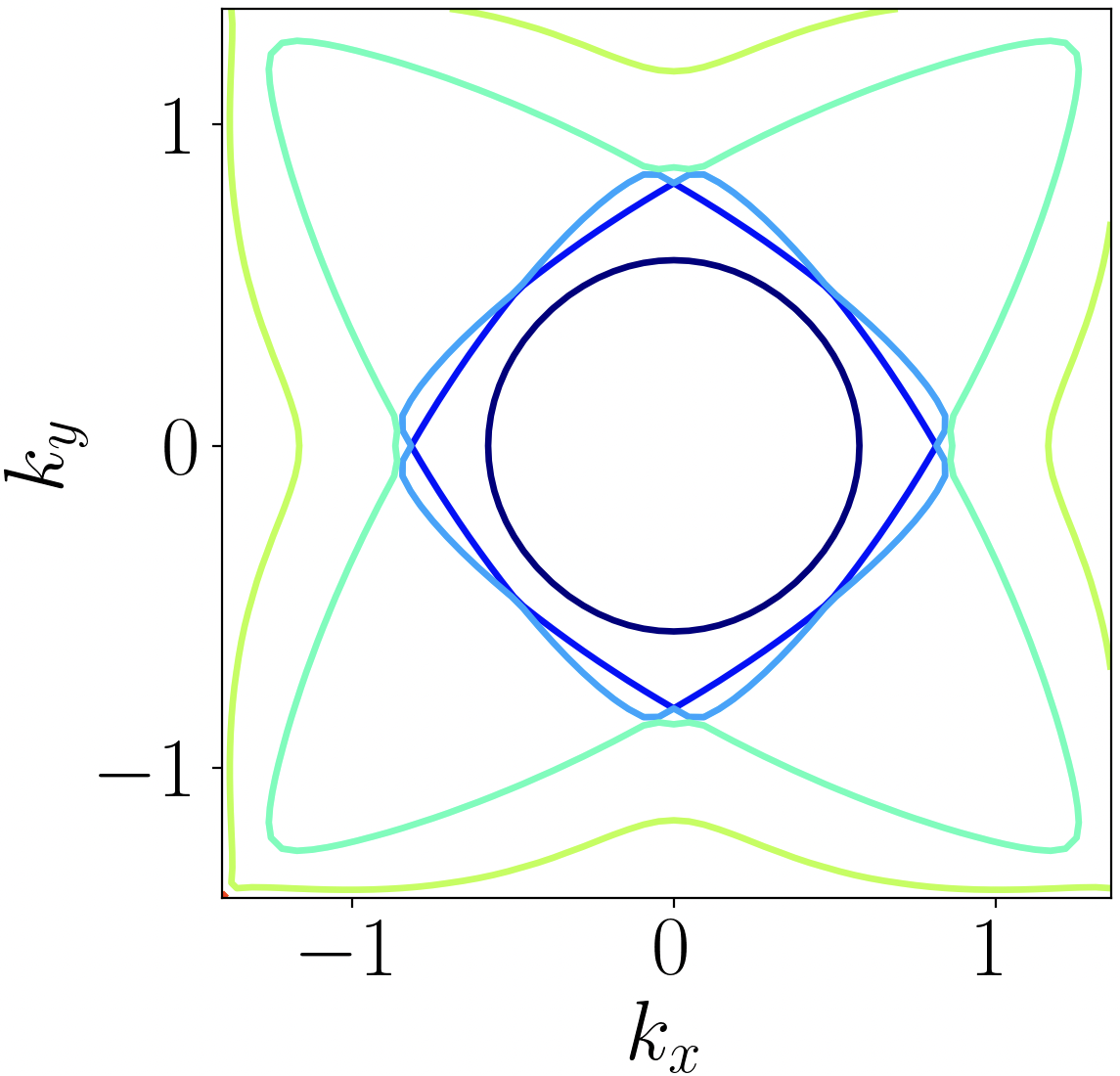} % 1145*1100
    \caption{$g_0=0.025$~eV.}
    \label{fig:FS_g025}
\end{subfigure}
\captionsetup{justification=raggedright}
\caption{Normal-state Fermi surfaces in the first moiré Brillouin zone for (a) $g_0=0$ and (b) $g_0=0.025$~eV.
An additional annular Fermi surface centered at the $\Gamma$ point emerges at panel (b).
The parameters are $\theta=53.13^{\circ}$, $t=0.153$~eV, $t^{\prime}=-0.45t$, $\mu=-1.3t$, $d=2.22a$, and $\rho=0.39a$, where a is the Cu-Cu lattice constant.}
\label{fig:FS}
\end{figure}
%---------------------------------------------------------------------------

%---------------------------------------------------------------------------
\begin{figure}[h]
%\centering
\hspace*{-0.6cm}
\includegraphics[width=0.8\textwidth]{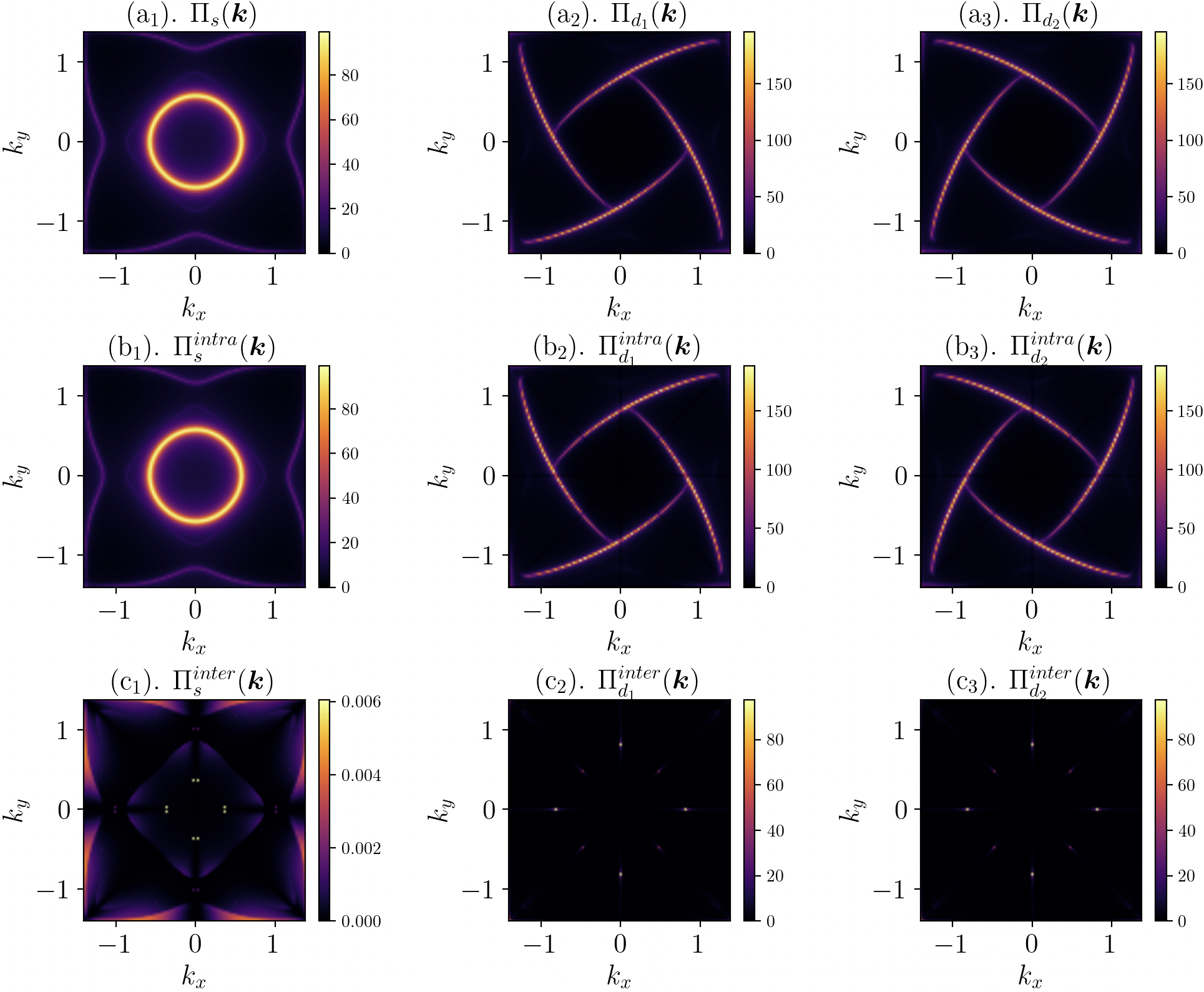}
\captionsetup{justification=raggedright}
\caption{Momentum-resolved pairing susceptibilities for the $s$-, $d_1$-, and $d_2$-wave channels at $g_0=0.025$~eV.
Panels (a)--(c) show the total, intraband, and interband contributions, respectively.
The susceptibility is predominantly intraband, while interband weight is confined to isolated Fermi-surface crossings.
Notably, the $\Gamma$-centered annular pocket carries substantial $s$-wave but negligible $d_{1,2}$-wave weight.
The parameters are the same as those in Fig.~\ref{fig:FS}.}
\label{fig:Pi_g025}
\end{figure}
%---------------------------------------------------------------------------

\subsection{Physical origin of the s-wave enhancement}
\label{s_enhancement}

To clarify the origin of the enhancement of the $s$-wave channel, we first examine the evolution of the normal-state Fermi surface with interlayer tunneling $g_0$. 
Fig.~\ref{fig:FS} compares the Fermi surfaces at $g_0=0$ and $g_0=0.025$~eV, 
the latter corresponding to the maximum of $T_c^s$ in Fig.~2 of the main text and Fig.~\ref{fig:Tc_g0_sd} here .
At $g_0=0.025$~eV, an additional annular Fermi surface emerges around the $\Gamma$ point, providing a new set of low-energy states for Cooper pairing.

The role of this additional Fermi surface can be seen from the momentum-resolved pairing susceptibility,
\begin{equation}
\Pi_{\alpha}(\boldsymbol{k})=\frac{T}{L^d} \sum_{\omega_n} \mathrm{Tr} \left[ \hat{\phi}_\alpha^{\dagger}(\boldsymbol{k}) \hat{\mathcal{G}}^0_{k} \hat{\phi}_\alpha(\boldsymbol{k}) \hat{\mathcal{G}}^{0,T}_{-k} \right] = \sum_{ij} \left|\hat{\Phi}_\alpha^{ij}(\boldsymbol{k})\right|^2 K_T^{ij},
\end{equation}
where $\hat{\Phi}_\alpha(\boldsymbol{k})=U_{\boldsymbol{k}}^{0\dagger} \hat{\phi}_\alpha(\boldsymbol{k}) U^0_{\boldsymbol{k}}$ is the pairing vertex projected onto the normal-state bands, and
\begin{equation}
K_T^{ij} = \frac{\tanh{\left(\hat{E}^{0i}_{\boldsymbol{k}}/2T\right)}+\tanh{\left(\hat{E}^{0j}_{\boldsymbol{k}}/2T\right)}}{2L^d(\hat{E}^{0i}_{\boldsymbol{k}}+\hat{E}^{0j}_{\boldsymbol{k}})}
\end{equation}
denotes the channel-independent Cooper kernel.
Therefore, for a given normal-state band structure, the channel dependence of $\Pi_{\alpha}(\boldsymbol{k})$ is encoded in $|\hat{\Phi}_\alpha(\boldsymbol{k})|^2$,
while $K_T$ primarily select states near the Fermi surface.
Fig.~\ref{fig:Pi_g025} shows the momentum-resolved susceptibilities for the $s$-, $d_1$-, and $d_2$-wave channels.
The newly formed $\Gamma$-centered annular Fermi surface carries a pronounced $s$-wave response, whereas the $d_{1,2}$-wave responses are strongly suppressed.
A decomposition into intraband and interband contributions further shows that $\Pi_{\alpha}(\boldsymbol{k})$ is dominated by intraband terms.
 
Motivated by the dominant intraband contribution, we thus examine the intraband pairing vertex $|\hat{\Phi}^{ii}_\alpha(\boldsymbol{k})|^2$ close to the Fermi surface.
In particular, $i=1$ denotes the band forming the newly emerged $\Gamma$-centered annular Fermi surface.
As shown in Fig.~\ref{fig:Phi11}, the $s$-wave intraband pairing weight is sizable and nearly uniform along the annular Fermi surface.
In contrast, the $d_{1,2}$-wave pairing weights exhibit the characteristic modulation of their respective form factors and are suppressed by more than two orders of magnitude relative to the $s$-wave weight.
The same contrast is evident from their angular dependence along the Fermi surface. % contour.

%---------------------------------------------------------------------------
\begin{figure}[h]
\centering
\begin{subfigure}{0.8\textwidth}
    % \centering
    \hspace*{0.36cm}
    \includegraphics[width=\textwidth]{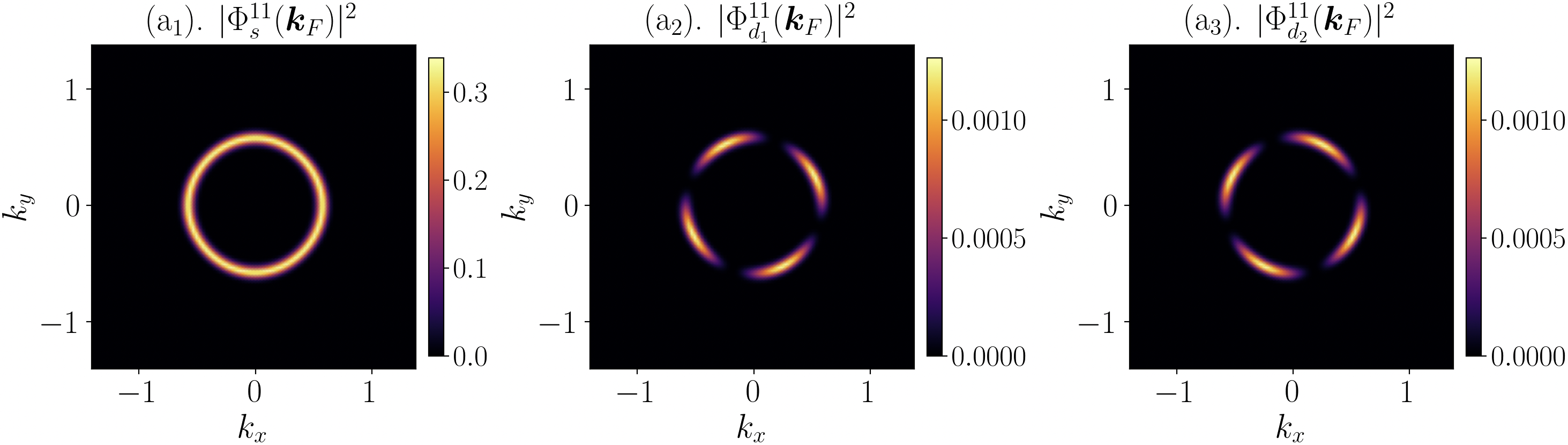} % 2926*830
    \label{subfig:11k}
\end{subfigure}
\hspace{0.06\textwidth}
\begin{subfigure}{0.7\textwidth}
    \centering
    \includegraphics[width=\textwidth]{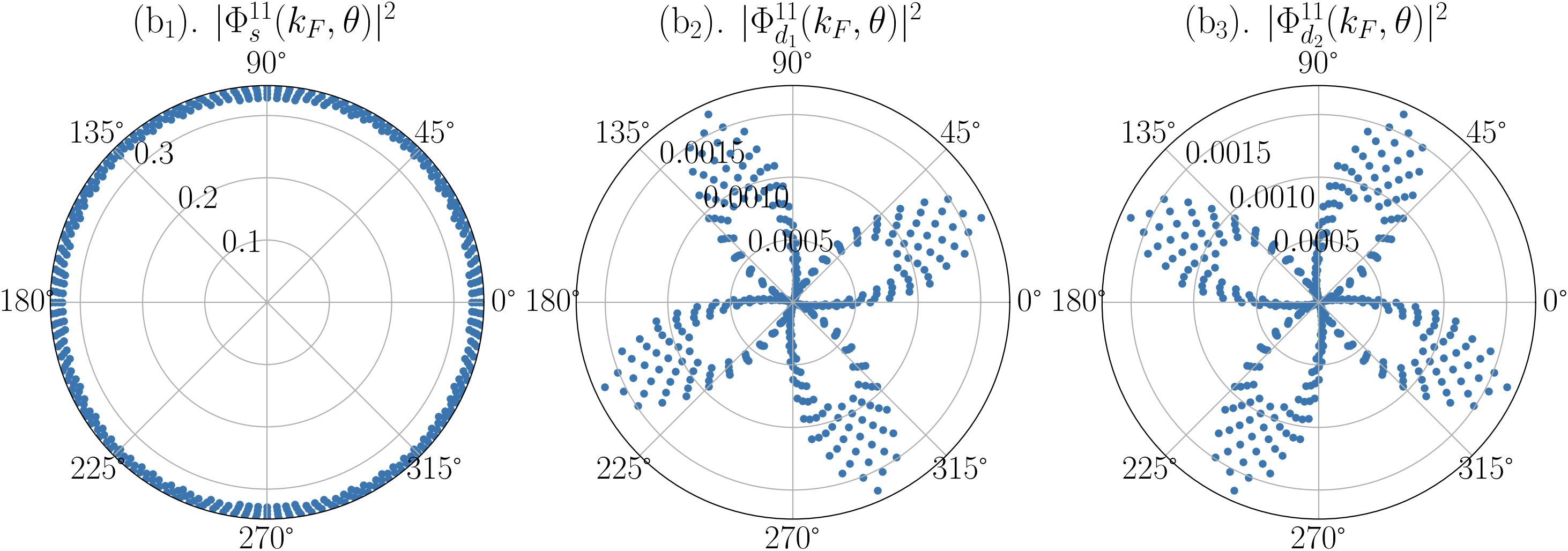} % 2925*1198
    \label{subfig:11theta}
\end{subfigure}
\captionsetup{justification=raggedright}
\caption{Intraband pairing vertices for the $s$-, $d_1$-, and $d_2$-wave channels on band 1, which forms the $\Gamma$-centered annular Fermi surface.
Panel (a) shows their momentum-space distributions, while panel (b) shows the angular dependence at fixed $|\boldsymbol{k}|=k_F$.
The $s$-wave weight is substantially larger than the $d_{1,2}$-wave weights, whose fourfold angular patterns reflect the corresponding, mutually rotated $d_{x^2-y^2}$-wave form factors.}
\label{fig:Phi11}
\end{figure}
%---------------------------------------------------------------------------

The strong suppression of the $d$-wave pairing vertex near $\Gamma$-point has a symmetry origin.
The normal-state Hamiltonian satisfies the $C_4$ covariance relation,
\begin{equation}
h^0(C_4\boldsymbol{k})=\mathcal{C}_4(\boldsymbol{k}) h^0(\boldsymbol{k}) \mathcal{C}_4^{\dagger}(\boldsymbol{k}),
\end{equation}
where $\mathcal{C}_4(\boldsymbol{k})$ represents the four-fold rotation in the multiband Bloch basis.
Since $\Gamma$ is invariant under $C_4$, namely $C_4\Gamma=\Gamma$, one has
\begin{equation}
\left[h^0(\Gamma), \mathcal{C}_4(\Gamma) \right]=0.
\end{equation}
For band $i$ crossing the $\Gamma$-point,
\begin{equation}
h^0(\Gamma)U_i(\Gamma)=E_i(\Gamma)U_i(\Gamma).
\end{equation}
Applying $\mathcal{C}_4(\Gamma)$ gives
\begin{align}
h^0(\Gamma) \left[\mathcal{C}_4(\Gamma) U_i(\Gamma)\right] = \mathcal{C}_4(\Gamma) \left[h^0(\Gamma) U_i(\Gamma)\right] = \mathcal{C}_4(\Gamma) \left[E_i(\Gamma) U_i(\Gamma)\right] = E_i(\Gamma) \left[\mathcal{C}_4(\Gamma) U_i(\Gamma)\right].
\end{align}
Therefore, $\mathcal{C}_4(\Gamma)U_i(\Gamma)$ is also an eigenstate of $h^0(\Gamma)$ with the same eigenvalue.
For a nondegenerate band at $\Gamma$, the corresponding Bloch eigenstate can be chosen as a $\mathcal{C}_4$ eigenstate,
\begin{equation}
\mathcal{C}_4(\Gamma)U_i(\Gamma) = \xi_i U_i(\Gamma), \quad \xi_i=\pm 1.
\end{equation}

The $s$- and $d$-wave pairing form factors transform differently under $C_4$, at $\Gamma$-point,
\begin{equation}
\hat{\phi}_{\alpha}(C_4\Gamma)=\hat{\phi}_{\alpha}(\Gamma)=\chi_{\alpha} \mathcal{C}_4(\Gamma) \hat{\phi}_{\alpha}(\Gamma) \mathcal{C}_4^{\dagger}(\Gamma),
\end{equation}
in which $\chi_s=1$ and $\chi_d=-1$.
The intraband pairing vertex satisfies
\begin{align}
\chi_{\alpha} \Phi^{ii}_{\alpha}(\Gamma) = U_i^{\dagger}(\Gamma) \chi_{\alpha} \hat{\phi}_{\alpha}(\Gamma) U_i(\Gamma) = U_i^{\dagger}(\Gamma) \mathcal{C}_4^{\dagger}(\Gamma) \hat{\phi}_{\alpha}(\Gamma) \mathcal{C}_4(\Gamma) U_i(\Gamma) = \xi_i^*U_i^{\dagger}(\Gamma) \hat{\phi}_{\alpha}(\Gamma) \xi_i U_i(\Gamma) = \Phi^{ii}_{\alpha}(\Gamma).
\end{align}
Consequently, the intraband $d$-wave pairing vertex satisfies $\Phi^{ii}_d(\Gamma)=-\Phi^{ii}_d(\Gamma)=0$,
while the $s$-wave vertex is not symmetry forbidden.
Furthermore, because both $s$- and $d$-wave pairings are even in momentum, the expansion around $\Gamma$ contains no linear term.
The leading $s$- and $d$-wave contributions therefore take the form
\begin{align}
\Phi_s^{ii}(\boldsymbol{k})&=\Phi_s^{ii}(0)+\mathcal{O}(k^2), \notag\\
\Phi_d^{ii}(\boldsymbol{k})&=A(k_x^2-k_y^2)+B(k_xk_y)+\mathcal{O}(k^4).
\end{align}
Hence, on a small $\Gamma$-centered Fermi surface, $|\Phi_d^{ii}|^2 \sim k_F^4$, whereas the $s$-wave pairing weight remains finite.
This symmetry-imposed suppression of the $d$-wave intraband vertex allows the newly emerged annular Fermi surface to contribute predominantly to the $s$-wave pairing susceptibility and provides a microscopic explanation for the strong enhancement of $T_c^s$ with introducing interlayer tunneling.

%%%%%%%%%%%%%%%%%%%%%%%%%%%%%%%%%%%%%%%%%%%%%%%%%%%%%%

\section{Ginzburg-Landau free energies}

%\begin{widetext}
\subsection{Four-component-pairing GL free energy with $s_1$-, $s_2$-, $d_1$-, and $d_2$-wave symmetries at a $45^\circ$ twist}
\label{app:GL45}
%---------------------------------------------------------------------------
\begin{figure}[h]
\centering
\includegraphics[width=0.29\textwidth]{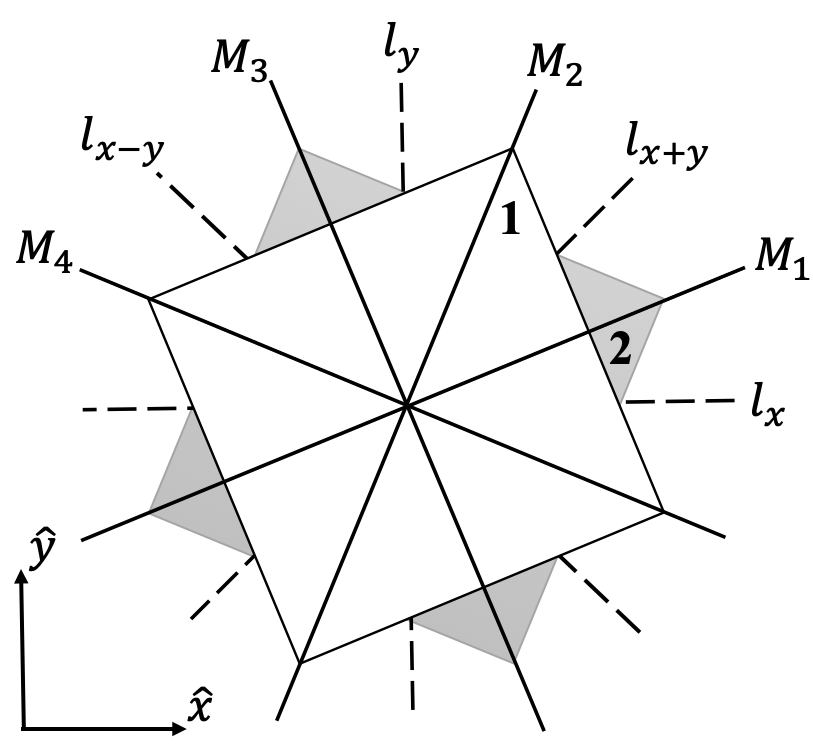}
\captionsetup{justification=raggedright}
\caption{Schematic plot of the $D_{4d}$ group of the twisted bilayer square lattice for twist angle $\theta=45^\circ$.
The $\hat{x}$ and $\hat{y}$-axes are defined as the two nearest-neighbor directions of the original untwisted lattice.
Layer 1 is rotated by $22.5^\circ$ around the $c$-axis, while layer 2 is rotated by $-22.5^\circ$.}
\label{fig:D4dgroup}
\end{figure}
%---------------------------------------------------------------------------
When the twist angle between the bilayer is $45^\circ$, the symmetry point group is $D_{4d}$, as shown in Fig.~\ref{fig:D4dgroup}.
It consists of four $90^\circ$-rotation operations around the $c$-axis, four $180^\circ$-rotation operations that interchange two layers, four mirror reflection operations, and four $S_8$ operations.
The intralayer rotations around the $c$-axis are represented as $C_4$, $C_2$, and $C_4^3$, corresponding to rotations around the $c$-axis by angles $90^\circ$, $180^\circ$, and $270^\circ$.
The interlayer rotations are denoted by $l_x$, $l_{x+y}$, $l_y$, and $l_{x-y}$, whose rotation axes lie in the plane midway between the two cuprate layers and point along the $\hat{x}$, $\hat{x}+\hat{y}$, $\hat{y}$ and $\hat{x}-\hat{y}$ directions, respectively.
The mirror reflection planes are determined by the planes spanned by the $c$-axis and the solid lines $M_1$, $M_2$, $M_3$, and $M_4$ in Fig.~\ref{fig:D4dgroup}.
Each $S_8$ operation consists of a $45^\circ$ rotation around the $c$-axis followed by inversion through the center.

We note that the two $d_{x^2-y^2}$-wave order parameters residing in each monolayer, $d_1, d_2$, constitude the $E_2$-representation of $D_{4d}$.
In addition, the symmetric combination of the $s$-wave components from the two layers, $s_1+s_2$, forms the $A_1$-representation,
while the antisymmetric combination, $s_1-s_2$, belongs to the $B_2$-representation of $D_{4d}$.

The terms in the free energy up to quadratic level are
\begin{align}
f_{1,2}^{(2)}&=|\psi_{s1} \pm \psi_{s2}|^2=\left(|\psi_{s1}|^2+|\psi_{s2}|^2\right) \pm \left(\psi_{s1}\psi_{s2}^*+\psi_{s1}^*\psi_{s2} \right),
\end{align}
and
\begin{equation}
f_3^{(2)}=|\psi_{d1}|^2+|\psi_{d2}|^2,
\end{equation}
where $\psi_{s1}$, $\psi_{s2}$ are the $s$-wave complex order parameters in layer 1 and layer 2,
and $\psi_{d1}$, $\psi_{d2}$ are the $d_{x^2-y^2}$-wave complex order parameters in layer 1 and layer 2, respectively.
The independent quadratic terms in free energy are
\begin{align}
f^{(2)}=\alpha_{s_{1,2}}\left(|\psi_{s1}|^2+|\psi_{s2}|^2\right)+k_{s_{1,2}}^{(2)}\left(\psi_{s1}\psi_{s2}^*+\psi_{s1}^*\psi_{s2} \right)+\alpha_{d_{1,2}}\left(|\psi_{d1}|^2+|\psi_{d2}|^2 \right).
\end{align}

Up to the quartic terms, the extra $D_{4d}$-invariant terms are
\begin{align}
f^{(4)}_{1,2}&=|\psi_{s1} \pm \psi_{s2}|^4=|\psi_{s1}|^4+|\psi_{s2}|^4+4|\psi_{s1}|^2|\psi_{s2}|^2 \pm 2\left(|\psi_{s1}|^2+|\psi_{s2}|^2 \right)\left(\psi_{s1}\psi_{s2}^*+\psi_{s1}^*\psi_{s2} \right)+\psi_{s1}^2\psi_{s2}^{*2}+\psi_{s1}^{*2}\psi_{s2}^2, \notag\\
f^{(4)}_3&=\left(\psi_{s1}+\psi_{s2}\right)^2\left(\psi_{s1}^*-\psi_{s2}^*\right)^2+h.c. =2\left(|\psi_{s1}|^4+|\psi_{s2}|^4-4|\psi_{s1}|^2|\psi_{s2}|^2+\psi_{s1}^2\psi_{s2}^{*2}+\psi_{s1}^{*2}\psi_{s2}^2\right), \notag\\
% f^{(4)}_3&=\left(\psi_{s1}+\psi_{s2}\right)^2\left(\psi_{s1}^*-\psi_{s2}^*\right)^2-h.c.=4\left(|\psi_{s1}|^2-|\psi_{s2}|^2 \right)\left(\psi_{s1}\psi_{s2}^*-\psi_{s1}^*\psi_{s2} \right), \notag\\
f^{(4)}_4&=\left(\psi_{s1} \pm \psi_{s2}\right)^2\left(\psi_{d1}^{*2}+\psi_{d2}^{*2}\right)+h.c. \notag\\
&=\left(\psi_{s1}^2+\psi_{s2}^2\right)\left(\psi_{d1}^{*2}+\psi_{d2}^{*2}\right) + \left(\psi_{s1}^{*2}+\psi_{s2}^{*2}\right)\left(\psi_{d1}^2+\psi_{d2}^2\right) \pm 2\left[\psi_{s1}\psi_{s2}\left(\psi_{d1}^{*2}+\psi_{d2}^{*2}\right) + \psi_{s1}^*\psi_{s2}^*\left(\psi_{d1}^2+\psi_{d2}^2\right)\right], \notag\\
f^{(4)}_5&=\left(\psi_{d1}^2+\psi_{d2}^2\right) \left(\psi_{d1}^{*2}+\psi_{d2}^{*2}\right)=|\psi_{d1}|^4+|\psi_{d2}|^4+\psi_{d1}^2\psi_{d2}^{*2}+\psi_{d1}^{*2}\psi_{d2}^2, \notag\\
f^{(4)}_6&=\left(|\psi_{d1}|^2+|\psi_{d2}|^2\right)^2=|\psi_{d1}|^4+|\psi_{d2}|^4+2|\psi_{d1}|^2|\psi_{d2}|^2, \notag\\
f^{(4)}_7&=|\psi_{s1}+\psi_{s2}|^2|\psi_{s1}-\psi_{s2}|^2=|\psi_{s1}|^4+|\psi_{s2}|^4-\left(\psi_{s1}^2\psi_{s2}^{*2}+\psi_{s1}^{*2}\psi_{s2}^2 \right), \notag\\
f^{(4)}_8&=|\psi_{s1} \pm \psi_{s2}|^2\left(|\psi_{d1}|^2+|\psi_{d2}|^2\right)=\left(|\psi_{s1}|^2+|\psi_{s2}|^2 \right)\left(|\psi_{d1}|^2+\psi_{d2}|^2\right) \pm \left(\psi_{s1}\psi_{s2}^*+\psi_{s1}^*\psi_{s2} \right)\left(|\psi_{d1}|^2+\psi_{d2}|^2\right).
\end{align}

The independent quartic terms in free energy are
\begin{align}
f_{ss}&=\frac{1}{2}\beta_{s_{1,2}}\left(|\psi_{s1}|^4+|\psi_{s2}|^4 \right)+\gamma_{s_{1,2}}|\psi_{s1}|^2|\psi_{s2}|^2+k_{s_{1,2}}^{(4)}\left(|\psi_{s1}|^2+|\psi_{s2}|^2 \right)\left(\psi_{s1}\psi_{s2}^*+\psi_{s1}^*\psi_{s2} \right) +g_{s_{1,2}}\left(\psi_{s1}^2\psi_{s2}^{*2}+\psi_{s1}^{*2}\psi_{s2}^2 \right), \notag\\
% &+ih_{s_{1,2}} \left(|\psi_{s1}|^2-|\psi_{s2}|^2 \right)\left(\psi_{s1}\psi_{s2}^*-\psi_{s1}^*\psi_{s2} \right), \notag\\
f_{dd}&=\frac{1}{2}\beta_{d_{1,2}}\left(|\psi_{d1}|^4+|\psi_{d2}|^4 \right)+\gamma_{d_{1,2}}|\psi_{d1}|^2|\psi_{d2}|^2+g_{d_{1,2}}\left(\psi_{d1}^2\psi_{d2}^{*2}+\psi_{d1}^{*2}\psi_{d2}^2\right),
\end{align}
and
\begin{align}
f_{sd}=~&\gamma_{s_{1,2}d_{1,2}}\left(|\psi_{s1}|^2+|\psi_{s2}|^2 \right)\left(|\psi_{d1}|^2+|\psi_{d2}|^2\right)
+g_{s_{1,2}d_{1,2}}\left[\left(\psi_{s1}^2+\psi_{s2}^2\right)\left(\psi_{d1}^{*2}+\psi_{d2}^{*2}\right)+\left(\psi_{s1}^{*2}+\psi_{s2}^{*2}\right)\left(\psi_{d1}^2+\psi_{d2}^2\right) \right] \notag\\
% &+ih_{s_{1,2}d_{1,2}}\left[\left(\psi_{s1}^2+\psi_{s2}^2\right)\left(\psi_{d1}^{*2}+\psi_{d2}^{*2}\right)-\left(\psi_{s1}^{*2}+\psi_{s2}^{*2}\right)\left(\psi_{d1}^2+\psi_{d2}^2\right) \right] + ic_{sd_{1,2}}\left[\psi_{s1}\psi_{s2}\left(\psi_{d1}^{*2}+\psi_{d2}^{*2}\right)-\psi_{s1}^*\psi_{s2}^*\left(\psi_{d1}^2+\psi_{d2}^2\right) \right]
&+k_{s_{1,2}d}\left(\psi_{s1}\psi_{s2}^*+\psi_{s1}^*\psi_{s2} \right)\left(|\psi_{d1}|^2+|\psi_{d2}|^2\right)
+k_{s_{1,2}^\prime d} \left[\psi_{s1}\psi_{s2}\left(\psi_{d1}^{*2}+\psi_{d2}^{*2}\right)+\psi_{s1}^*\psi_{s2}^*\left(\psi_{d1}^2+\psi_{d2}^2\right) \right].
\end{align}
Therefore, when the twist angle between the two layers is $\theta=45^\circ$, the full four-component free energy including $s_1$-, $s_2$-, $d_1$-, and $d_2$-wave components up to the quartic terms can be written as
\begin{equation}
F_{\text{full}}=f^{(2)}+f_{ss}+f_{dd}+f_{sd}.
\label{eq:F45full}
\end{equation}

%------------------------------------------------
\begin{comment}
If we adopt the approximation used in the main text, the contribution from the $s_1-s_2$ term is approximately zero, i.e., $\psi_{s1}=\psi_{s2}=\psi_s/2$, and $\psi_{d1}=\psi_1$, $\psi_{d2}=\psi_2$, then Eq.~\eqref{eq:F45full} is reduced to Eq.~\eqref{eq:F} in the main text, where the correspondences between the coefficients are given by
\begin{align}
& \alpha_s=\frac{1}{2}\left[\alpha_{s_{1,2}}+k_{s_{1,2}}^{(2)} \right], ~ \beta_s=\frac{1}{8}\left[\beta_{s_{1,2}}+\gamma_{s_{1,2}}+4k_{s_{1,2}}^{(4)}+2g_{s_{1,2}} \right], ~ \alpha_d=\alpha_{d_{1,2}}, ~ \beta_d=\beta_{d_{1,2}}, \notag\\
& \gamma_{dd}=\gamma_{d_{1,2}}, ~ g_{dd}=g_{d_{1,2}}, ~ \gamma_{sd}=\frac{1}{2}\left(\gamma_{s_{1,2}d_{1,2}}+k_{s_{1,2}d} \right), ~ g_{sd}=\frac{1}{4}\left(2g_{s_{1,2}d_{1,2}}+k_{s_{1,2}^\prime d} \right).
\end{align}
\end{comment}
%------------------------------------------------

%%%%%%%%%%%%%%%%%%%%%%%%%%%%%%%%%%%%%%%%%%%%%%%%%

\subsection{Degenerate pairing configurations of the reduced three-component-pairing GL free energy at a $45^{\circ}$ twist}

%-----------------------------------------------------------------------------------
\begin{figure}[h]
\centering
\begin{subfigure}{0.49\textwidth}
    \centering
    \includegraphics[width=\textwidth]{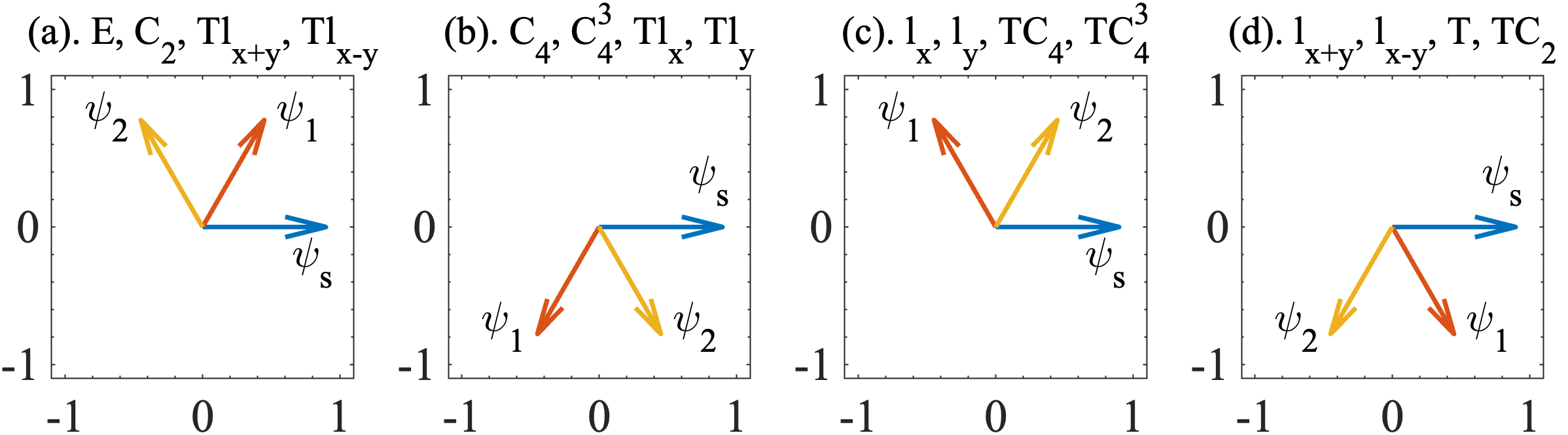} % 2396*664
    \label{subfig:TD4d_dds0}
\end{subfigure}
\hspace{-0.1cm}
\begin{subfigure}{0.49\textwidth}
    \centering
    \includegraphics[width=\textwidth]{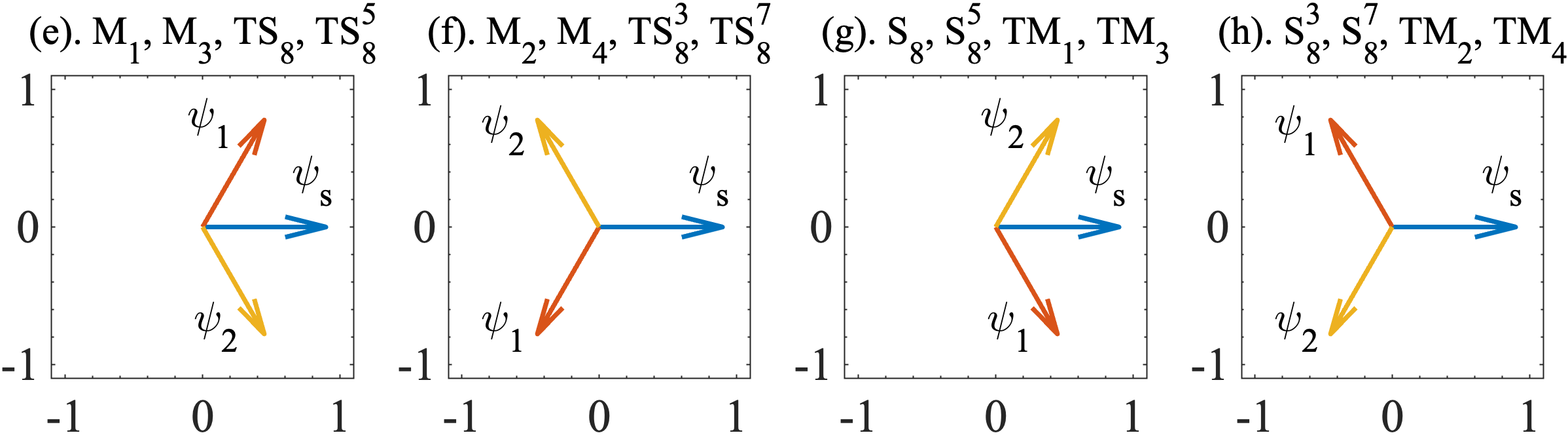} % 2396*664
    \label{subfig:TD4d_dds1}
\end{subfigure}
\captionsetup{justification=raggedright}
\caption{Degenerate configurations of the three-component pairing $s + d_1 e^{i\phi_1} + d_2 e^{i\phi_2}$ at $\theta=45^{\circ}$.
The parameters are chosen as $\alpha_s=\alpha_d=-N_F$, $\beta_s=2N_F/T_c^2$, $\beta_d=N_F/T_c^2$, $\gamma_{sd}=0.5N_F/T_c^2$, $\gamma_{dd}=-N_F/T_c^2$, $g_{dd}=1.5N_F/T_c^2$, and $g_{sd}=1.5N_F/T_c^2$.}
\label{fig:TD4d_dds}
\end{figure}
%---------------------------------------------------------------------------------

By minimizing the three-component-pairing GL free energy at $\theta=45^{\circ}$, we obtain eight symmetry-related degenerate ground states, as shown in Fig.~\ref{fig:TD4d_dds}.
Fig.~\ref{fig:TD4d_dds}~(a) shows one representative configuration, for which the relative phases of the $d_{1,2}$-wave order parameters with respect to the $s$-wave order parameter are $\phi_1=\pi/3$ and $\phi_2=2\pi/3$, satisfying $\phi_1+\phi_2=\pi$, and the unbroken symmetry group of this state is $\{E,C_2,\mathcal{T}l_{x\pm y}\}$.
Symmetry operations that can transform the configuration in Fig.~\ref{fig:TD4d_dds}~(a) to (b)-(h) are indicated on top of each subfigure.

The phase constraint in Fig.~\ref{fig:TD4d_dds}~(a) follows from the invariance of the pairing state under the combined antiunitary operations $\mathcal{T}l_{x\pm y}$.
The twofold interlayer rotations $l_{x\pm y}$, defined in Fig.~\ref{fig:D4dgroup}, interchange the two layers and transform the pairing components as
$d_1\to -d_2$ and $d_2\to -d_1$, while $\mathcal{T}$ acts by complex conjugation.
Starting from $s+d_1e^{i\phi_1}+d_2e^{i\phi_2}$, the action of $l_{x\pm y}$ gives $s+d_2e^{i(\phi_1-\pi)}+d_1e^{i(\phi_2-\pi)}$.
Subsequent application of $\mathcal{T}$ yields $s+d_2 e^{i(\pi-\phi_1)}+d_1 e^{i(\pi-\phi_2)}$.
The invariance under $\mathcal{T}l_{x\pm y}$ then implies $\pi-\phi_2=\phi_1, \pi-\phi_1=\phi_2$, or equivalently, $\phi_1+\phi_2=\pi$.
Defining $\phi=\phi_1-\phi_2$, the three-component pairing gap function in Fig.~\ref{fig:TD4d_dds}~(a) can then be written as $s+i(d_1 e^{i\phi/2}+d_2 e^{-i\phi/2})$.

For the configuration in Fig.~\ref{fig:TD4d_dds}~(b), the phase constraints are $\phi_1+\phi_2=-\pi$ and $\phi_1-\phi_2=\phi$, leading to $s-i(d_1e^{i\phi/2}+d_2e^{-i\phi/2})$.
Compared with panels~(a) and (b), the pairing configurations in panels~(c) and (d) differ only by an interchange of the phases of $d_1$ and $d_2$.
Therefore, their pairing states can be written as $s+i(d_1 e^{-i\phi/2}+d_2 e^{i\phi/2})$ and $s-i(d_1e^{-i\phi/2}+d_2e^{i\phi/2})$, respectively.
For Fig.~\ref{fig:TD4d_dds}~(e), $\phi_1+\phi_2=0$, giving $s+(d_1e^{i\phi/2}+d_2e^{-i\phi/2})$, whereas for Fig.~\ref{fig:TD4d_dds}~(f), $\phi_1+\phi_2=2\pi$, yielding $s-(d_1e^{i\phi/2}+d_2e^{-i\phi/2})$.
Similarly, the pairing configurations in Figs.~\ref{fig:TD4d_dds}~(g) and (h) take the form $s+(d_1e^{-i\phi/2}+d_2e^{i\phi/2})$ and $s-(d_1e^{-i\phi/2}+d_2e^{i\phi/2})$, respectively.

Consequently, the eight degenerate ground-state pairing configurations can be compactly expressed as
\begin{equation}
s+\eta_1i^{\eta_2} \left(d_1e^{i\eta_3\phi/2}+d_2e^{-i\eta_3\phi/2} \right),
\label{eq:eight_sdd}
\end{equation}
where $\eta_{1,3}=\pm 1$ and $\eta_2=0,1$.
Here, $\eta_1i^{\eta_2} \in \{ \pm 1, \pm i \}$ specifies the four possible overall relative phases of the $d$-wave sector with respect to the $s$-wave component,
while $\eta_3=\pm 1$ distinguishes the two configurations related by interchanging the phases of $d_1$ and $d_2$.

%%%%%%%%%%%%%%%%%%%%%%%%%%%%%%%%%%%%%%%%%%%%%%%%%

\subsection{Four-component-pairing GL free energy with $s_1$-, $s_2$-, $d_1$-, and $d_2$-wave symmetries at  general twist angles}
\label{app:GLgeneral}

%---------------------------------------------------------------------------
\begin{figure}[h]
\centering
\includegraphics[width=0.27\textwidth]{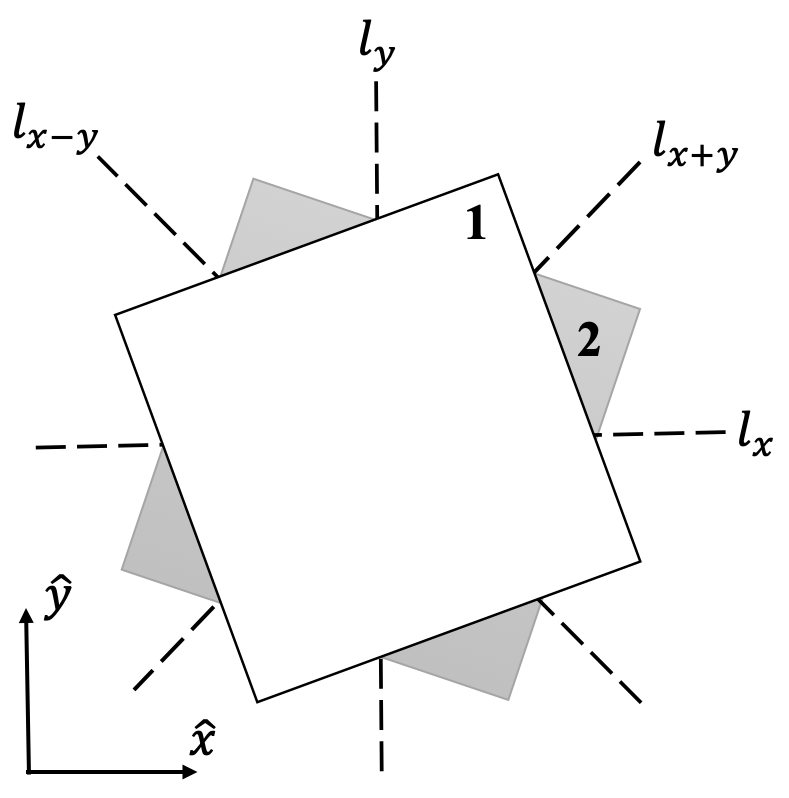}
\captionsetup{justification=raggedright}
\caption{Schematic plot of the $D_4$ group of the twisted bilayer square lattice.
In this case, layer 1 is rotated by $\theta$ around the $c$-axis, while layer 2 is rotated by $-\theta$, where $\theta \neq 0, 22.5^\circ$.}
\label{fig:D4group}
\end{figure}
%---------------------------------------------------------------------------

For twist angles away from $45^\circ$, the point-group symmetry of the system is reduced to $D_4$, as shown in Fig.~\ref{fig:D4group}.
The $D_4$ group contains four $90^\circ$ rotations about the $c$-axis and four $180^\circ$ rotations that interchange the two layers, denoted by $l_x$, $l_{x+y}$, $l_y$, and $l_{x-y}$.
Under the representations of $D_4$, the two layer-resolved $d_{x^2-y^2}$-wave pairing components, $(d_1,d_2)$, form a reducible representation, which decomposes into two irreducible representations as
$(d_1,d_2)=B_1\oplus B_2$.
Similarly, the two layer-resolved $s$-wave pairing components decompose as
$(s_1,s_2)=A_1\oplus A_2$.

The quadratic terms in the free energy in this case are
\begin{align}
f_{1,2}^{(2)\prime}&=|\psi_{s1} \pm \psi_{s2}|^2=\left(|\psi_{s1}|^2+|\psi_{s2}|^2\right) \pm \left(\psi_{s1}\psi_{s2}^*+\psi_{s1}^*\psi_{s2} \right), \notag\\
f_{3,4}^{(2)\prime}&=|\psi_{d1} \pm \psi_{d2}|^2=\left(|\psi_{d1}|^2+|\psi_{d2}|^2\right) \pm \left(\psi_{d1}\psi_{d2}^*+\psi_{d1}^*\psi_{d2} \right),
\end{align}
where $\psi_{s1}$, $\psi_{s2}$,  $\psi_{d1}$, $\psi_{d2}$ are still the $s$-wave and $d_{x^2-y^2}$-wave complex order parameters in layer 1 and layer 2, respectively.
The independent quadratic free energy in this case can be written as
\begin{align}
f^{(2)\prime}=\alpha_{s_{1,2}}^\prime \left(|\psi_{s1}|^2+|\psi_{s2}|^2\right)+k_{s_{1,2}}^{(2)\prime}\left(\psi_{s1}\psi_{s2}^*+\psi_{s1}^*\psi_{s2} \right)+\alpha_{d_{1,2}}^\prime \left(|\psi_{d1}|^2+|\psi_{d2}|^2 \right)+k_{d_{1,2}}^{(2)}\left(\psi_{d1}\psi_{d2}^*+\psi_{d1}^*\psi_{d2} \right).
\end{align}

Up to the quartic terms, the extra $D_{4d}$-invariant terms are
\begin{align}
f^{(4)\prime}_{1,2}&=|\psi_{s1} \pm \psi_{s2}|^4=|\psi_{s1}|^4+|\psi_{s2}|^4+4|\psi_{s1}|^2|\psi_{s2}|^2 \pm 2\left(|\psi_{s1}|^2+|\psi_{s2}|^2 \right)\left(\psi_{s1}\psi_{s2}^*+\psi_{s1}^*\psi_{s2} \right)+\psi_{s1}^2\psi_{s2}^{*2}+\psi_{s1}^{*2}\psi_{s2}^2, \notag\\
f^{(4)\prime}_3&=\left(\psi_{s1}+\psi_{s2}\right)^2\left(\psi_{s1}^*-\psi_{s2}^*\right)^2+h.c. =2\left(|\psi_{s1}|^4+|\psi_{s2}|^4-4|\psi_{s1}|^2|\psi_{s2}|^2+\psi_{s1}^2\psi_{s2}^{*2}+\psi_{s1}^{*2}\psi_{s2}^2\right), \notag\\
f^{(4)\prime}_4&=\left(\psi_{s1}+\eta_1\psi_{s2}\right)^2\left(\psi_{d1}^*+\eta_2\psi_{d2}^*\right)^2 + h.c. \notag\\
&=\left[\left(\psi_{s1}^2+\psi_{s2}^2\right)\left(\psi_{d1}^{*2}+\psi_{d2}^{*2}\right) + \left(\psi_{s1}^{*2}+\psi_{s2}^{*2}\right)\left(\psi_{d1}^2+\psi_{d2}^2\right)\right]+2\eta_1\left[\psi_{s1}\psi_{s2}\left(\psi_{d1}^{*2}+\psi_{d2}^{*2}\right) + \psi_{s1}^*\psi_{s2}^*\left(\psi_{d1}^2+\psi_{d2}^2\right) \right] \notag\\
& \quad +2\eta_2\left[\left(\psi_{s1}^2+\psi_{s2}^2 \right) \psi_{d1}^*\psi_{d2}^* + \left(\psi_{s1}^{*2}+\psi_{s2}^{*2} \right) \psi_{d1}\psi_{d2} \right]+4\eta_1\eta_2\left[\psi_{s1}\psi_{s2}\psi_{d1}^*\psi_{d2}^* + \psi_{s1}^*\psi_{s2}^*\psi_{d1}\psi_{d2}\right], \notag\\
f^{(4)\prime}_{5,6}&=|\psi_{d1} \pm \psi_{d2}|^4=|\psi_{d1}|^4+|\psi_{d2}|^4+4|\psi_{d1}|^2|\psi_{d2}|^2\pm2\left(|\psi_{d1}|^2+|\psi_{d2}|^2 \right)\left(\psi_{d1}\psi_{d2}^*+\psi_{d1}^*\psi_{d2} \right)+\psi_{d1}^2\psi_{d2}^{*2}+\psi_{d1}^{*2}\psi_{d2}^2, \notag\\
f^{(4)\prime}_7&=\left(\psi_{d1}+\psi_{d2}\right)^2 \left(\psi_{d1}^*-\psi_{d2}^{*}\right)^2+h.c.=2\left(|\psi_{d1}|^4+|\psi_{d2}|^4-4|\psi_{d1}|^2|\psi_{d2}|^2+\psi_{d1}^2\psi_{d2}^{*2}+\psi_{d1}^{*2}\psi_{d2}^2\right), \notag\\
% f^{(4)\prime}_8&=\left(\psi_{d1}+\psi_{d2}\right)^2 \left(\psi_{d1}^*-\psi_{d2}^{*}\right)^2-h.c.=4\left(|\psi_{d1}|^2-|\psi_{d2}|^2 \right)\left(\psi_{d1}\psi_{d2}^*-\psi_{d1}^*\psi_{d2} \right), \notag\\
f^{(4)\prime}_8&=|\psi_{s1}+\psi_{s2}|^2|\psi_{s1}-\psi_{s2}|^2=|\psi_{s1}|^4+|\psi_{s2}|^4-\left(\psi_{s1}^2\psi_{s2}^{*2}+\psi_{s1}^{*2}\psi_{s2}^2 \right), \notag\\
f^{(4)\prime}_9&=|\psi_{s1}+\eta_1\psi_{s2}|^2|\psi_{d1}+\eta_2\psi_{d2}|^2 \notag\\
&=\left(|\psi_{s1}|^2+|\psi_{s2}|^2\right)\left(|\psi_{d1}|^2+|\psi_{d2}|^2\right)+\eta_1\left(\psi_{s1}\psi_{s2}^*+\psi_{s1}^*\psi_{s2} \right)\left(|\psi_{d1}|^2+|\psi_{d2}|^2\right) \notag\\
& \quad +\eta_2\left(\psi_{d1}\psi_{d2}^*+\psi_{d1}^*\psi_{d2} \right)\left(|\psi_{s1}|^2+|\psi_{s2}|^2\right)+\eta_1\eta_2\left(\psi_{s1}\psi_{s2}^*+\psi_{s1}^*\psi_{s2} \right)\left(\psi_{d1}\psi_{d2}^*+\psi_{d1}^*\psi_{d2} \right), \notag\\
f^{(4)\prime}_{10}&=|\psi_{d1}+\psi_{d2}|^2|\psi_{d1}-\psi_{d2}|^2=|\psi_{d1}|^4+|\psi_{d2}|^4-\left(\psi_{d1}^2\psi_{d2}^{*2}+\psi_{d1}^{*2}\psi_{d2}^2 \right),
\end{align}
where $\eta_{1,2}=\pm1$.
The independent quartic terms in free energy in this case are
\begin{align}
f^{\prime}_{ss}&=\frac{1}{2}\beta^\prime_{s_{1,2}}\left(|\psi_{s1}|^4+|\psi_{s2}|^4 \right)
+\gamma^\prime_{s_{1,2}}|\psi_{s1}|^2|\psi_{s2}|^2
+k_{s_{1,2}}^{(4)\prime}\left(|\psi_{s1}|^2+|\psi_{s2}|^2 \right)\left(\psi_{s1}\psi_{s2}^*+\psi_{s1}^*\psi_{s2} \right)
+g^\prime_{s_{1,2}}\left(\psi_{s1}^2\psi_{s2}^{*2}+\psi_{s1}^{*2}\psi_{s2}^2 \right), \notag\\
f^\prime_{dd}&=\frac{1}{2}\beta^\prime_{d_{1,2}}\left(|\psi_{d1}|^4+|\psi_{d2}|^4 \right)
+\gamma^\prime_{d_{1,2}}|\psi_{d1}|^2|\psi_{d2}|^2
+k_{d_{1,2}}^{(4)}\left(|\psi_{d1}|^2+|\psi_{d2}|^2 \right)\left(\psi_{d1}\psi_{d2}^*+\psi_{d1}^*\psi_{d2} \right) +g^\prime_{d_{1,2}}\left(\psi_{d1}^2\psi_{d2}^{*2}+\psi_{d1}^{*2}\psi_{d2}^2\right),
%&+ih^\prime_{d_{1,2}} \left(|\psi_{d1}|^2-|\psi_{d2}|^2 \right)\left(\psi_{d1}\psi_{d2}^*-\psi_{d1}^*\psi_{d2} \right), \notag\\
\end{align}
and
\begin{align}
f_{sd}^\prime=~&\gamma^\prime_{s_{1,2}d_{1,2}}\left(|\psi_{s1}|^2+|\psi_{s2}|^2 \right)\left(|\psi_{d1}|^2+|\psi_{d2}|^2\right)+g_{s_{1,2}d_{1,2}}^\prime \left[\left(\psi_{s1}^2+\psi_{s2}^2\right)\left(\psi_{d1}^{*2}+\psi_{d2}^{*2}\right)+\left(\psi_{s1}^{*2}+\psi_{s2}^{*2}\right)\left(\psi_{d1}^2+\psi_{d2}^2\right) \right] \notag\\
% &+ih_{s_{1,2}d_{1,2}}^\prime \left[\left(\psi_{s1}^2+\psi_{s2}^2\right)\left(\psi_{d1}^{*2}+\psi_{d2}^{*2}\right)-\left(\psi_{s1}^{*2}+\psi_{s2}^{*2}\right)\left(\psi_{d1}^2+\psi_{d2}^2\right) \right] +ic_{sd_{1,2}}^\prime \left[\psi_{s1}\psi_{s2}\left(\psi_{d1}^{*2}+\psi_{d2}^{*2}\right)-\psi_{s1}^*\psi_{s2}^*\left(\psi_{d1}^2+\psi_{d2}^2\right) \right]
&+k^\prime_{s_{1,2}d}\left(\psi_{s1}\psi_{s2}^*+\psi_{s1}^*\psi_{s2} \right)\left(|\psi_{d1}|^2+|\psi_{d2}|^2\right)
+k_{s_{1,2}^\prime d}^\prime \left[\psi_{s1}\psi_{s2}\left(\psi_{d1}^{*2}+\psi_{d2}^{*2}\right)+\psi_{s1}^*\psi_{s2}^*\left(\psi_{d1}^2+\psi_{d2}^2\right) \right] \notag\\
&+k_{sd_{1,2}}\left(|\psi_{s1}|^2+|\psi_{s2}|^2\right)\left(\psi_{d1}\psi_{d2}^*+\psi_{d1}^*\psi_{d2} \right)
+k_{s_{1,2}d_{1,2}}\left(\psi_{s1}\psi_{s2}^*+\psi_{s1}^*\psi_{s2} \right)\left(\psi_{d1}\psi_{d2}^*+\psi_{d1}^*\psi_{d2} \right) \notag\\
&+k_{sd_{1,2}^\prime} \left[\left(\psi_{s1}^2+\psi_{s2}^2\right)\psi_{d1}^*\psi_{d2}^*+\left(\psi_{s1}^{*2}+\psi_{s2}^{*2}\right)\psi_{d1}\psi_{d2} \right] + k\left(\psi_{s1}\psi_{s2}\psi_{d1}^*\psi_{d2}^*+\psi_{s1}^*\psi_{s2}^*\psi_{d1}\psi_{d2} \right).
% &+ih_{s_{1,2}d}\left[\left(\psi_{s1}^2+\psi_{s2}^2\right)\psi_{d1}^*\psi_{d2}^*+\left(\psi_{s1}^{*2}+\psi_{s2}^{*2}\right)\psi_{d1}\psi_{d2} \right]+ih\left(\psi_{s1}\psi_{s2}\psi_{d1}^*\psi_{d2}^*-\psi_{s1}^*\psi_{s2}^*\psi_{d1}\psi_{d2} \right).
\end{align}
Therefore, when the twist angle between the two layers is a general angle which is not equal to $45^\circ$, the full four-component free energy including $s_1$-, $s_2$-, $d_1$-, and $d_2$-wave components up to the quartic terms can be written as
\begin{equation}
F_{\text{full}}^\prime=f^{(2)\prime}+f_{ss}^{\prime}+f_{dd}^\prime+f_{sd}^\prime.
\label{eq:Fgeneralfull}
\end{equation}

\begin{comment}
If we assume $\psi_{s1}=\psi_{s2}=\psi_s/2$ again, Eq.~\eqref{eq:Fgeneralfull} is reduced to Eq.~\eqref{eq:Fprime}, and the corresponding coefficients are given by
\begin{align}
&\alpha_s^\prime=\frac{1}{2}\left[\alpha^\prime_{s_{1,2}}+k_{s_{1,2}}^{(2)\prime} \right], ~
\beta_s^\prime=\frac{1}{8}\left[\beta^\prime_{s_{1,2}}+\gamma^\prime_{s_{1,2}}+4k_{s_{1,2}}^{(4)\prime}+2g^\prime_{s_{1,2}} \right], \notag\\
&\alpha_d^\prime=\alpha^\prime_{d_{1,2}}, ~
k_{dd}=k_{d_{1,2}}^{(2)}, ~
\beta_d^\prime=\beta^\prime_{d_{1,2}}, ~
\gamma^\prime_{dd}=\gamma^\prime_{d_{1,2}}, ~
g^\prime_{dd}=g^\prime_{d_{1,2}}, ~
k_{dd}^\prime=k_{d_{1,2}}^{(4)},
\end{align}
and
\begin{align}
\gamma^\prime_{sd}=\frac{1}{2}\left(\gamma^\prime_{s_{1,2}d_{1,2}}+k^\prime_{s_{1,2}d} \right), ~
g^\prime_{sd}=\frac{1}{4}\left(2g^\prime_{s_{1,2}d_{1,2}}+k^\prime_{s_{1,2}^\prime d} \right), ~
k_{sd}=\frac{1}{2}\left(k_{sd_{1,2}}+k_{s_{1,2}d_{1,2}}\right), ~
k_{sd}^\prime=\frac{1}{4}\left(2k_{s_{1,2}d}+k \right).
\end{align}
\end{comment}

%\end{widetext}

%%%%%%%%%%%%%%%%%%%%%%%%%%%%%%%%%%%%%%%%%%%%%%%%%

\subsection{Reduced three-component-pairng GL free energy analysis at general twist angles}
\label{app:general}

When the order parameters $s_1$ and $s_2$ are approximately equal,
we can adopt the symmetric combination $s=s_1+s_2$ in the free-energy analysis, and omit the antisymmetric $(s_1-s_2)$-term.
The reduced three-component GL free energy respecting the $U(1)$ gauge, time-reversal ($\mathcal{T}$), and $D_4$ symmetries up to the quartic order can be written as
\begin{equation}
F^\prime=F_d^{(0)\prime}+F_s^{(0)\prime}+F_{dd}^\prime+F_{sd}^\prime,
\label{eq:Fprime}
\end{equation}
in which $F_d^{(0)\prime}=\alpha_d^{\prime} \left( |\psi_1|^2+|\psi_2|^2 \right) + \beta_d^{\prime} \left( |\psi_1|^4+|\psi_2|^4 \right)/2$, $F_s^{(0)\prime}=\alpha_s^{\prime}|\psi_{s}|^2+\beta_s^{\prime}|\psi_{s}|^4/2$, and
\begin{align}
F_{dd}^\prime=&~\gamma_{dd}^{\prime}|\psi_1|^2|\psi_2|^2+k_{dd}^{\prime}\left(|\psi_1|^2+|\psi_2|^2\right) \left(\psi_1\psi_2^*+\psi_1^*\psi_2\right)
+ g_{dd}^{\prime}\left(\psi_1^2\psi_2^{*2}+\psi_1^{*2}\psi_2^2\right)+k_{dd}\left(\psi_1\psi_2^*+\psi_1^*\psi_2\right), \notag\\
F_{sd}^\prime=&~\gamma_{sd}^{\prime}|\psi_s|^2 \left(|\psi_1|^2+|\psi_2|^2 \right)
+ g_{sd}^{\prime} \left[\psi_s^2 \left(\psi_1^{*2}+\psi_2^{*2} \right)+\psi_s^{*2} \left(\psi_1^2+\psi_2^2 \right)\right] \notag\\
&+ k_{sd}|\psi_s|^2 \left(\psi_1\psi_2^*+\psi_1^*\psi_2\right)
+ k_{sd}^{\prime} \left(\psi_s^2 \psi_1^*\psi_2^*+\psi_s^{*2} \psi_1\psi_2 \right),
\end{align}
where $\psi_1$, $\psi_2$ and $\psi_s$ denote the complex order parameters of $d_1$-, $d_2$-, and $s$-wave order parameters, respectively.

Notice that the $d$-wave order parameters change sign under a $90^\circ$ rotation.
Therefore, when the twist angle of layer 2 increases from $0^\circ$ to $90^\circ$ while keeping layer 1 fixed,
$\psi_1$ stays unchanged and $\psi_2$ transits to  $-\psi_2$.
This implies that the coefficients of the terms containing $(\psi_1\psi_2^* + h.c.)$ and $(\psi_s^2\psi_1^*\psi_2^* + h.c.)$ appearing in the free energy must change sign under a $90^\circ$ rotation and vanish at a rotation of $45^\circ$.
To satisfy these conditions, the simplest choice is 
\begin{align}
\left(k_{dd}, k_{dd}^\prime, k_{sd}, k_{sd}^\prime \right)
=\left(k_{dd}^0, k_{dd}^{\prime 0}, k_{sd}^0, k_{sd}^{\prime 0} \right) \cos{(2\theta)},
\end{align}
where $\theta$ is the twist angle between two layers.
Moreover, to ensure that the phases of $\psi_1$ and $\psi_2$ are equal when the twist angle is $0^\circ$,
the parameters should be chosen as $k_{dd}^0, k_{dd}^{\prime 0}, k_{sd}^0, k_{sd}^{\prime 0}< 0$.

Setting the phase of $\psi_s$ to be zero, we obtain the phase-sensitive term in Eq. \eqref{eq:Fprime} as
\begin{equation}
F^\prime_{\phi}=g_0 \cos{\left(2\phi_1-2\phi_2 \right)}
+ g_1 \cos{\left(2\phi_1 \right)}+g_2 \cos{\left(2\phi_2 \right)}
%&+h_1\sin{\left(2\phi_1\right)}+h_2\sin{\left(2\phi_2\right)} \notag\\
+ k_0\cos{\left(\phi_1-\phi_2\right)}+k_1 \cos{\left(\phi_1+\phi_2\right)},
%&+h_0\sin{\left(\phi_1-\phi_2\right)}+h_0^\prime \sin{\left(\phi_1+\phi_2\right)},
\label{eq:Fphiprime}
\end{equation}
in which
\begin{align}
&g_0=2g_{dd}^\prime|\psi_1|^2|\psi_2|^2, \quad
g_{1}=2g_{sd}^\prime|\psi_s|^2|\psi_{1}|^2, \quad
g_{2}=2g_{sd}^\prime|\psi_s|^2|\psi_{2}|^2, \notag\\
&k_0=2|\psi_1||\psi_2|\left[k_{dd}+k_{dd}^\prime \left(|\psi_1|^2+|\psi_2|^2\right)+k_{sd}|\psi_s|^2\right], \quad
k_1=2k_{sd}^\prime|\psi_s|^2|\psi_1||\psi_2|.
\end{align}
The parameters $g_{dd}^\prime>0$ and $g_{sd}^\prime>0$ favor $\phi_1$, $\phi_2$, and $\phi_1-\phi_2$ to all tend toward $\pi/2$.
Restricting $\theta$ to $(0^\circ, 180^\circ)$,
when $\theta<45^\circ$ or $\theta>135^\circ$, the conditions $k_0, k_1<0$ favor $\phi_1 \pm \phi_2$ to tend toward 0;
when $45^\circ<\theta_0<135^\circ$, the conditions $k_0, k_1>0$ favor $\phi_1 \pm \phi_2$ to tend toward $\pm \pi$.

%---------------------------------------------------------------------
% \subsection{Topological frustrated three-component pairing}
By fixing the twist angle between two layers at $\theta_1=2\arctan{(2/5)} \approx 43.6^\circ$ and $\theta_2=2\arctan{(1/2)} \approx 53.13^\circ$, and minimizing the free energy in Eq. \eqref{eq:Fprime},
we obtain the pairing configurations shown in Fig. \ref{fig:TD4dds} (a) and Fig. \ref{fig:TD4dds} (b), respectively.
In both cases, the four pairing configurations are energetically degenerate.

%-----------------------------------------------------------------------------------
\begin{figure}[htbp]
    \centering
    \begin{subfigure}[b]{0.48\textwidth}
        \centering
        \includegraphics[width=\textwidth]{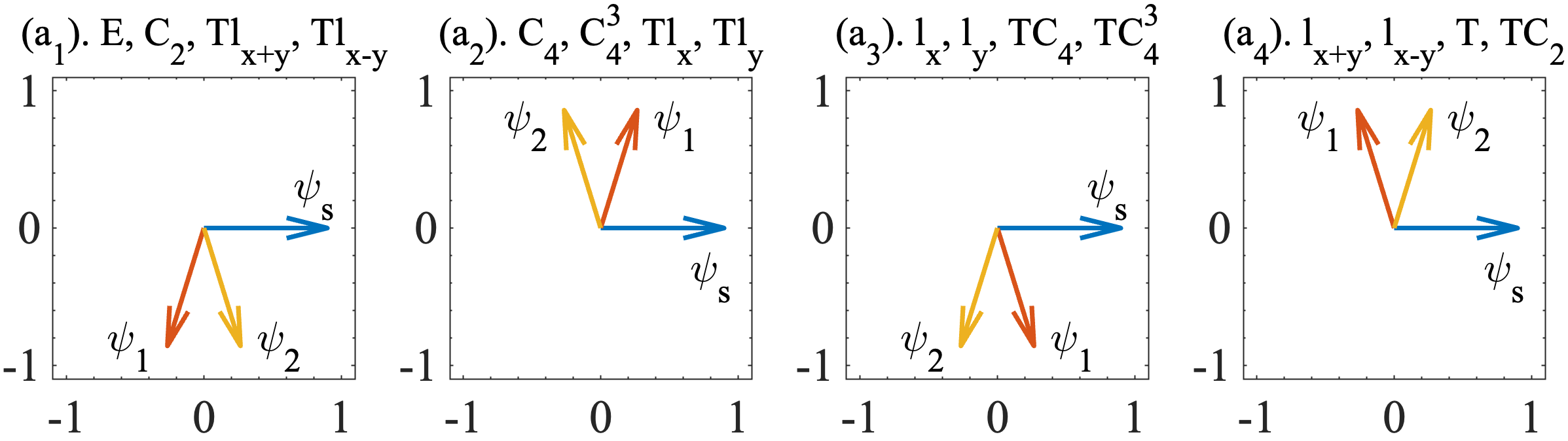}
        \caption{$\theta_1=2\arctan{(2/5)} \approx 43.6^\circ$}
        \label{fig:TD4dds_a}
    \end{subfigure}
    \hfill
    \begin{subfigure}[b]{0.48\textwidth}
        \centering
        \includegraphics[width=\textwidth]{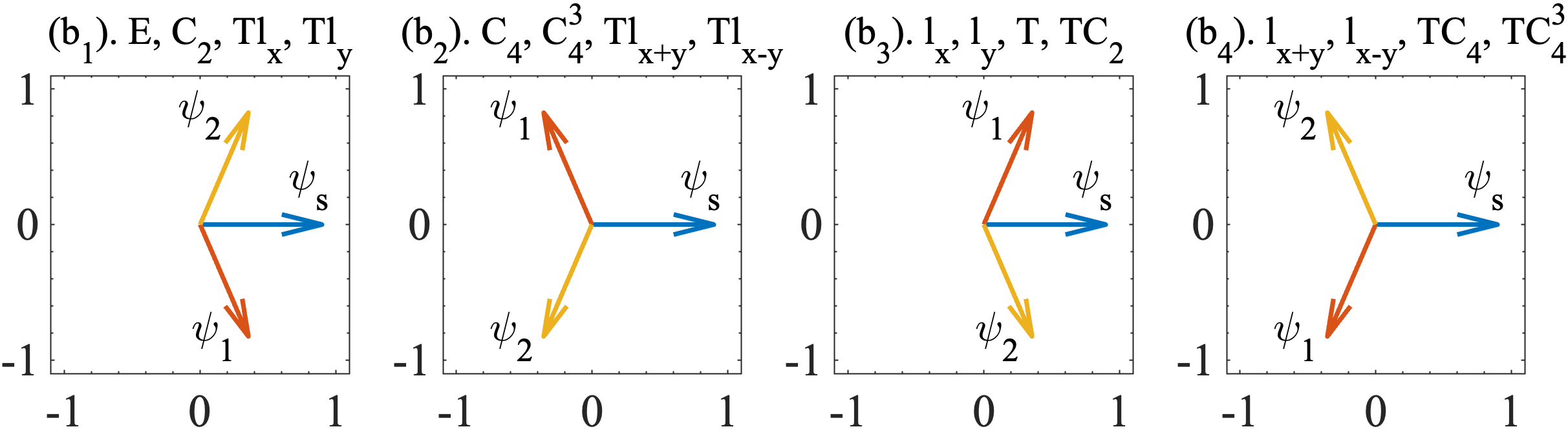}
        \caption{$\theta_2=2\arctan{(1/2)} \approx 53.13^\circ$}
        \label{fig:TD4dds_b}
    \end{subfigure}
    \captionsetup{justification=raggedright}
    \caption{Degenerate configurations of the three-component pairing $s + d_1 e^{i\phi_1} + d_2 e^{i\phi_2}$ at two different twist angles.
    The parameters in Eq.~\eqref{eq:Fprime} common to both angles are chosen as
    $\alpha_s=\alpha_d=-N_F$, $\beta_s=2N_F/T_c^2$, $\beta_d=N_F/T_c^2$,
    $\gamma_{sd}=0.5N_F/T_c^2$, $\gamma_{dd}=-N_F/T_c^2$,
    and $g_{dd}=g_{sd}=1.5N_F/T_c^2$.
    The remaining parameters are
    $k_{dd}^0=k_{dd}^{\prime 0}=k_{sd}^0=-10N_F/T_c^2$ and $k_{sd}^{\prime 0}=-5N_F/T_c^2$ for panel (a),
    while $k_{dd}^0=k_{dd}^{\prime 0}=k_{sd}^0=k_{sd}^{\prime 0}=-N_F/T_c^2$ for (b).}
    \label{fig:TD4dds}
\end{figure}
%-----------------------------------------------------------------------------------

The obtained order parameter magnitudes in both Fig.~\ref{fig:TD4dds} (a) and Fig.~\ref{fig:TD4dds} (b) are identically
$|\psi_s|=|\psi_1|=|\psi_2|=1k_BT_c$, 
with the phases $\phi_1=1.404\pi$, $\phi_2=1.596\pi$ in panel (a) and $\phi_1=1.63\pi$, $\phi_2=0.37\pi$ in panel (b).
The additional linear Josephson coupling terms introduced in Eq.~\eqref{eq:Fprime} alter the phase competition among $\psi_s$, $\psi_1$, and $\psi_2$,
consequently, the phase difference $\left| \phi_1-\phi_2 \right|$ deviates from $\pi/3$ or $2\pi/3$ even when $g_{sd}^{\prime}=g_{dd}^{\prime}$.
Notably, the $s+d_1e^{i\phi_1}+d_2e^{i\phi_2}$ pairing state is stabilized within an intermediate  finite range of $g_{dd}/g_{sd}$, beyond which the pairing evolves into $s+i(d_1\pm d_2)$ or $d_1+e^{i\phi}d_2$,
where $\phi=\phi_1-\phi_2$ can deviate from $\pm \pi/2$ due to the presence of terms involving linear Josephson couplings between $d_1$ and $d_2$.

% \subsection{Symmetry breaking pattern}
% \label{subsubsec:sym_breaking_general_twist}

Nevertheless, specific phase constraints between $\phi_1$ and $\phi_2$ are strictly preserved as a consequence of the remaining unbroken symmetries.
Under the $D_4$ point group, the three-component pairing gap function remains invariant under the $C_2$ rotation around the $c$-axis and specific twofold interlayer rotations combined with $\mathcal{T}$.
Specially, for the configuration in Fig.~\ref{fig:TD4dds}~(a), the constraint $\phi_1+\phi_2=\pi$ is protected by the unbroken symmetry group $\left \{ E, C_2, \mathcal{T}l_{x+y}, \mathcal{T}l_{x-y} \right \}$.
In contrast, for the configuration in Fig.~\ref{fig:TD4dds}~(b), the phase constraint is $\phi_1+\phi_2=0$, which is stabilized by the unbroken symmetry group $\left \{ E, C_2, \mathcal{T}l_x, \mathcal{T}l_y \right \}$.

Despite these differences in specific symmetry elements, both unbroken groups are equivalent to $D_2$.
Consequently, the two configurations share the identical symmetry-breaking pattern, $D_4 \times \mathbb{Z}_2^\mathcal{T} \to D_2$.
Since $|D_4 \times \mathbb{Z}^\mathcal{T}_2| / |D_2| = 4$, there are four degenerate solutions of the ground state pairing configurations in both cases.
% Symmetry operations that generate the degenerate configurations from panel (a) are indicated on top of the subfigures,
Symmetry operations that can transform the pairing configuration in Fig.~\ref{fig:TD4dds} (a$_1$) and Fig.~\ref{fig:TD4dds} (b$_1$) to the corresponding degenerate configuration are indicated on top of each subfigure in Fig. \ref{fig:TD4dds},
where $E$ denotes the identity operation, and other symmetry operations are the same as those in Fig.~\ref{fig:D4group}.
In addition to $\mathcal{T}$-breaking, this topologically nontrivial three-component pairing pattern also spontaneously breaks the intralayer $C_4$ rotational symmetry down to $C_2$, thereby exhibiting a nematic superconducting nature.

%%%%%%%%%%%%%%%%%%%%%%%%%%%%%%%%%%%%%%%%%%%%%%%

\section{Derivation of self-consistent mean field equations}
\label{app:self_consistent}

Including electron-electron interactions described by an extended Hubbard model with onsite repulsion and nearest-neighbor attraction to the normal-state noninteracting Hamiltonian \cite{Can2021SM}, the full Hamiltonian is given by
\begin{equation}
H=-\sum_{ij,\sigma l}t_{ij}\left(c_{i\sigma l}^\dagger c_{j\sigma l}+h.c. \right) -\mu \sum_{i\sigma l}n_{i\sigma l}
-\sum_{ij\sigma}\left(g_{ij}c_{i\sigma 1}^\dagger c_{j\sigma 2}+h.c. \right) + \sum_{ij,l}V_{ij}n_{il} n_{jl},
\label{eq:Hfull}
\end{equation}
where $l=1,2$ are layer indices; $t_{ij}$ includes the intralayer hoppings $t$, $t^\prime$;
$g_{ij}$ is the interlayer tunneling given by Eq.~\eqref{eq:gij};
and $V_{ij}$ is the density-density interaction with onsite repulsion $U$ and nearest-neighbor attraction $V$.

Retaining only the spin-singlet, nearest-neighbor attractive pairing terms, and expressing the theory in the form of the coherent state path integral, the corresponding quantum partition function takes the form:
 \begin{equation}
 \mathcal{Z}=\int D\left[\bar{\psi}, \psi \right] \exp{\left(-S\left[\bar{\psi}, \psi \right]\right)},
 \label{eq:Z}
 \end{equation}
 where $\psi$ denote Grassmann fields,
 and $S\left[\bar{\psi}, \psi \right]$ is the action corresponding to Eq.~\eqref{eq:Hfull}.
 Noting that exchanging the indices $i$ and $j$ leaves the sum over $\langle ij \rangle$ invariant, the partition function for the pairing term can be written as
\begin{equation}
\mathcal{Z}_{int}=\int D\left[\bar{\psi}, \psi \right] \exp{\left(V\int_0^\beta d\tau \sum_{\langle ij \rangle l} \bar{\psi}_{i\uparrow l} \bar{\psi}_{j\downarrow l} \psi_{j\downarrow l} \psi_{i\uparrow l}\right)},
\label{eq:Zint}
\end{equation}
where $V$ is a positive constant.
Using the Hubbard–Stratonovich decoupling and carry out the Grassmann–Gaussian integral, Eq. \eqref{eq:Z} can be written as
\begin{equation}
\mathcal{Z}=\int D[\bar{\Delta}, \Delta] \exp{\left(-S_\text{eff}\right)},
\end{equation}
where the effective action, $S_{\text{eff}}$, for $\Delta_{ij,l}$ is
 \begin{equation}
 S_{\text{eff}} = -\sum_{\boldsymbol{k}n} \text{Tr}~ \text{ln}~ \hat{\mathcal{G}}^{-1} + \frac{\beta}{V} \sum_{\langle ij \rangle l}\Delta_{ij,l} \Delta_{ij,l}^*,
 \label{eq:Seff}
 \end{equation}
in which $\hat{\mathcal{G}}^{-1}=i\omega_n\hat{I}-h_{\boldsymbol{k}}$, $h_{\boldsymbol{k}}$ is the BdG Hamiltonian obtained by Fourier transforming Eq.~\eqref{eq:Hfull}, and $\Psi_{\boldsymbol{k}n}$ is given by Eq. \eqref{eq:Psi_a}. % in Sec.~\ref{app:derive_linearized_gap}.
 To evaluate the saddle point condition for Eq.~\eqref{eq:Seff}, let $\partial S_{\text{eff}}/\partial \Delta_{ij,a}^*=0$.
 After performing the Matsubara sum, this simplifies to the self-consistent mean field equation,
 \begin{equation}
 \Delta_{ij,l}=-V \sum_{\boldsymbol{k}} \text{Tr} \left[ \frac{\partial h_{\boldsymbol{k}}}{\partial \Delta_{ij,l}^*} U_{\boldsymbol{k}} n_F\left(E_{\boldsymbol{k}} \right) U_{\boldsymbol{k}}^\dagger \right],
 \label{eq:Delta_ijl}
 \end{equation}

 %-----------------------------------------------------------------------------------
\begin{figure}[h]
    \centering
    \begin{subfigure}{0.49\textwidth}
        \centering
       \includegraphics[width=\textwidth]{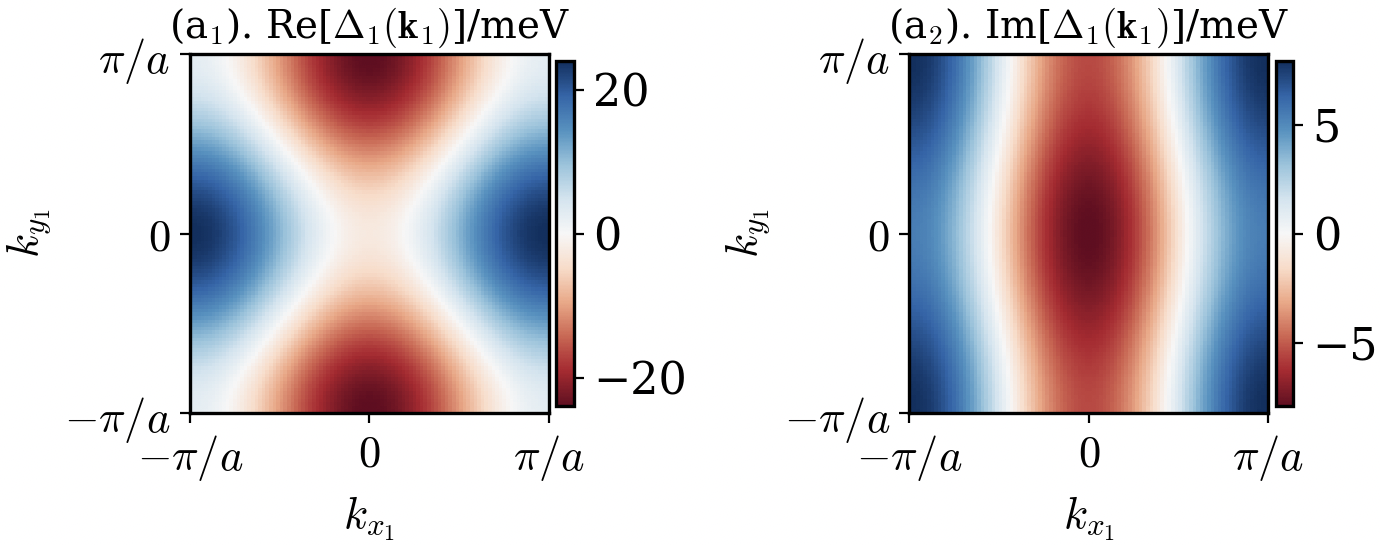}
        \caption{}
        \label{fig:subDelta1}
    \end{subfigure}
    \hfill
    \begin{subfigure}{0.49\textwidth}
        \centering
        \includegraphics[width=\textwidth]{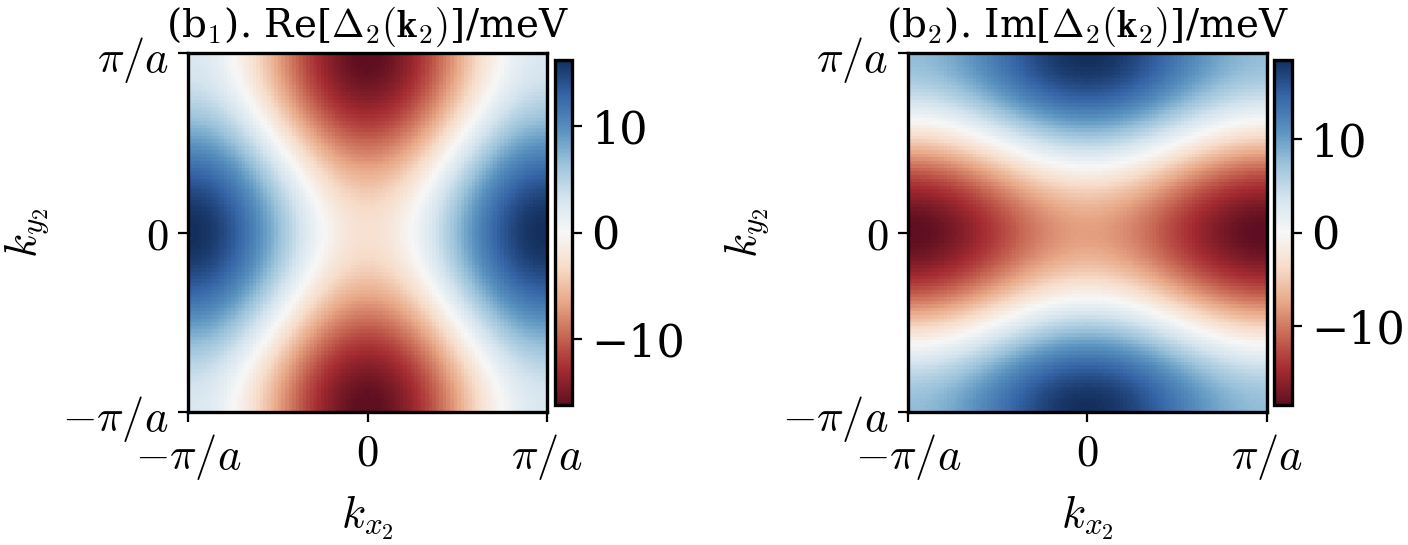}
        \caption{}
        \label{fig:subDelta2}
    \end{subfigure}
    \captionsetup{justification=raggedright}
    \caption{Real and imaginary parts of the pairing order parameters in momentum space for (a) layer 1 and (b) layer 2 with the twist vector $\boldsymbol{v}=(2,5)$ and twist angle $\theta=2\arctan(2/5)=43.6^\circ$.
The parameters in Eq. \eqref{eq:Hfull} are chosen as $t=0.153$~eV, $t^\prime=-0.45t$, $\mu=-1.05t$, $g_0=0.036$~eV,  $V=0.146$~eV,  $U=0$, $d=2.22a$, and $\rho=0.39a$, where $a$ is the lattice constant.
The temperature is set to be $T=0$~K.}
\label{fig:Deltak}
\end{figure}
%---------------------------------------------------------------------------------

Solving the real-space bond order parameters $\Delta_{ij,l}$ self-consistently according to Eq.~\eqref{eq:Delta_ijl} and performing Fourier transform within each individual layer,
we obtain the momentum-space pairing functions for two layers, $\Delta_1(\boldsymbol{k}_1)$ and $\Delta_2(\boldsymbol{k}_2)$, as shown in Fig.~\ref{fig:Deltak},
Here $\boldsymbol{k}_1=(k_{x_1},k_{y_1})=\hat{R}(\theta/2)(k_x,k_y)$ and $\boldsymbol{k}_2=(k_{x_2},k_{y_2})=\hat{R}(-\theta/2)(k_x,k_y)$ are the momenta defined in the rotated coordinate frames of the two layers, respectively,
in which the momentum $(k_x,k_y)$ is expressed in the unrotated reference frame, and $\hat{R}(\pm\theta/2)$ denotes the rotation matrix.

%%%%%%%%%%%%%%%%%%%%%%%%%%%%%%%%%%%%%%%%%%%%%%%%%%%%%%%

\section{Projections and quantifications of pairing functions}

\subsection{Projections to various pairing channels}
\label{app:D4projection}

Since the pairing functions generally contain admixtures of several symmetry channels, we decompose them into components according to the irreducible representations of the $D_4$ point group,
which allows us to identify which pairing channels are present in the numerical self-consistent solution.

%---------------------------------------------------
\begin{table}[h!]
\centering
\begin{tabular}{|c|c|c|c|c|c|c|}
    \hline
    $D_4$ & 1 & 2 & 4 & $2_{100}$ & $2_{1-10}$ & functions \\
    \hline
    $A_1$ & 1 & 1 & 1 & 1 & 1 & $x^2+y^2$ \\
    \hline
    $A_2$ & 1 & 1 & 1 & -1 & -1 & $2xy\left(x^2-y^2 \right)$ \\
    \hline
    $B_1$ & 1 & 1 & -1 & 1 & -1 & $x^2-y^2$ \\
    \hline
    $B_2$ & 1 & 1 & -1 & -1 & 1 & $2xy$ \\
    \hline
    $E$ & 2 & -2 & 0 & 0 & 0 & $(x, y)$ \\
    \hline
\end{tabular}
\caption{Character table of the $D_4$ group}
\label{tab:D4}
\end{table}
%------------------------------------------------------

For the momentum-dependent pairing functions $\Delta_l(\boldsymbol{k})$, the projection onto the irreducible representation $\Gamma$ is constructed using the standard group-theoretical projection operator,
where $l=1,2$ are the layer indices.
For the one-dimensional representations $\Gamma=A_1,A_2,B_1,B_2$, the projected component can be written as
\begin{equation}
f^\Gamma(\boldsymbol{k})=\frac{d_\Gamma}{|D_4|}\sum_{R\in D_4}\chi^\Gamma(R)\Delta_{R(l)}\left(R^{-1}\boldsymbol{k_l}\right),
\label{eq:projection_1D}
\end{equation}
where $|D_4|=8$ is the order of the group, $d_\Gamma=1$ is the dimension of the corresponding irreducible representation, $R$ runs over all symmetry operations in $D_4$, and $\chi^\Gamma(R)$ is the character of $R$ in the representation $\Gamma$.
$R(l)$ denotes the action of $R$ on the layer degree of freedom.
For the in-plane operations, the layer index of $\Delta_l(\boldsymbol{k})$ is invariant, $R(l)=l$,
whereas for operations accompanied by a layer exchange, the index is flipped such that $R(1)=2$ and $R(2)=1$.
The character table and the relevant symmetry operations are listed in Table~\ref{tab:D4}.
The obtained pairing components for the $A_1$-, $A_2$-, $B_1$-, and $B_2$-representations are denoted as $f^{A_1}(\boldsymbol{k})$, $f^{A_2}(\boldsymbol{k})$, $f^{B_1}(\boldsymbol{k})$, and $f^{B_2}(\boldsymbol{k})$, respectively.
On the other hand, for the two-dimensional irreducible representation $E$, the projection should in general be understood as a projection onto the corresponding two-dimensional representation space rather than as a single scalar component. 
In practice, this means that one projects $f(\boldsymbol{k})$ onto a chosen basis of the $E$ representation and examines the resulting two-component coefficients. 
We find that the projection onto the $E$ sector, $f^{E}(\boldsymbol{k})$, vanishes within numerical accuracy.

%-----------------------------------------------------------------------------------
\begin{figure}[htb]
    \centering
    \begin{subfigure}[b]{0.995\textwidth}
        \centering
        \includegraphics[width=\textwidth]{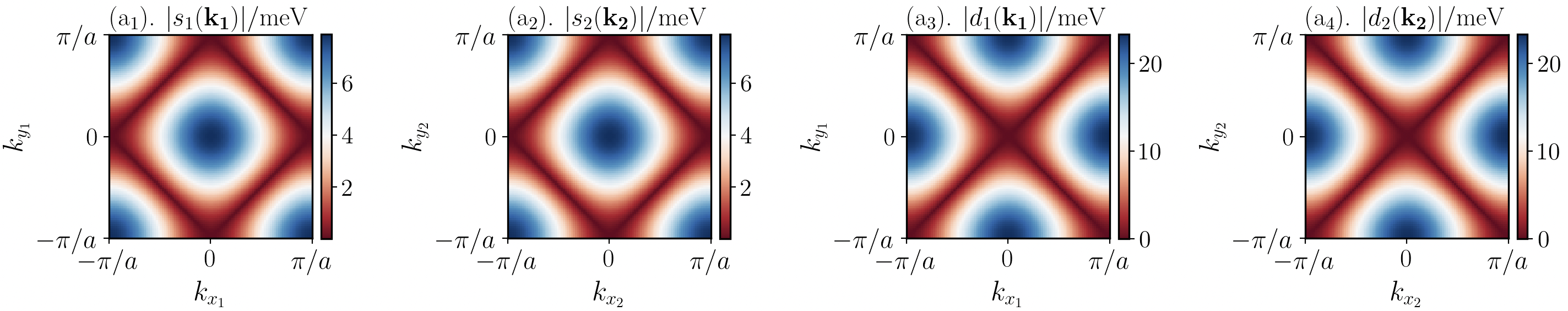}
        \caption{}
        \label{fig:sub_abs}
    \end{subfigure}
    \hfill
    \begin{subfigure}[b]{1\textwidth}
        \centering
        \includegraphics[width=\textwidth]{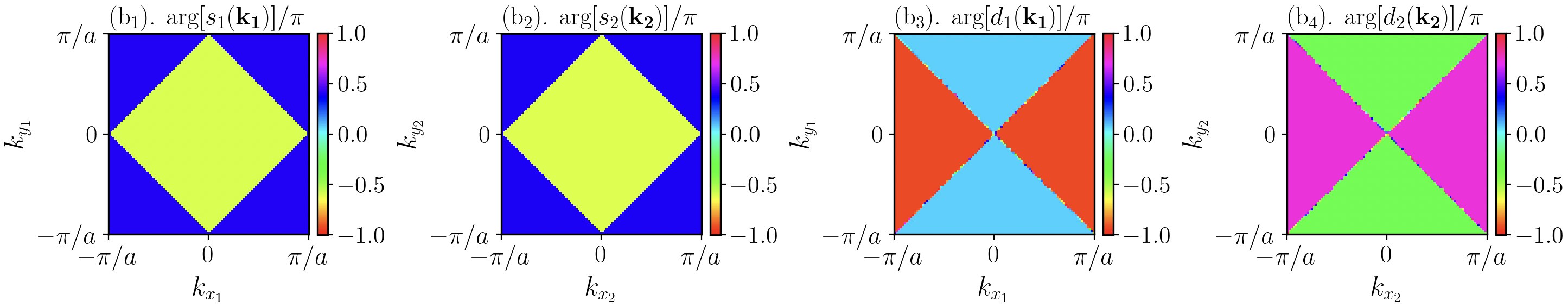}
        \caption{}
        \label{fig:sub_phase}
    \end{subfigure}
    \captionsetup{justification=raggedright}
    \caption{Plots of the (a) absolute values and (b) phases of $s_1(\boldsymbol{k}_1)$, $s_2(\boldsymbol{k}_2)$, $d_1(\boldsymbol{k}_1)$, and $d_2(\boldsymbol{k}_2)$ within the first Brillouin zone.
The parameters used to generate this figure are the same as those in Fig.~\ref{fig:Deltak}.
Here, $s_{1,2}$ and $d_{1,2}$ are defined in the rotated coordinate frames of layer 1,2, respectively,
with $\boldsymbol{k}_1 = R(\theta/2)\boldsymbol{k}$, and $\boldsymbol{k}_2 = R(-\theta/2)\boldsymbol{k}$.
Combining the amplitude and phase distributions, the basis functions for $s_{1,2}$ and $d_{1,2}$ take the form $f^{s_{1,2}}=\cos(k_{x_{1,2}} a) + \cos(k_{y_{1,2}} a)$, and $f^{d_{1,2}}=\cos(k_{x_{1,2}} a) - \cos(k_{y_{1,2}} a)$,
and the phases are spatially uniform across the Brillouin zone.}
\label{fig:s1s2d1d2}
\end{figure}
%---------------------------------------------------------------------------------

Although the components classified by the irreducible representations $A_1$, $A_2$, $B_1$, and $B_2$ provide the symmetry-adapted decomposition of the pairing gap function, it is often more physically transparent to express the pairing structure in terms of layer-resolved $s$-wave and $d$-wave components. 
These layer-resolved functions, denoted by $s_1(\boldsymbol{k}_1)$, $d_1(\boldsymbol{k}_1)$, $s_2(\boldsymbol{k}_2)$, and $d_2(\boldsymbol{k}_2)$, are defined with respect to the local rotated coordinate frames of layer 1 and layer 2, where $\boldsymbol{k}_1=R(\theta/2)\boldsymbol{k}$ and $\boldsymbol{k}_2=R(-\theta/2)\boldsymbol{k}$. 
They do not by themselves form irreducible representations of the global $D_4$ symmetry, but can be reconstructed from the symmetry-resolved components obtained above.
Specifically, the $A_1$ and $A_2$ sectors correspond to the symmetric and antisymmetric combinations of the layer-resolved $s$-wave components, while the $B_1$ and $B_2$ sectors correspond to the symmetric and antisymmetric combinations of the layer-resolved $d_{x^2-y^2}$-wave components. 
The relations are
\begin{align}
& f^{A_1}(\boldsymbol{k})=s_1(\boldsymbol{k}_1)+s_2(\boldsymbol{k}_2), \quad
f^{A_2}(\boldsymbol{k})=s_1(\boldsymbol{k}_1)-s_2(\boldsymbol{k}_2), \notag\\
&f^{B_1}(\boldsymbol{k})=d_1(\boldsymbol{k}_1)+d_2(\boldsymbol{k}_2), \quad
f^{B_2}(\boldsymbol{k})= d_1(\boldsymbol{k}_1)-d_2(\boldsymbol{k}_2).
\label{eq:layer_irrep_relation}
\end{align}

The layered resolved pairing components  can be obtained by first performing projections to the pairing gap functions in Fig. \ref{fig:Deltak} and then applying Eq. (\ref{eq:layer_irrep_relation}). 
The resulting magnitudes  and phases of $s_1(\boldsymbol{k}_1)$, $d_1(\boldsymbol{k}_1)$, $s_2(\boldsymbol{k}_2)$, and $d_2(\boldsymbol{k}_2)$ are presented in Fig. \ref{fig:s1s2d1d2}.
Although each subfigure in Fig. \ref{fig:s1s2d1d2}(b) seems to show two different phases, the two phases in fact differ only by $\pi$,
which is crucial for enforcing transformation properties in different channels, especially $d_1(\boldsymbol{k}_1)$ and $d_2(\boldsymbol{k}_1)$ in Fig. \ref{fig:s1s2d1d2} (b$_3$,~b$_4$) which change sign under $90^\circ$ rotation.
Absorbing the $\pi$-phase difference in each subfigure into the amplitude function in the corresponding subfigure in Fig.~\ref{fig:s1s2d1d2}~(a), 
the superconducting phases of $s_1$, $s_2$, $d_1$, $d_2$ become uniform across the Brillouin zone, given by $\phi_{s_1}=-0.596\pi$, $\phi_{s_2}=-0.604\pi$, $\phi_{d_1}=0.071\pi$, and $\phi_{d_2}=-0.271\pi$.

%---------------------------------------------------------------------------
\begin{figure}[h]
\centering
\includegraphics[width=0.275\textwidth]{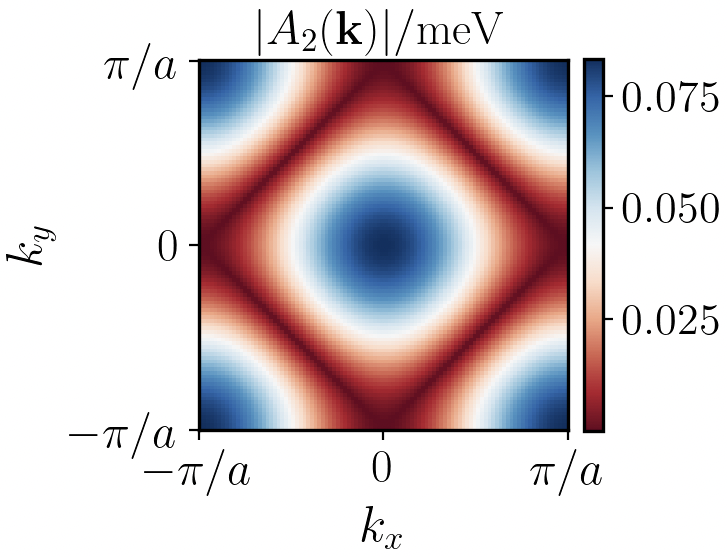}
\captionsetup{justification=raggedright}
\caption{Absolute value of $f^{A_2}(\boldsymbol{k})$ within the first Brillouin zone.
The parameters used to generate this figure are the same as those in Fig.~\ref{fig:Deltak}.}
\label{fig:A2}
\end{figure}
%---------------------------------------------------------------------------

The absolute value of $f^{A_2}(\boldsymbol{k})$ is shown in Fig.~\ref{fig:A2}.
As can be seen, the maximum magnitude of $f^{A_2}(\boldsymbol{k})$ is much smaller than that of $s_{1,2}(\boldsymbol{k_{1,2}})$ and $d_{1,2}(\boldsymbol{k_{1,2}})$, implying that the $(s_1-s_2)$-pairing channel can be neglected.

%%%%%%%%%%%%%%%%%%%%%%%%%%%%%%%%%%%%%%%%%%%%%%%%%

\subsection{Quantifying pairing strengths in different channels}
\label{app:Quantify}

Unlike the superconducting phases, the magnitudes of different pairing components $|s_1(\boldsymbol{k}_1)|$, $|d_1(\boldsymbol{k}_1)|$, $|s_2(\boldsymbol{k}_2)|$, and $|d_2(\boldsymbol{k}_2)|$ are not homogenous in the Brillouin zone, making them not directly available for a comparison of relative strengths. 
To quantify the contributions from different pairing channels, we determine their fractions in the superfluid condensate.
%For notational convenience, we define $F(\boldsymbol{k}_l)$ ($l=1,2$) as
%\begin{eqnarray}
%F(\boldsymbol{k}_l)=F_l(\boldsymbol{k}),
%\end{eqnarray}
%in which $\boldsymbol{k}_l$ are related to $\boldsymbol{k}$ via  $\boldsymbol{k}_1=\hat{R}(\theta/2) \boldsymbol{k}$, $\boldsymbol{k}_2=\hat{R}(-\theta/2) \boldsymbol{k}$. 
%%%%%%%%%%%%%%%%%%%%%%%%%%%%%%%%%%%%%%%%%%%%%%%
%\subsection{Quantification of pairing strengths through decomposition of pairing wave function}
In conventional $s$-wave superconductors, the Cooper pair densities $N_0/\Omega$ can be related to the (un-normalized) pairing wave function $F(\boldsymbol{k})= \langle c_{\boldsymbol{k}\uparrow}c_{-\boldsymbol{k}\downarrow} \rangle$ via
\begin{equation}
\frac{N_0}{\Omega}=\sum_{\boldsymbol{k}} |F(\boldsymbol{k})|^2=\frac{\pi}{4}\Delta N(0),
\label{eq:condensate_s}
\end{equation}
where $\Omega$ is the volume of the system, $\Delta$ is the pairing strength and $N(0)$ is the density of states at Fermi energy. 
As a result, up to an overall numerical factor, the pairing strength can be characterized by the condensate density $N_0/\Omega$. 

In the present situation, the layer resolved pairing wave function in real space can be defined as
\begin{eqnarray}
F_l(\boldsymbol{r}_i-\boldsymbol{r}_j) = \langle c_{i\uparrow l}c_{j\downarrow l} \rangle,
\end{eqnarray}
where $l=1,2$ is the layer index, and $\boldsymbol{r}_i,\boldsymbol{r}_j$ are the position vectors of sites $i,j$.
Then $F_l(\boldsymbol{k}_l)$ can be obtained from Fourier transforming $F_l(\boldsymbol{r}_i-\boldsymbol{r}_j)$. 

Starting from $F_l(\boldsymbol{k}_l)$,  
the component $F^\Gamma(\boldsymbol{k})$ of the pairing wave function in channel $\Gamma$ ($\Gamma=A_{1},A_2,B_{1},B_2$) can be obtained by projection, using the same group-theoretical method as Eq. (\ref{eq:projection_1D}).  
In analogy with Eq. (\ref{eq:condensate_s}), the pairing strength $\Delta_\Gamma$ can be quantified as 
\begin{eqnarray}
\Delta_{\Gamma}=\frac{\sum_{\boldsymbol{k}} \left|F^\Gamma(\boldsymbol{k}) \right|^2}{\pi N(0)/4}.
\end{eqnarray}
To be proved below,
\begin{eqnarray}
\sum_{l=1,2}\sum_{\boldsymbol{k}} |F_l(\boldsymbol{k})|^2= \sum_{\Gamma}\sum_{\boldsymbol{k}} \left|F^\Gamma(\boldsymbol{k}) \right|^2,
\label{eq:completeness_F}
\end{eqnarray}
hence $\Delta_{\Gamma}$ can be used to quantitatively characterize the fraction of Cooper pairs in pairing channel $\Gamma$  in the superfluid.
The pairing strengths in the layer-resolved basis channels $s_1,s_2,d_1,d_2$ can be obtained as
\begin{align}
\left|\psi_{s_1} \right|e^{i\phi_{s_1}}&=\frac{1}{2} (\Delta_{A_1}e^{i\phi_{A_1}}+\Delta_{A_2}e^{i\phi_{A_2}}), \quad \left|\psi_{s_2} \right|e^{i\phi_{s_2}}=\frac{1}{2} (\Delta_{A_1}e^{i\phi_{A_1}}-\Delta_{A_2}e^{i\phi_{A_2}}), \notag\\
\left|\psi_{d_1} \right|e^{i\phi_{d_1}}&=\frac{1}{2} (\Delta_{B_1}e^{i\phi_{B_1}}+\Delta_{B_2}e^{i\phi_{B_2}}), \quad \left|\psi_{d_2} \right|e^{i\phi_{d_2}}=\frac{1}{2} (\Delta_{B_1}e^{i\phi_{B_1}}-\Delta_{B_2}e^{i\phi_{B_2}}),
\label{eq:define_psi}
\end{align}
in which $\phi_{\Gamma}$ is the phase of the pairing channel $\Gamma$.

%--------------------------------------------------------------------------------------------------------

%\textit{Proof of completeness relation in Cooper pair density.} --
%\label{app:proof_Cooper_density}
Let $X=\{(l,\boldsymbol{k})\mid l=1,2\}$, and regard the pairing amplitude as a function on this set, we have
\begin{equation}
F(l,\boldsymbol{k})\equiv F_l(\boldsymbol{k})
=
\langle c_{l,\boldsymbol{k}\uparrow}c_{l,-\boldsymbol{k}\downarrow}\rangle.
\end{equation}
Since the point group $D_4$ may interchange the two layers, its action is naturally defined on the extended variable $(l,\boldsymbol{k})$ rather than on $\boldsymbol{k}$ alone. Thus, for $R\in D_4$, we write
\begin{equation}
R:(l,\boldsymbol{k})\mapsto R(l,\boldsymbol{k}),
\end{equation}
where some group elements preserve the layer index while others exchange $l=1$ and $l=2$,
which induces a representation on the function space over $X$,
\begin{equation}
[T(R)f](x)=f(R^{-1}x), \qquad x\in X.
\end{equation}
We equip this space with the inner product
\begin{equation}
\langle f,g\rangle=\sum_{l=1}^2\sum_{\boldsymbol{k}} f(l,\boldsymbol{k})^*\,g(l,\boldsymbol{k}).
\end{equation}
Assuming that the momentum set is invariant under the $D_4$ action, $T(R)$ is unitary, $\langle T(R)f,T(R)g\rangle=\langle f,g\rangle$.

The projection onto an irreducible representation $\Gamma$ is defined by
\begin{equation}
P_\Gamma=\frac{d_\Gamma}{|D_4|}
\sum_{R\in D_4}\chi^\Gamma(R)^*\,T(R),
\end{equation}
and the corresponding symmetry-resolved component is $F^\Gamma=P_\Gamma F$.
Note that $F^\Gamma$ is obtained by mixing the values of $F$ on different layers whenever the group action exchanges $(1,\boldsymbol{k})$ and $(2,\boldsymbol{k})$. Therefore, $F^\Gamma$ should be viewed as a projected component in the full representation space over $X$, rather than as an object carrying a fixed layer index.

By the standard orthogonality relations of finite-group representations,
$P_\Gamma P_{\Gamma'}=\delta_{\Gamma\Gamma'}P_\Gamma$,
and since $T(R)$ is unitary, the projectors are Hermitian,
$P_\Gamma^\dagger=P_\Gamma$.
Therefore, different symmetry components are orthogonal:
\begin{equation}
\langle F^\Gamma,F^{\Gamma'}\rangle
=
\langle P_\Gamma F,P_{\Gamma'}F\rangle
=
\langle F,P_\Gamma P_{\Gamma'}F\rangle
=
\delta_{\Gamma\Gamma'}\langle F,P_\Gamma F\rangle
=
\delta_{\Gamma\Gamma'}\langle F^\Gamma, F^\Gamma \rangle.
\end{equation}
In particular, $\langle F^\Gamma,F^{\Gamma'}\rangle=0$ for $\Gamma\neq\Gamma'$.

The irreducible representations of $D_4$ consist of four one-dimensional irreducible representations $A_1$, $A_2$, $B_1$, and $B_2$, together with one two-dimensional irreducible representation $E$. If the pairing amplitude has no component in the $E$ sector, then
\begin{equation}
F=\sum_{\Gamma\in\{A_1,A_2,B_1,B_2\}}F^\Gamma.
\end{equation}
Using the orthogonality of different irreducible components, one finds
\begin{align}
\sum_{l=1}^2\sum_{\boldsymbol{k}}|F_l(\boldsymbol{k})|^2=\langle F,F\rangle
=\left\langle \sum_\Gamma F^\Gamma,\sum_{\Gamma'}F^{\Gamma'}\right\rangle
=\sum_{\Gamma,\Gamma'}\langle F^\Gamma,F^{\Gamma'}\rangle
=\sum_\Gamma \langle F^\Gamma,F^\Gamma\rangle.
\end{align}
Therefore,
\begin{equation}
\sum_{l=1}^2\sum_{\boldsymbol{k}}|F_l(\boldsymbol{k})|^2
=
\sum_{\Gamma\in\{A_1,A_2,B_1,B_2\}}\|F^\Gamma\|^2,
\end{equation}
where $\|F^\Gamma\|^2\equiv \langle F^\Gamma,F^\Gamma\rangle$.

To simplify the notation, one may introduce a reduced momentum-space function $F^\Gamma(\boldsymbol{k})$.
By adopting a proper normalization that preserves the norm of the projected component in the full $(l,\boldsymbol{k})$ space,
one have $\sum_{\boldsymbol{k}}|F^\Gamma(\boldsymbol{k})|^2=\|F^\Gamma\|^2$,
which corresponds to Eq.~\eqref{eq:completeness_F}.
%Without such a definition, the most precise form of the identity is

For example, at $\theta=43.6^\circ$ with other parameters detailed in Fig.~\ref{fig:Deltak}, % in Sec.~\ref{app:self_consistent},
the resulting pairing strengths in different channels in unit of meV are $|\psi_{A_1}|=3.593$, $|\psi_{A_2}|=0.046$, $|\psi_{B_1}|=16.133$, $|\psi_{B_2}|=9.608$, $|\psi_{s_{1,2}}|=1.797$, and $|\psi_{d_{1,2}}|=9.389$,
while the corresponding phases are given by $\phi_{A_1}=-0.6\pi$, $\phi_{A_2, B_1}=-0.1\pi$, $\phi_{B_2}=0.4\pi$, $\phi_{s_1}=-0.596\pi$, $\phi_{s_2}=-0.604\pi$, $\phi_{d_1}=0.071\pi$, and $\phi_{d_2}=-0.271\pi$.
% \subsubsection{Quantification of pairing strengths}
% The full four-component pairing gap function, comprising $s_{1,2}$- and $d_{1,2}$-waves, can be represented as $\hat{\Delta}=\psi_{s_1}\hat{s}_1+\psi_{s_2}\hat{s}_2+\psi_{d_1}\hat{d}_1+\psi_{d_2}\hat{d}_2$.
Notice that $|\psi_{A_2}|$ is negligibly small, allowing us to omit the $(s_1-s_2)$-pairing.

%%%%%%%%%%%%%%%%%%%%%%%%%%%%%%%%%%%%%%%%%%%%%%%%%%%%

\subsection{Origin of the suppression of the $s_1-s_2$ channel}

Consider a continuum model in which the two layers are rotated by $\pm \theta/2$, the superconducting gap in each layer contains both $s$- and $d_{x^2-y^2}$-wave components,
\begin{equation}
\Delta_{\boldsymbol{k_1}}=s_1+d_1\cos(2\alpha_{\boldsymbol{k}}-\theta),
\quad
\Delta_{\boldsymbol{k_2}}=s_2+d_2\cos(2\alpha_{\boldsymbol{k}}+\theta),
\label{eq:Deltak12}
\end{equation}
where the $s$-wave form factor has been approximated as unity. The effective 2-band Hamiltonian can be written as
\begin{equation}
h_{\boldsymbol{k}}=
\begin{pmatrix}
\xi_{\boldsymbol{k}}&\Delta_{\boldsymbol{k_1}}&g(\boldsymbol{k})&0\\
\Delta_{\boldsymbol{k_1}}^*&-\xi_{\boldsymbol{k}}&0&-g(\boldsymbol{k})\\
g(\boldsymbol{k})&0&\xi_{\boldsymbol{k}}&\Delta_{\boldsymbol{k_2}}\\
0&-g(\boldsymbol{k})&\Delta_{\boldsymbol{k_2}}^*&-\xi_{\boldsymbol{k}}
\end{pmatrix},
\label{eq:hk_2band}
\end{equation}
where $g(\boldsymbol{k})$ is the interlayer tunneling.
Expanding the BdG free energy,
\begin{equation}
\mathcal{F}_{\text{BdG}}=-2k_BT\sum_{\boldsymbol{k},\lambda} \ln\left[2\cosh\left(\frac{E_{\boldsymbol{k}\lambda}}{2k_BT}\right)\right] +\mathcal{F}_{\text{int}},
\label{eq:FBdG}
\end{equation}
in powers of $(s_1,d_1,s_2,d_2)$ generates the complete multicomponent GL free energy. % with the quadratic interlayer part
%\begin{equation}
%\mathcal F_{\text{inter}}^{(2)}= \left(B_{ss}s_1s_2^*+B_{dd}d_1d_2^*+B_{sd}s_1d_2^*+B_{ds}d_1s_2^*\right) +c.c..
%\label{eq:F2_int}
%\end{equation}
To isolate the coefficients of $s_1s_2^*+c.c.$ ($B_s$) and $s_1^2s_2^{*2}+c.c.$ ($C_s$), we  set $d_{1,2}=0$ and take the derivatives of $\mathcal{F}_{\text{BdG}}$ with respect to $s_{1,2}$. % at vanishing order parameters. % set $d_1=d_2=0$ temporarily and; which does not amount to neglecting the $d$-wave components in the complete GL theory.
Writing $s_1=\rho_s$ and $s_2=\rho_s e^{i\phi_s}$ with $\phi_s=\phi_{s2}-\phi_{s1}$, the positive BdG eigenenergies become
\begin{equation}
E_{\lambda}^2=\xi^2+g^2+\rho_s^2+2\lambda g \sqrt{\xi^2+\rho_s^2\sin^2\left(\phi_s/2\right)}, \quad \lambda=\pm1.
\label{eq:E_lambda}
\end{equation}

Expanding the energy-resolved free energy to fourth order in $\rho_s$ gives
\begin{equation}
\delta\mathcal{F}_{\xi}=K_{1,s}(\xi)\rho_s^2\cos\phi_s+K_{2,s}(\xi)\rho_s^4\cos2\phi_s +\cdots.
\label{eq:F_xi}
\end{equation}
Defining $e_\lambda=|\xi+\lambda g|$, $t_\lambda=\tanh\left(e_\lambda/2k_BT\right)$, and $u_\lambda=\text{sech}^2\left(e_\lambda/2k_BT\right)$,
the first- and second-harmonic kernels are
\begin{equation}
K_{1,s}(\xi)=\sum_{\lambda=\pm} \frac{\lambda g\,t_\lambda}{4\xi e_\lambda}, \quad
K_{2,s}(\xi)=\sum_{\lambda=\pm}\left[-\frac{g^2u_\lambda}{128k_BT\,\xi^2e_\lambda^2} + \frac{\lambda g t_\lambda}{64\xi^3e_\lambda} + \frac{g^2t_\lambda}{64\xi^2e_\lambda^3}\right],
\label{eq:K1sK2s}
\end{equation}
as shown in Fig.~\ref{fig:Bs_Cs}, in which $K_{1,s}$ is negative throughout the integration range. % the singularities at $\xi=0,g$ are removable after summing over $\lambda$.
The GL coefficients are then obtained from
\begin{equation}
B_s=\mathcal{N}\int_0^{\epsilon_c}K_{1,s}(\xi)d\xi<0, \quad
C_s=\mathcal{N}\int_0^{\epsilon_c}K_{2,s}(\xi)d\xi,
\label{eq:BC_s}
\end{equation}
where $\mathcal N>0$ contains the density of states and convention-dependent numerical factors common to the two coefficients.
The corresponding phase-dependent contribution is
\begin{equation}
\mathcal{F}_s(\phi_s)=2B_s\rho_s^2\cos\phi_s + 2C_s\rho_s^4\cos2\phi_s.
\label{eq:Fs_phi}
\end{equation}

%------------------------------------------------
\begin{figure}[htb]
    \centering
    \includegraphics[width=0.6\textwidth]{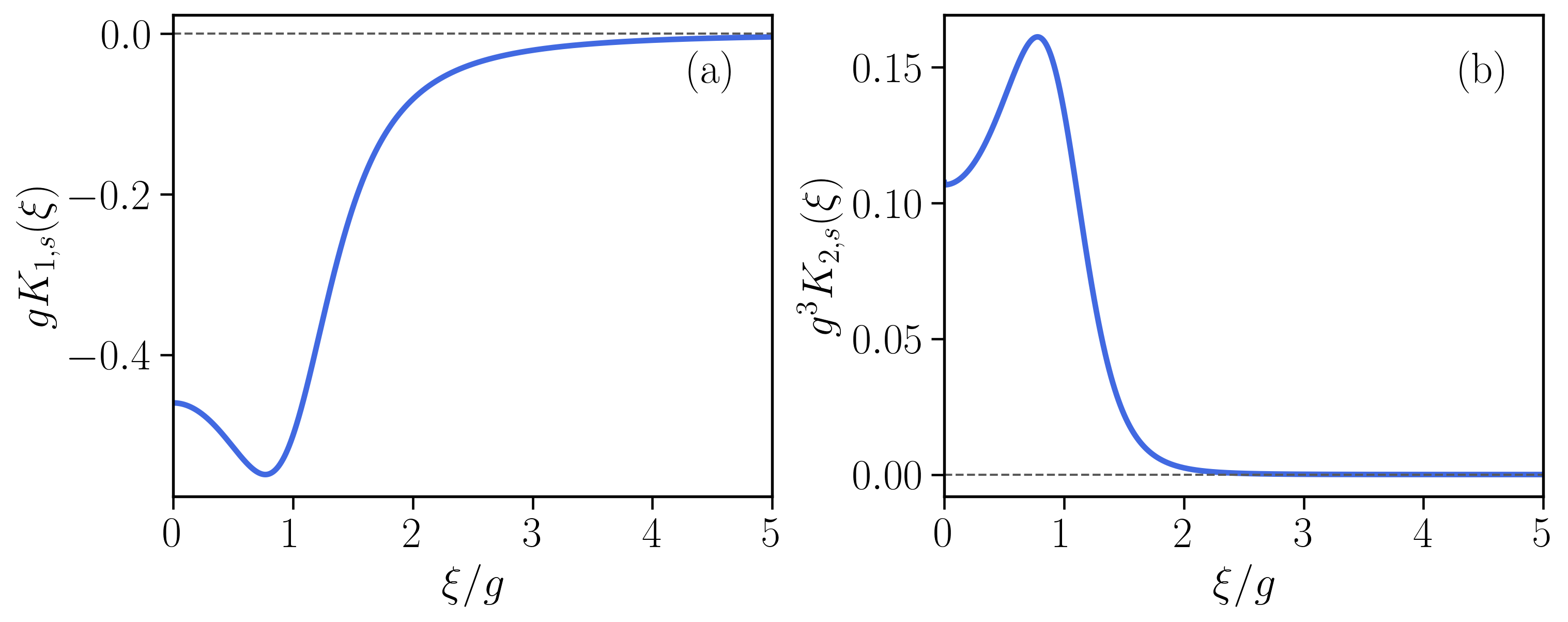}
    \captionsetup{justification=raggedright}
    \caption{Energy integrands associated with the first- and second-harmonic interlayer $s$-wave couplings, (a) $gK_{1,s}(\xi)$ and (b) $g^3K_{2,s}(\xi)$.
    Throughout the integration range, $K_{1,s}(\xi)<0$, whereas $K_{2,s}(\xi)>0$.} % and approximately follows $K^2_{1,s}$.
    \label{fig:Bs_Cs}
\end{figure}
%----------------------------------------------

% For $k_BT=0.2g$ and $\epsilon_c=5g$, the dimensionless integrals are approximately
More importantly, the second-harmonic contribution to the free energy carries an additional factor of $\rho_s^2$.
To quantify its magnitude relative to the first-harmonic Josephson coupling, we define the dimensionless ratio
\begin{equation}
\eta_s \equiv \frac{4C_s\rho_s^2}{|B_s|}=\frac{4\rho_s^2 \int_0^{\epsilon_c}K_{2,s}(\xi)d\xi}{\left|\int_0^{\epsilon_c}K_{1,s}(\xi)d\xi\right|} \simeq 0.86\left(\frac{\rho_s}{g}\right)^2,
\end{equation}
where we take $k_B T=0.2g$ and $\epsilon_c=5g$, respectively, corresponding to physically motivated estimates of the temperature and cutoff energy scales \cite{Can2021SM}.
The factor of 4 arises from minimizing $\mathcal{F}_s$, rather than the bare energy-amplitude ratio $C_s\rho_s^2/|B_s|$, directly controls the stability of the phase-locked solution $\phi_s=0$.
As shown in Fig.~\ref{fig:sd1d2_T}, in the parameter regime of interest, the $s$-wave order-parameter amplitude satisfies approximately $\rho_s/g \lesssim 1/6$, giving $\eta_s \lesssim 0.024$.
Therefore, the second-harmonic contribution of the interlayer $s$-wave coupling is parametrically small compared with the first-harmonic term.
Since $B_s<0$, the $B_s(s_1s_2^*+s_1^*s_2)$ term pins the relative phase close to $\phi_{s2}-\phi_{s1}=0$, with only small corrections from the second-harmonic terms.
Consequently, the antisymmetric combination $s_1-s_2$ is negligibly small, implying $A_2 \simeq 0$.
By contrast, the first-harmonic coupling of the the $d$-wave pairing is $\theta$-dependent, $B_d\propto\cos(2\theta)$, and hence becomes strongly suppressed as $\theta$ approaches $45^\circ$.
In this regime, the first-harmonic term no longer dominates the higher-harmonic couplings, so the relative phase $\phi_{d2}-\phi_{d1}$ need not be pinned to zero.
These estimates therefore justify neglecting the $s_1-s_2$ channel and retaining only the symmetric channel $s=s_1+s_2$ as a well-controlled approximation.

%%%%%%%%%%%%%%%%%%%%%%%%%%%%%%%%%%%%%%%%%%%%%%%%%%%%%

\section{Temperature dependence of order parameters for various interlayer tunneling strength}
\label{app:sd1d2T_vec}

To investigate the evolution of the $s$-, $d_1$-, and $d_2$-wave pairing order parameters with temperature,
we solve the self-consistent equations from $T=0$~K up to the bulk critical temperature  $T_c$.
Focusing on two specific twist angles, $\theta=53.13^\circ$ and $\theta=43.6^\circ$,
the temperature dependence of the pairing strengths and phases of the $s$-, $d_1$-, and $d_2$-wave order parameters calculated under several representative sets of $(g_0, \mu)$ parameters, are presented in Fig.~\ref{fig:sd1d2_T}. % and Fig.~\ref{fig:sd1d2_T_vec12}, respectively.
Here, Regions I$_0$ and I$_{-4}$ represent the topologically trivial ($C=0$) and nontrivial ($C=-4$) sector of the frustrated $s+d_1e^{i\phi_1}+d_2e^{i\phi_2}$ pairing state, respectively.
Region II denotes the topologically nontrivial $d_1e^{i\phi_1}+d_2e^{i\phi_2}$ pairing states, while Region IV corresponds to the pure $d_1\pm d_2$ states.
Most notably, the topologically nontrivial $s+d_1e^{i\phi_1}+d_2e^{i\phi_2}$ state can be stabilized at both twist angles and can persist down to $g_0=0.02$~eV upon tuning the chemical potential and temperature.

Across all parameter sets, the $s$-, $d_1$-, and $d_2$-wave pairing strengths exhibit standard BCS-like square-root behavior toward their respective critical temperatures, $T_c^s$ and $T_c^d$.
Since $T_c^s<T_c^d$ in these parameter regimes, the $s$-wave component is suppressed more rapidly than the $d_{1,2}$-wave component as temperature increases, driving the frustrated three-component state from its topologically trivial sector into the nontrivial one.
Upon reaching $T_c^s$, the $s$-wave component vanishes, leaving a topologically nontrivial $d_1e^{i\phi_1}+d_2e^{i\phi_2}$ state over a subsequent temperature window, where the relative phase $\left|\phi_1-\phi_2\right|$ is not necessarily locked to $\pi/2$.
With further heating, the relative phase eventually approaches $0$ or $\pi$, and the system enters the $d_1\pm d_2$ state before superconductivity disappears at $T_c$.

% Furthermore, the bulk superconducting critical temperature of the twisted bilayer system matches the monolayer critical temperature $T_c$.
Crucially, comparison between the two twist angles in Figs.~\ref{fig:sd1d2_T}~(a,b) and \ref{fig:sd1d2_T}~(c,d) shows that both $T_c^s$ and the temperature span of the topologically nontrivial frustrated three-component phase are generally larger at $\theta=43.6^\circ$ than at $\theta=53.13^\circ$.
This indicates that as the twist angle approaches $45^\circ$, the topologically compatible $s$-wave pairing component is enhanced,
thereby stabilizing and expanding the frustrated three-component pairing state $s+d_1e^{i\phi_1}+d_2e^{i\phi_2}$ without destroying the topological nature.

%-----------------------------------------------------------------------------------
\FloatBarrier
\begin{figure}[p]
    \centering
     \begin{subfigure}{0.81\textwidth}
        \centering
        \includegraphics[width=\textwidth]{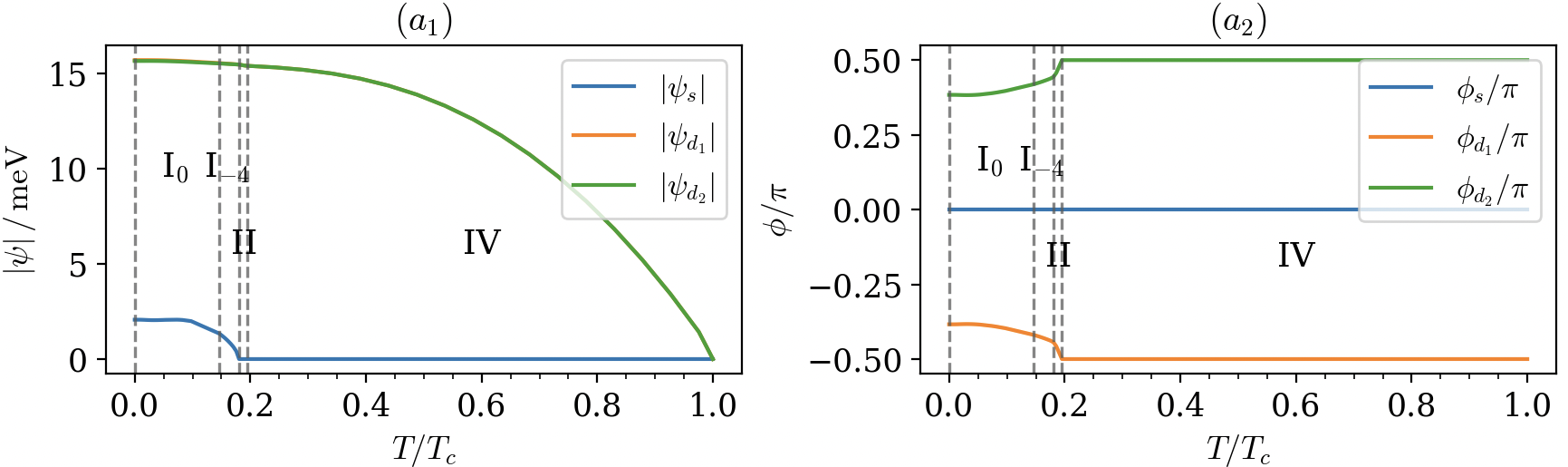} % 1731*516
        \caption{$\theta=53.13^{\circ}$, $g_0=0.02$~eV, $\mu=-1.34t$.}
        \label{subfig:vec12g020}
    \end{subfigure}
    \hfill
    \begin{subfigure}{0.81\textwidth}
        \centering
        \includegraphics[width=\textwidth]{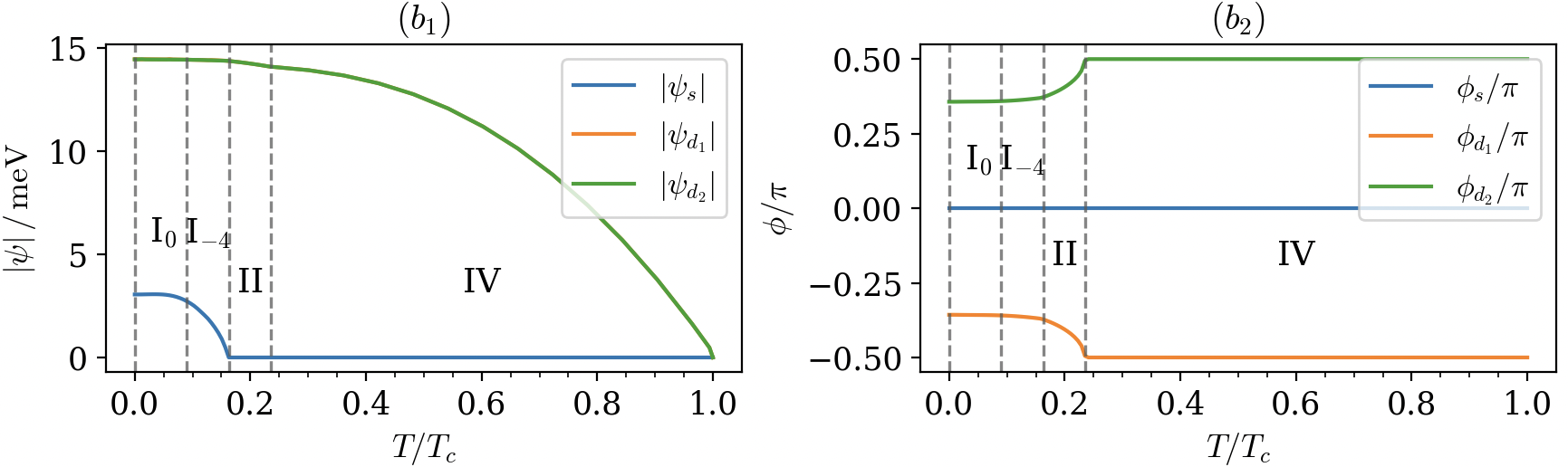} % 1731*516
        \caption{$\theta=53.13^{\circ}$, $g_0=0.024$~eV, $\mu=-1.2t$.}
        \label{subfig:vec12g024}
    \end{subfigure}
    \hfill
    \begin{subfigure}{0.81\textwidth}
        \centering
        \includegraphics[width=\textwidth]{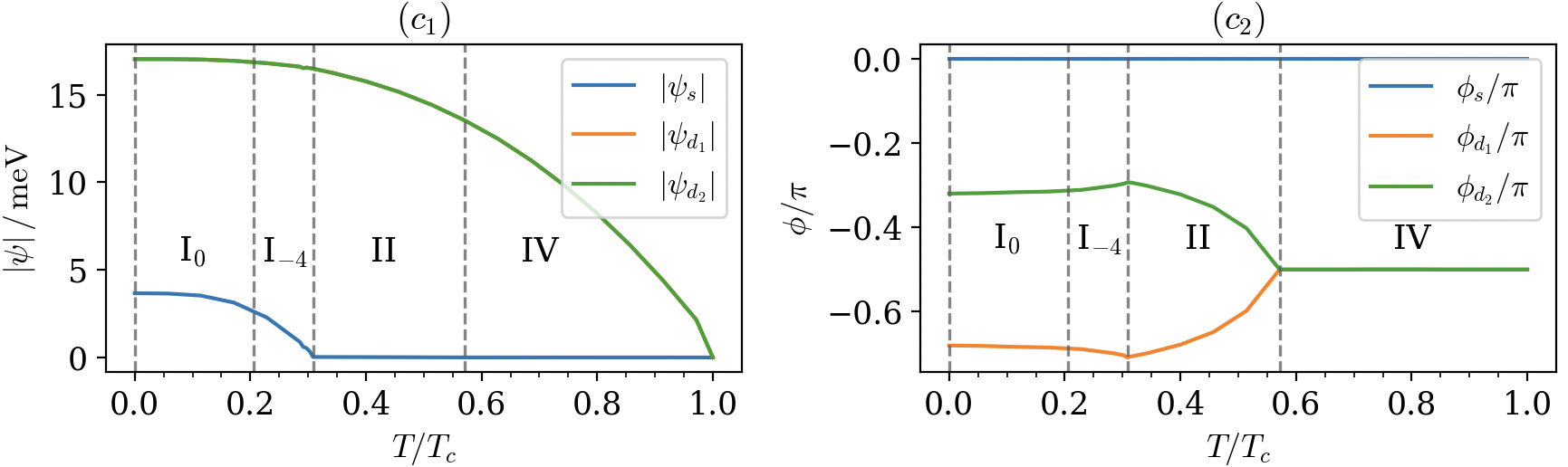} % 1731*516
        \caption{$\theta=43.6^{\circ}$, $g_0=0.024$~eV, $\mu=-1.25t$.}
        \label{subfig:vec25g024}
    \end{subfigure}
    \hfill
    \begin{subfigure}{0.81\textwidth}
        \centering
        \includegraphics[width=\textwidth]{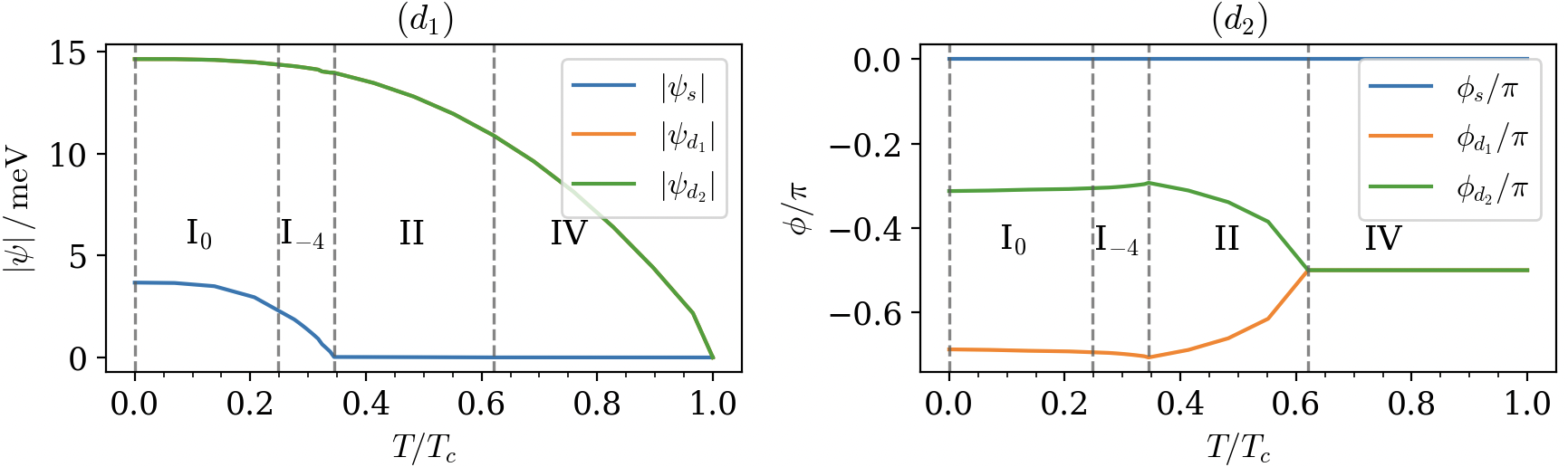} % 1731*516
        \caption{$\theta=43.6^{\circ}$, $g_0=0.028$~eV, $\mu=-1.15t$.}
        \label{subfig:vec25g028}
    \end{subfigure}
\captionsetup{justification=raggedright}
\caption{Temperature dependence of the amplitudes and phases of the $s$-, $d_1$-, and $d_2$-wave pairing order parameters at $\theta=53.13^\circ$ and $\theta=43.6^\circ$.
Vertical dashed lines mark the boundaries between the different pairing regimes.
The parameters are $t=0.153$~eV, $t^\prime=-0.45t$, $U=0$, $V=0.146$~eV, $d=2.22a$, and $\rho=0.39a$, while $g_0$ and $\mu$ are specified in the corresponding subfigure captions.
Increasing the interlayer tunneling strength $g_0$ enhances the $s$-wave pairing, as reflected by the simultaneous increase of both $T_c^s/T_c$ and $\left|\psi_s \right|/\left|\psi_{d_{1,2}} \right|$.}
    % This enhancement leads to the emergence of the topologically nontrivial three-component pairing state, $s+d_1e^{i\phi_1}+d_2e^{i\phi_2}$, over a broader temperature range.
\label{fig:sd1d2_T}
\end{figure}
\FloatBarrier
%-----------------------------------------------------------------------------------------------

%%%%%%%%%%%%%%%%%%%%%%%%%%%%%%%%%%%%%%%%%%%%%%%%

\section{Further characterization of the superconducting phase diagrams}

\subsection{Chern-number characterization}

%-----------------------------------------------------------------------------------
\begin{figure}[htb]
    \centering
    \begin{subfigure}{0.31\textwidth}
        \hspace*{-0.36cm}
        \includegraphics[width=\textwidth]{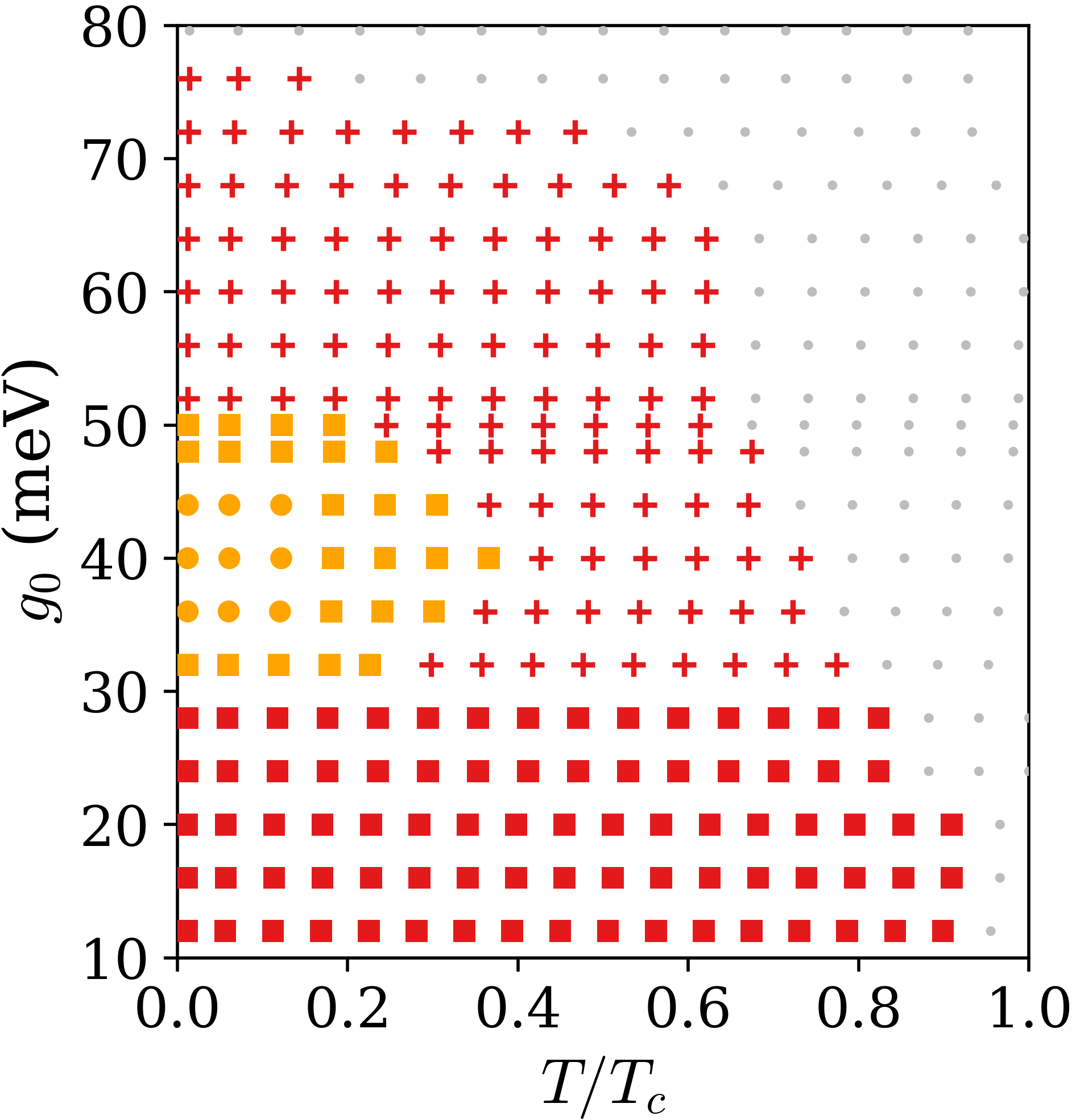} % 1885*1970
        \caption{$\theta=43.6^\circ$, $\mu=-0.9t$.}
        \label{subfig:v25mu-0.9chern}
    \end{subfigure}
    \hfill
    %\hspace*{0.6cm}
    \begin{subfigure}{0.31\textwidth}
       \hspace*{-0.36cm}
        \includegraphics[width=\textwidth]{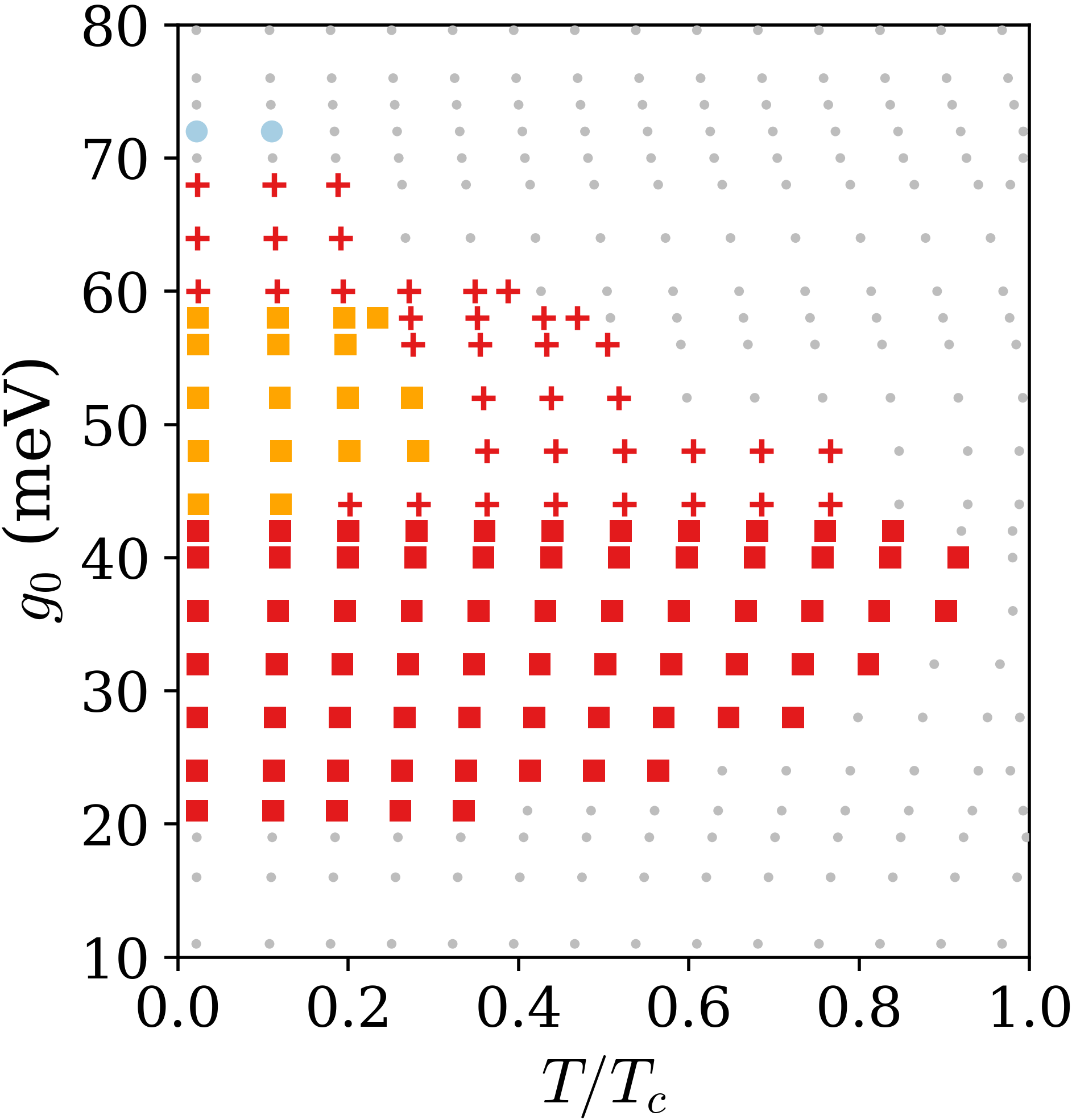} % 1885*1970
        \caption{$\theta=53.13^\circ$, $\mu=-0.5t$.}
        \label{subfig:v12mu-0.5chern}
    \end{subfigure}
    \hfill
    %\hspace*{1.1cm}
    \begin{subfigure}{0.31\textwidth}
         \hspace*{-0.6cm}
        \includegraphics[width=\textwidth]{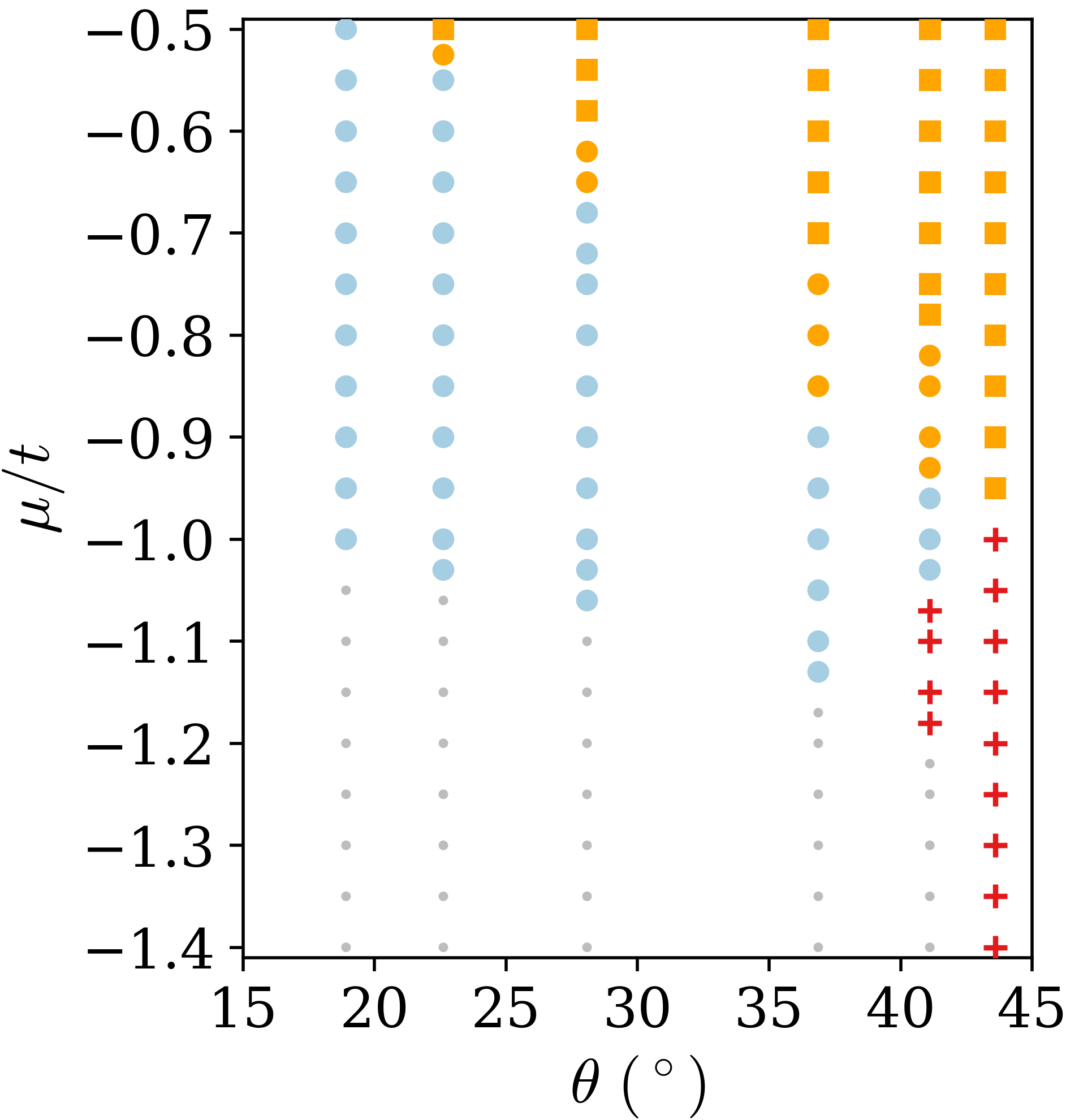} % 1885*1981
        \caption{$g_0=50$meV, $T=0$K.}
        \label{subfig:mu_theta_chern}
    \end{subfigure}
    \captionsetup{justification=raggedright}
    \caption{Chern-number characterization of the superconducting phase diagrams at the sampled parameter points corresponding to Fig.~1 in the main text.
    Orange, red, blue, and gray represent the $s+d_1e^{i\phi_1}+d_2e^{i\phi_2}$, $d_1e^{i\phi_1}+d_2e^{i\phi_2}$, $s+i(d_1\pm d_2)$, and $d_1\pm d_2$ states,
    while squares, circles, and plus signs denote $C=-4$, $C=0$, and $C=-2$, respectively.}
\label{fig:PhaseDiagramChern}
\end{figure}
%---------------------------------------------------------------------------------

To further characterize the topology of the superconducting phases in Fig.~1 in the main text, we calculate the Chern number $C$ at each sampled point in the parameter space, as shown in Fig.~\ref{fig:PhaseDiagramChern}.
The resulting Chern numbers confirm that the $s+d_1e^{i\phi_1}+d_2e^{i\phi_2}$ state hosts both topologically trivial and nontrivial sectors,
with the $C=-4$ nontrivial sector stabilized over a broad parameter range and separated from the trivial sector by bulk-gap closing boundaries.
Notably, this frustrated three-component pairing state can also support $C=-2$ at other parameter values, although such a sector is absent among the sampled points shown here.
The chiral and topologically nontrivial $d_1e^{i\phi_1}+d_2e^{i\phi_2}$ state carries $C=-4$ or $C=-2$,
with the former generally favored at smaller $g_0$ and the latter at larger $g_0$, while their precise boundary depends on the remaining parameters, such as the twist angle and chemical potential.
In contrast, the $s+i(d_1\pm d_2)$ state is topologically trivial with $C=0$.
However, the $d_1\pm d_2$ state is gapless, so a global Chern number is not well defined; its data points are therefore displayed with smaller markers in Fig.~\ref{fig:PhaseDiagramChern} and should not be interpreted as a conventional gapped topological invariant.
These results provide the topological characterization underlying the phase diagrams presented in the main text.

%----------------------------------------------------------------------------------------------------------------------------

\subsection{Topological transition in the frustrated three-component superconducting state}
\label{subsec:topo_transition}

We now examine how the topologically trivial and nontrivial sectors are connected in the frustrated three-component $s+d_1e^{i\phi_1}+d_2e^{i\phi_2}$ state across a representative temperature-driven topological transition.
The self-consistent solutions are topologically trivial with $C=0$ at $T=16$~K, whereas the $C=-4$ phase is established at $T=18$~K.
% A local two-band analysis identifies this process as a simultaneous Dirac-mass inversion at the four valleys.

% ============================================================
% occupied Berry curvature at T = 16 and 18 K
% ============================================================
To uncover the origin of this change, we first resolve the Chern-number evolution in momentum space,
\begin{equation}
C=\frac{1}{2\pi} \int_{\text{BZ}}d^2k \Omega_{\text{occ}}(\boldsymbol{{k}}),
\label{eq:chern_occ}
\end{equation}
where $\Omega_{\text{occ}}(\boldsymbol{{k}})$ is the occupied-space Berry curvature, defined as
% The momentum-space redistribution of Berry curvature provides a direct connection between these four Dirac valleys and the change of the Chern number. We define the Berry curvature summed over all occupied BdG states as
\begin{equation}
\Omega_{\text{occ}}(\boldsymbol{k})=\sum_{n\in{\text{occ}}}\Omega_n(\mathbf{k}).
\end{equation}
Fig.~\ref{fig:Omega_occ} compares $\Omega_{\text{occ}}(\boldsymbol{k})$ at $T=16$~K and $18$~K, on the two sides of the transition.
Remarkably, the Berry-curvature distribution over most of the Brillouin zone remains nearly unchanged, while the substantial redistribution is concentrated in four well-localized momentum-space regions.
This suggests that the change of the Chern number originates from four isolated low-energy sectors rather than from a broad reconstruction of the occupied BdG bands.
% The Berry-curvature distribution away from the four low-energy valleys remains nearly unchanged, while a pronounced redistribution occurs in the immediate vicinity of each $\boldsymbol{k}_{D,i}$.

%-----------------------------------------------
\begin{figure}[htb]
    \centering
    \begin{subfigure}{0.3\textwidth}
        \centering
        \includegraphics[width=\textwidth]{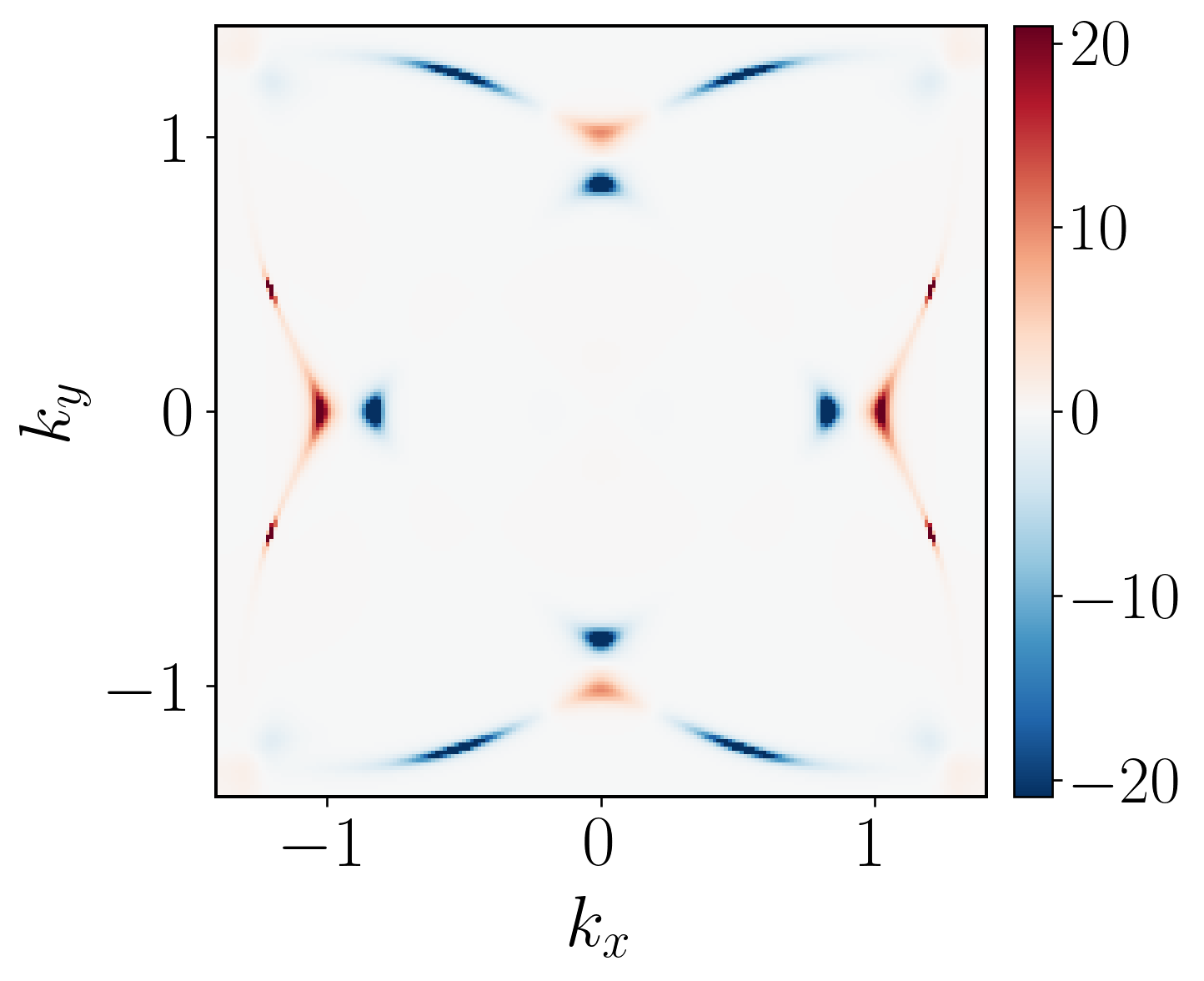}
        \caption{$T=16$~K.}
        \label{subfig:Omega16}
        \end{subfigure}
    \hspace{0.6cm}
    \begin{subfigure}{0.3\textwidth}
        \centering
        \includegraphics[width=\textwidth]{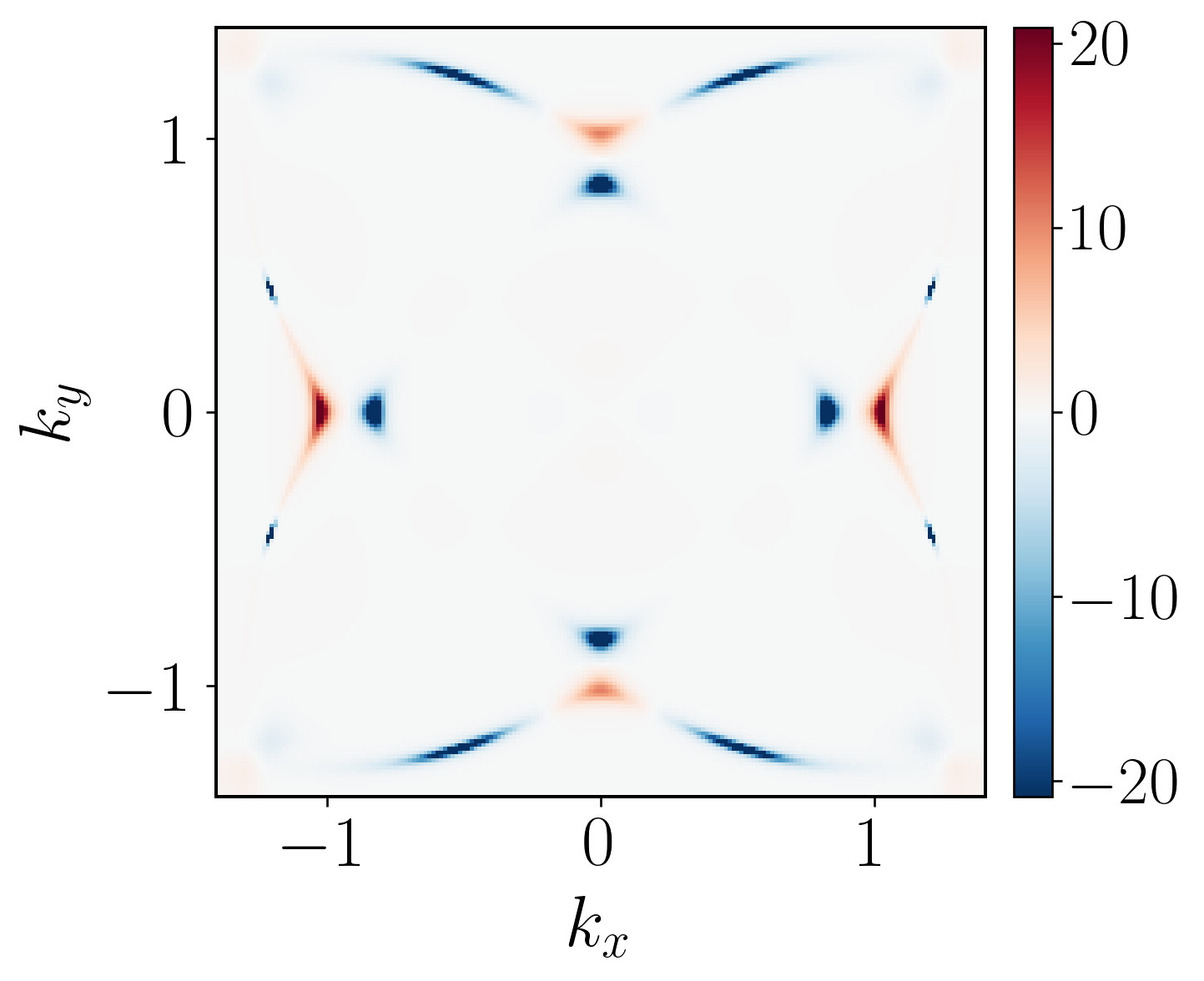}
        \caption{$T=18$~K.}
        \label{subfig:Omega18}
    \end{subfigure}
    \captionsetup{justification=raggedright}
    \caption{Occupied-state Berry curvature at (a) $T=16$~K and (b) $T=18$~K.
    The Berry curvature over most of the Brillouin zone is almost unchanged across the transition, while four localized momentum-space regions undergo a pronounced redistribution.}
    % The Berry curvature away from the four Dirac valleys is almost unchanged, whereas each Dirac region undergoes a substantial redistribution.} %  across the transition
    % Integrating the Berry curvature over a small disk around each valley gives a change $\Delta C_{\rm Dirac}\simeq-1$, while the background contribution remains essentially constant.
    \label{fig:Omega_occ}
\end{figure}
%-----------------------------------------------

To quantify this observation, we introduce four small circular regions $D_i$ of radius $R$ centered on the four momentum-space regions responsible for the Chern-number change and define
% we divide the Brillouin zone into four small circular regions $D_i$ of radius $R$ centered at the four Dirac valleys, and the complementary background region. Define the corresponding partial Berry-curvature contributions as
\begin{align}
C_{D_i}&=\frac{1}{2\pi} \int_{D_i} d^2k \Omega_{\text{occ}}(\boldsymbol{k}), \notag\\
C_{\text{back}}&=\frac{1}{2\pi} \int_{{\text{BZ}} \setminus \bigcup_i D_i} d^2k \Omega_{\text{occ}}(\boldsymbol{k}).
\end{align}
These quantities are not separately quantized Chern numbers; rather, they provide a momentum-resolved decomposition of the total topological invariant.
At $T=16$~K, we obtain $C_{D_i}=0.51$ and $C_{\text{back}}=-2.04$, leading to $C_{\text{back}}+\sum_i C_{D_i}=0$.
In contrast, at $T=18$~K, $C_{D_i}=-0.49$, and $C_{\text{back}}=-2.04$, yielding $C_{\text{back}}+\sum_i C_{D_i}=-4$.
Therefore, within numerical accuracy, each of the four localized regions contributes a change $\Delta C_{D_i}=-1$,
and the total Chern-number change is $\Delta C=\sum_i \Delta C_{D_i}= -4$.
% The evolution from approximately $+1/2$ to $-1/2$ in each Dirac region is also the characteristic Berry-flux transfer associated with a massive Dirac fermion whose mass changes sign.

% ============================================================
% BdG minimum-gap map at T = 17.08 K
% ============================================================

%------------------------------------------------
\begin{figure}[htb]
    \centering
    \includegraphics[width=0.31\textwidth]{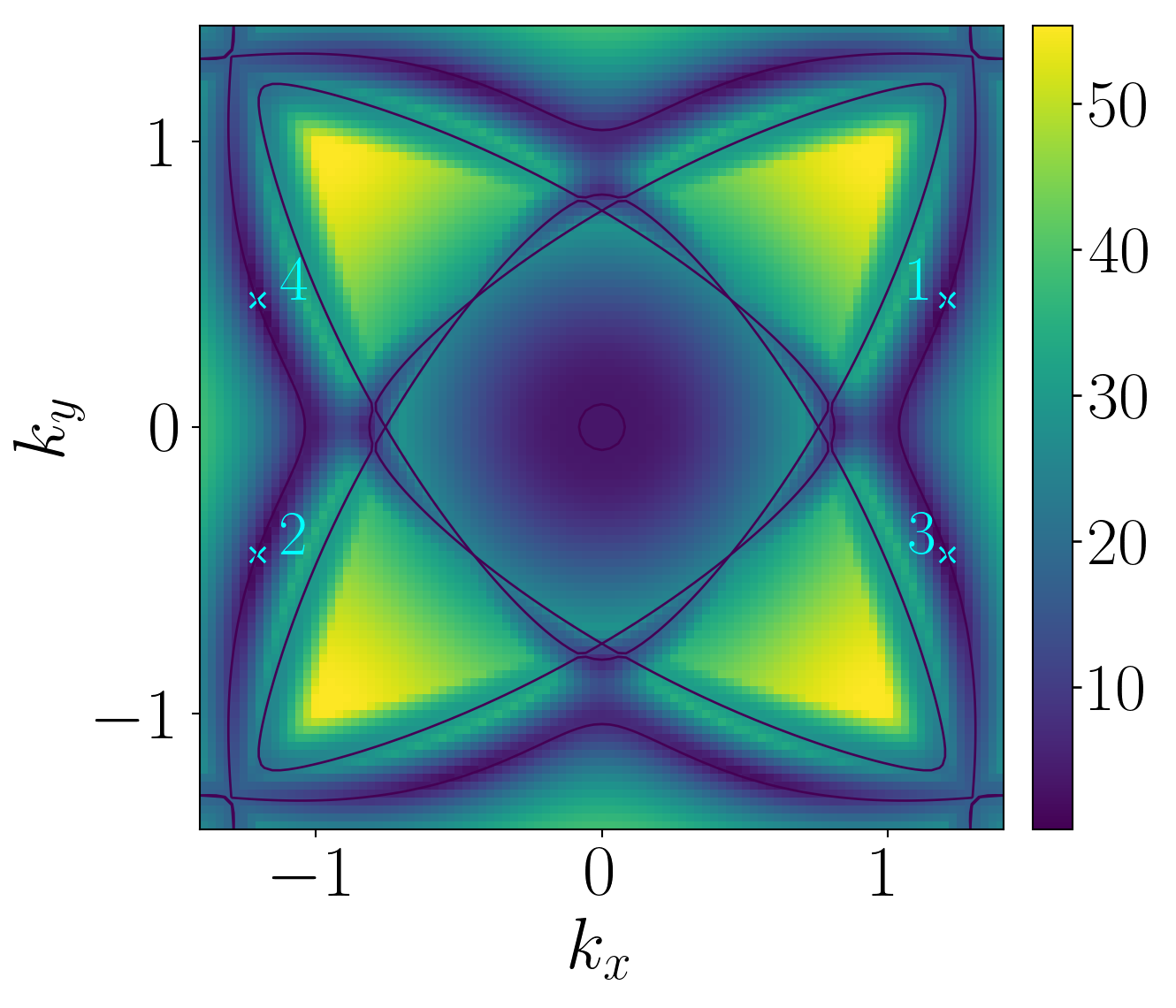}
    \captionsetup{justification=raggedright}
    \caption{Minimum positive BdG excitation energy at $T=17.08$~K. The four low-energy valleys $\boldsymbol{k}_{D,i}$ are marked by ``$\times$'' and label $i$.}
    \label{fig:dirac_gap_map}
\end{figure}
%----------------------------------------------

% The change of the Chern number is mediated by low-energy BdG valleys whose gaps close and reopen upon increasing temperature. % four symmetry-related
To locate these four momentum-space regions more precisely, we examine the minimum positive BdG excitation energy, $E_+(\boldsymbol{k})=\min_{n>0} E^{n}(\boldsymbol{k})$, at a temperature close to the transition. % near-transition temperature
As shown in Fig.~\ref{fig:dirac_gap_map}, the gap map at $T=17.08$~K exhibits four pronounced low-energy valleys centered at $\boldsymbol{k}_i$ ($i=1,\ldots,4$),
which coincide with the centers of the four momentum-space regions where the Berry curvature is redistributed across the transition.
The four momenta form a symmetry-related quartet consistent with the unbroken symmetry group $\{E,C_2,\mathcal{T}l_x,\mathcal{T}l_y\}$, % associated with
and the same four-valley structure persists at the other temperatures considered here, with the valley centers moving only smoothly. %  considered here; with temperature
Therefore, the bulk-gap closing responsible for the topological transition is localized at four unbroken-symmetry-related regions of the Brillouin zone.

% ============================================================
% anisotropy schematic and 2-band spectra: T = 16, 17.08, 18 K
% ============================================================

The local dispersion around each valley is described by the two BdG bands closest to zero energy.
For the $i$'th valley, we project the full BdG Hamiltonian, $H$, onto the corresponding two-dimensional low-energy subspace,
\begin{equation}
h_i(\boldsymbol{q},T)=P_i(T)^\dagger H\bigl(\boldsymbol{k}_i(T)+\boldsymbol{q},T\bigr) P_i(T),
\end{equation}
where $P_i=(U_{-,\boldsymbol{k}_i},U_{+,\boldsymbol{k}_i})$ is constructed from the negative- and positive-energy states closest to zero at the valley center. % (|u_{-,i}\rangle,|u_{+,i}\rangle)
The resulting two-band Hamiltonian can be written as
\begin{equation}
h_i(\boldsymbol{q},T)=d_{0,i}(\boldsymbol{q},T)\sigma_0+\boldsymbol{d}_i(\boldsymbol{q},T)\cdot\boldsymbol{\sigma}.
\label{eq:two_band_dirac}
\end{equation}
To characterize the momentum dependence, we define
\begin{equation}
\boldsymbol{b}_{x,i}=\left. \frac{\partial\boldsymbol{d}_i}{\partial k_x} \right|_{\boldsymbol{k}_{D,i}}, \quad
\boldsymbol{b}_{y,i}=\left. \frac{\partial\boldsymbol{d}_i}{\partial k_y} \right|_{\boldsymbol{k}_{D,i}}, \quad
\hat{\mathbf{n}}_i=\frac{\boldsymbol{b}_{x,i} \times \boldsymbol{b}_{y,i}}{|\boldsymbol{b}_{x,i} \times \boldsymbol{b}_{y,i}|},
\end{equation}
where $\hat{\mathbf{n}}_i$ is the unit vector perpendicular to the local Dirac plane in pseudo-spin space.
The signed Dirac mass is then defined from the full self-consistent BdG Hamiltonian as
\begin{equation}
m_i(T)=\boldsymbol{d}_i\bigl(\mathbf{k}_{D,i}(T),T\bigr)\cdot\hat{\boldsymbol{n}}_i(T).
\label{eq:signed_dirac_mass}
\end{equation}
% Importantly, both $\boldsymbol{k}_{D,i}(T)$ and the local two-band frame are re-evaluated independently at every self-consistent temperature; no interpolation between different self-consistent solutions is involved in the calculation of $m_i(T)$.

A singular-value decomposition of the momentum derivatives determines the two principal momentum directions, which we denote by $q_{\text{fast}}$ and $q_{\text{slow}}$.
Near the valley center, the spectrum takes the anisotropic massive-Dirac form
\begin{equation}
E_{\pm}(\boldsymbol{q},T) \simeq d_0 \pm \sqrt{ v_{\text{fast}}^2 q_{\text{fast}}^2 + v_{\text{slow}}^2 q_{\text{slow}}^2 + m_i^2(T)}.
\label{eq:massive_dirac_spectrum}
\end{equation}
% with $v_{\rm fast}\simeq 2.06\times10^{-1}$ and $v_{\rm slow}\simeq1.81\times10^{-2}$ in the units used for momentum, $v_{\rm fast}/v_{\rm slow}\simeq11.4$.
Fig.~\ref{fig:dirac_spectra} shows the corresponding local two-band spectra at $T=16.0$, $17.08$, and $18.0$~K.
As temperature increases, the gap decreases continuously, becomes extremely small near $17.08$~K, and subsequently reopens.
At the actual gap-closing temperature, $m_i=0$, and Eq.~\ref{eq:massive_dirac_spectrum} reduces to an anisotropic massless Dirac cone. %  the mass term vanishes
However, the two characteristic velocities remain nearly unchanged across the transition, with a strong anisotropy. % velocity
Therefore, the gap closing is controlled by the Dirac mass rather than by a collapse of the Dirac velocity or a quadratic band touching. % the emergence of
% Therefore, the transition is not associated with a vanishing Dirac velocity, but with the mass of a Dirac dispersion. % or a quadratic band touching; an otherwise well-defined; a sign reversal of

%------------------------------------------------
\begin{figure}[t]
    \centering
    \begin{subfigure}{0.31\textwidth}
    \centering
    \includegraphics[width=\textwidth]{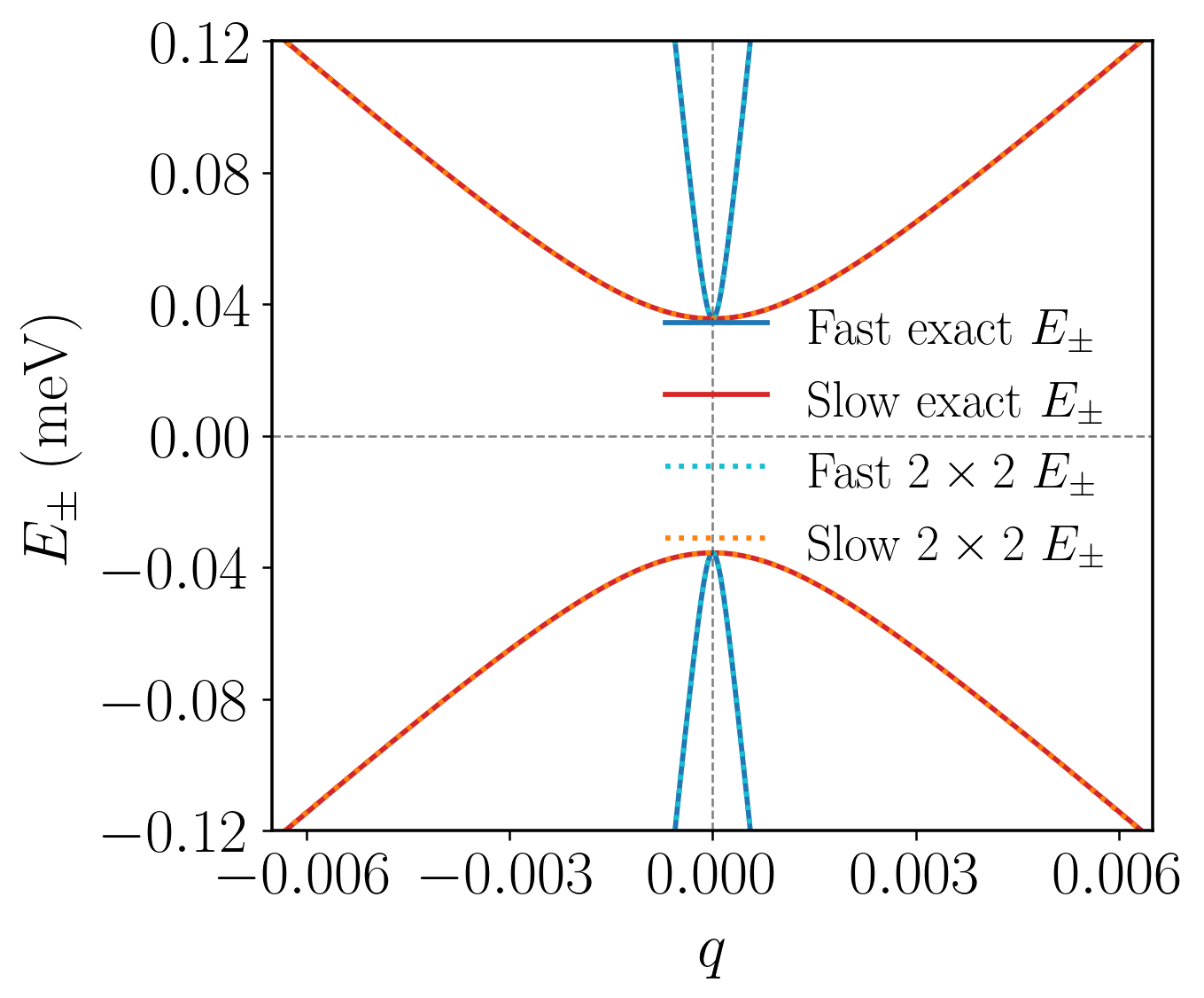}
    \caption{$T=16$~K.}
    \label{subfig:Dirac16}
    \end{subfigure}
\hfill
    \begin{subfigure}{0.31\textwidth}
    \centering
    \includegraphics[width=\textwidth]{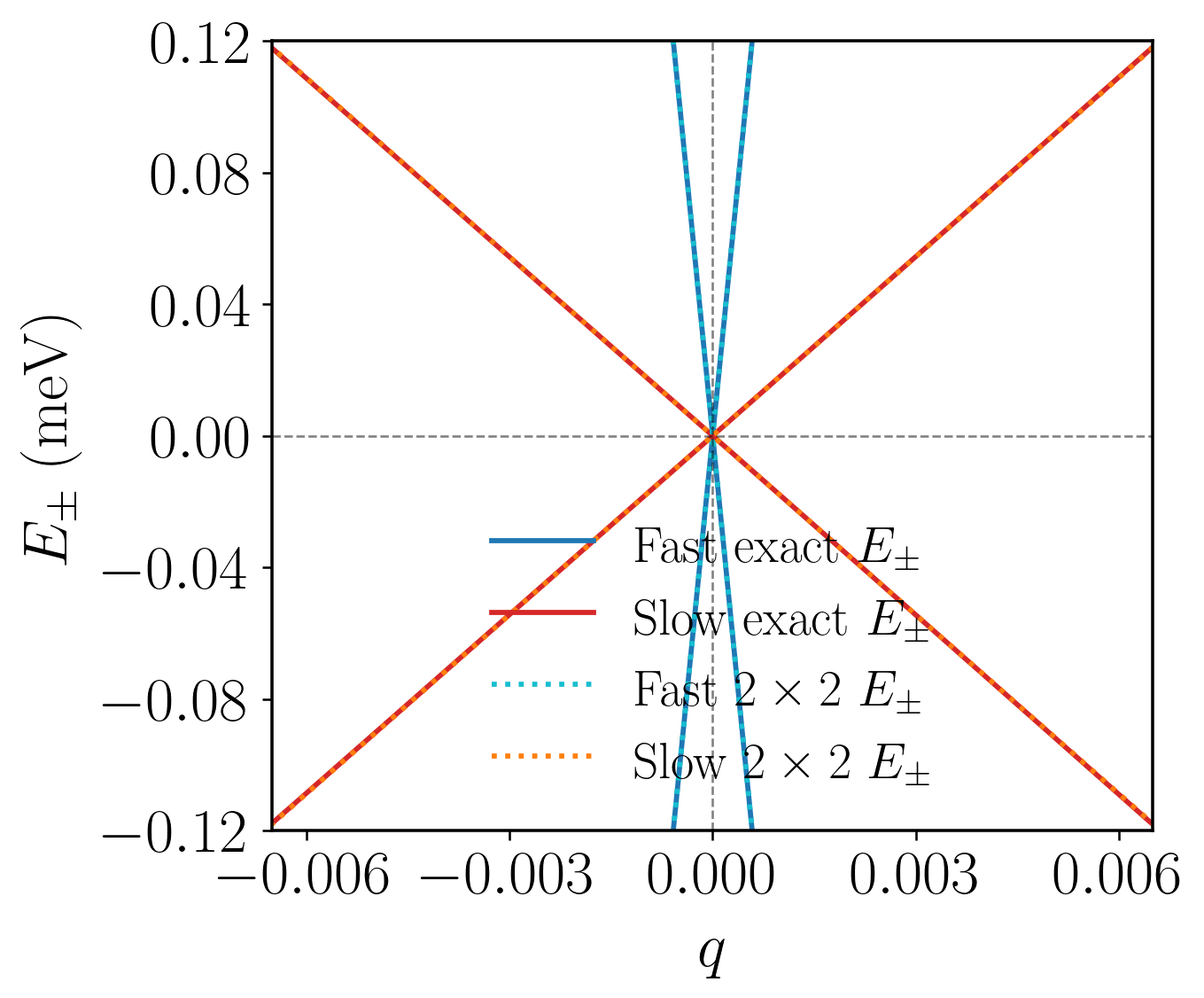}
    \caption{$T=17.08$~K.}
    \label{subfig:Dirac1708}
    \end{subfigure}
\hfill
    \begin{subfigure}{0.31\textwidth}
    \centering
    \includegraphics[width=\textwidth]{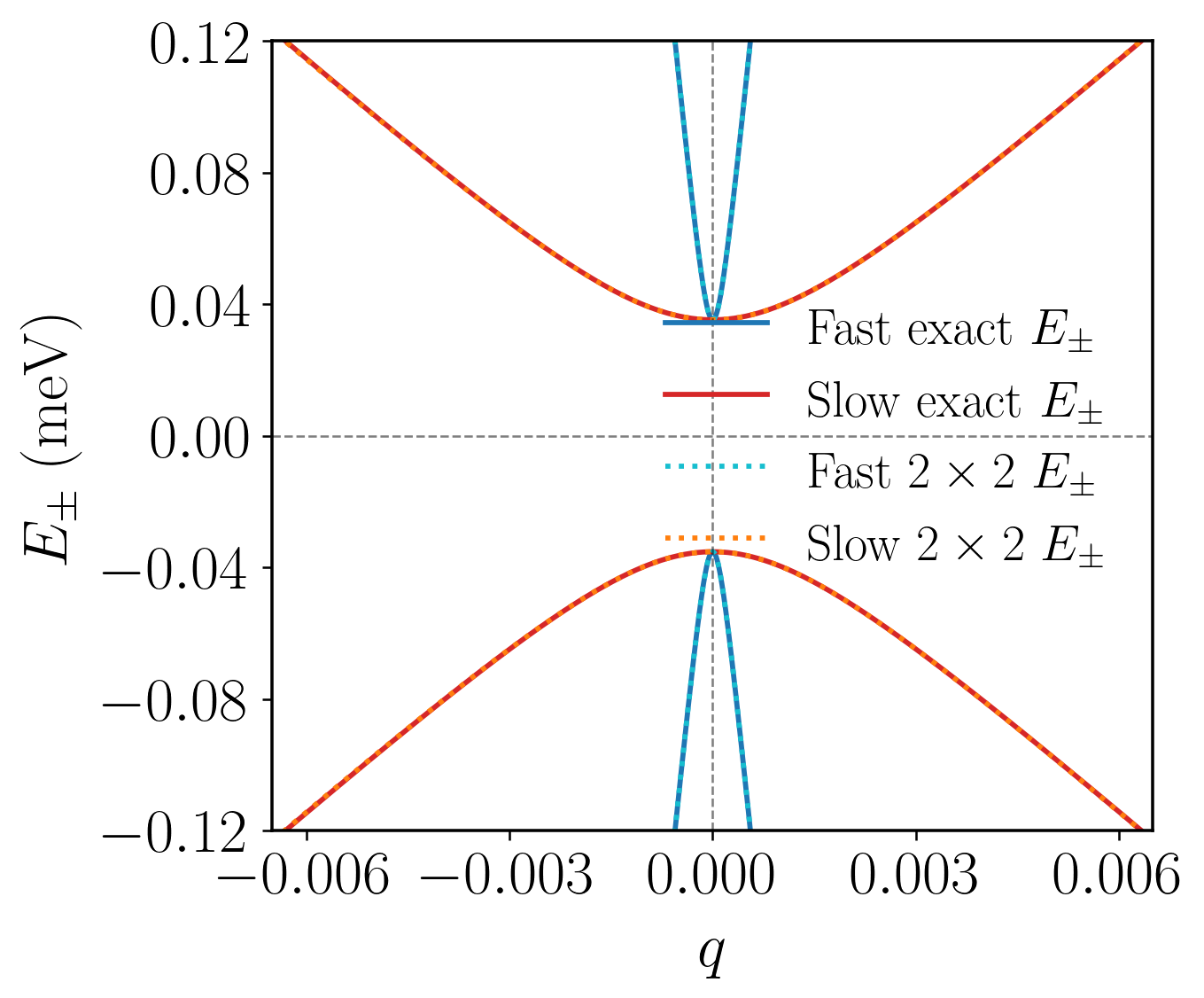}
    \caption{$T=18$~K.}
    \label{subfig:Dirac18}
    \end{subfigure}
\captionsetup{justification=raggedright}
% \hspace{0.06cm}
\caption{Evolution of the local low-energy spectrum near the four symmetry-related BdG valleys.
The anisotropic principal axes of the Dirac dispersion are indicated together with the two central BdG bands. % at $T=16.0$, $17.0$, $17.08$, $17.09$, $17.2$, and $18.0$~K.
The gap closes and reopens while the two principal Dirac velocities remain essentially unchanged, demonstrating a Dirac-mass inversion.}
\label{fig:dirac_spectra}
\end{figure}
%----------------------------------------------------

% ============================================================
% T dependence of signed Dirac mass
% ============================================================

% The same transition is directly resolved by tracking the signed Dirac mass ($m_i$) defined in Eq.~\ref{eq:signed_dirac_mass}.
To resolve the change of $m_i$ more accurately, we calculate it for each self-consistent solution within a narrower temperature window ($17.0$~K$\sim 17.2$~K) around the phase transition.
As shown in Fig.~\ref{fig:m_T}, the masses of all four symmetry-related valleys evolve identically and cross zero within the same narrow temperature interval.
In particular, $m_i(17.08~\text{K})=-1.86\times10^{-4}~\text{meV}$, whereas $m_i(17.09~\text{K})=1.72\times10^{-4}~\text{meV}$, directly locating the topological transition.
Together with the nearly temperature-independent Dirac velocities, this establishes the transition as a simultaneous mass inversion of four symmetry-related anisotropic Dirac fermions.
For a massive Dirac cone, reversing the sign of the mass transfers one unit of Chern number between the two topological sectors.
Here each of the four valleys carries local topological charge $Q_i=-1$, consistent with the numerically observed $\Delta C_{D_i}=-1$, and therefore $\Delta C=-4$.
% The sign reversal of each mass is accompanied by the local topological charge $Q_i=-1$, so that the four inversions collectively produce $\Delta C=-4$.

%---------------------------------------
\begin{figure}[t]
    \centering
    \includegraphics[width=0.36\textwidth]{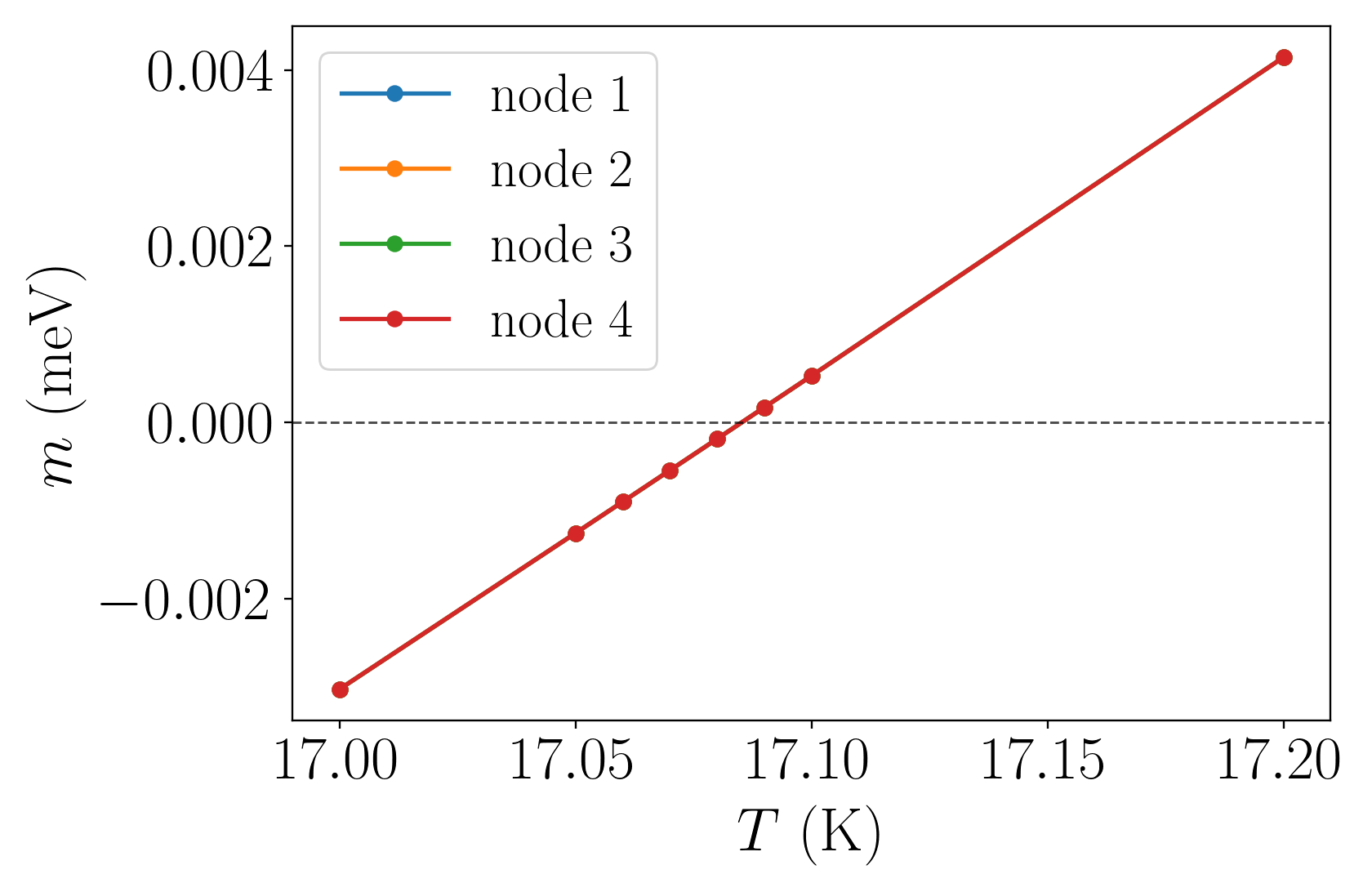}
    \captionsetup{justification=raggedright}
    \caption{Signed Dirac masses of the four unbroken-symmetry-related valleys as a function of temperature.
    The masses of all four nodes evolve identically and change sign between $T=17.08$~K and $T=17.09$~K.}
    \label{fig:m_T}
\end{figure}
%---------------------------------------

To further identify which component of the frustrated three-component order parameter is primarily responsible for the Dirac-mass inversion,
we decompose $\Delta m_i$ across the transition into contributions from the individual pairing channels.
After separating the variations of the complex pairing coefficients into amplitude and phase parts, we find that the change of the $s$-wave amplitude provides the dominant contribution driving the Dirac mass through zero, whereas the $d$-wave components mainly give small corrections of the opposite sign.
This identifies the evolution of the $s$-wave amplitude as the primary pairing degree of freedom controlling the topological transition.

% ==================================================================
% band-5 winding at T = 16, 17.08, 18 K,  whole complex trajectory + zoom near origin
% ==================================================================

A complementary global characterization can be obtained from the superconducting pairing projected onto the normal-state Fermi surfaces.
The effective pairing evaluated along the Fermi surface of band $m$ can be written as
\begin{equation}
\Delta_m(\boldsymbol{k})=U_{m,\boldsymbol{k}}^{0,\dagger} \hat{\Delta}(\boldsymbol{k}) U_{m,-\boldsymbol{k}}^{0,*},
\end{equation}
Parameterizing the Fermi surface by an angle $\theta\in[0,2\pi)$ gives,
\begin{equation}
\Delta_m(\theta)=|\Delta_m(\theta)|e^{i\varphi_m(\theta)},
\end{equation}
and the corresponding phase winding is defined as
\begin{equation}
w_m=\frac{1}{2\pi} \oint_{\text{FS}_m} \frac{\partial \varphi_m(\theta)}{\partial_\theta} d\theta.
\label{eq:FS_winding}
\end{equation}
In the weak-pairing limit, the BdG Chern number is related to the pairing windings through $C=-\sum_m w_m$.

The resulting winding numbers of the seven normal-state Fermi surfaces (band 1--7) are $\{0,-2,6,-2,-2,2,-2\}$ at $T=16$~K, and $\{0,-2,6,-2,2,2,-2\}$ at $T=18$~K. % summarized as
Notably, the only change across the transition occurs on band 5, where $w_5$ switches from $-2$ to $2$.
Since the four low-energy BdG valleys are located on this band, the resulting change $\Delta w_5=4$ accounts for the Chern-number change $C=0\rightarrow -4$.
Fig.~\ref{fig:w5} displays the complex trajectory of $\Delta_5(\theta)$ at $T=16$ and $18$~K, together with magnified views near the origin.
Since the Fermi surface is inversion-symmetric with respect to the origin, points on the same Fermi surface satisfy $\boldsymbol{k}(\theta+\pi)=-\boldsymbol{k}(\theta)$.
Moreover, the superconducting order parameter possesses $C_2$ symmetry, namely, $\Delta(\boldsymbol{k})=\Delta(-\boldsymbol{k})$.
It therefore follows that
\begin{equation}
\Delta_m[\boldsymbol{k}(\theta+\pi)]=\Delta_m[-\boldsymbol{k}(\theta)]=\Delta_m[\boldsymbol{k}(\theta)].
\end{equation}
Consequently, the trajectories of $\Delta_m(\theta)$ over the intervals $\theta \in [0,\pi)$ and $\theta \in [\pi,2\pi)$ completely overlap.
Although the apparent winding number around the origin in Fig.~\ref{fig:w5} seems to be $\mp 1$, the trajectory is in fact traversed twice in the same direction as $\theta$ varies from $0$ to $2\pi$.
The actual winding number should thus be $\mp 2$.
With increasing temperature, the trajectory is continuously deformed through the origin, allowing its winding number to change from $-2$ to $+2$.
The passage through the origin is the Fermi-surface counterpart of the BdG gap closing at the four Dirac valleys.

%---------------------------------------------
\begin{figure}[t]
    \centering
    \begin{subfigure}{0.28\textwidth}
        \centering
        \includegraphics[width=\textwidth]{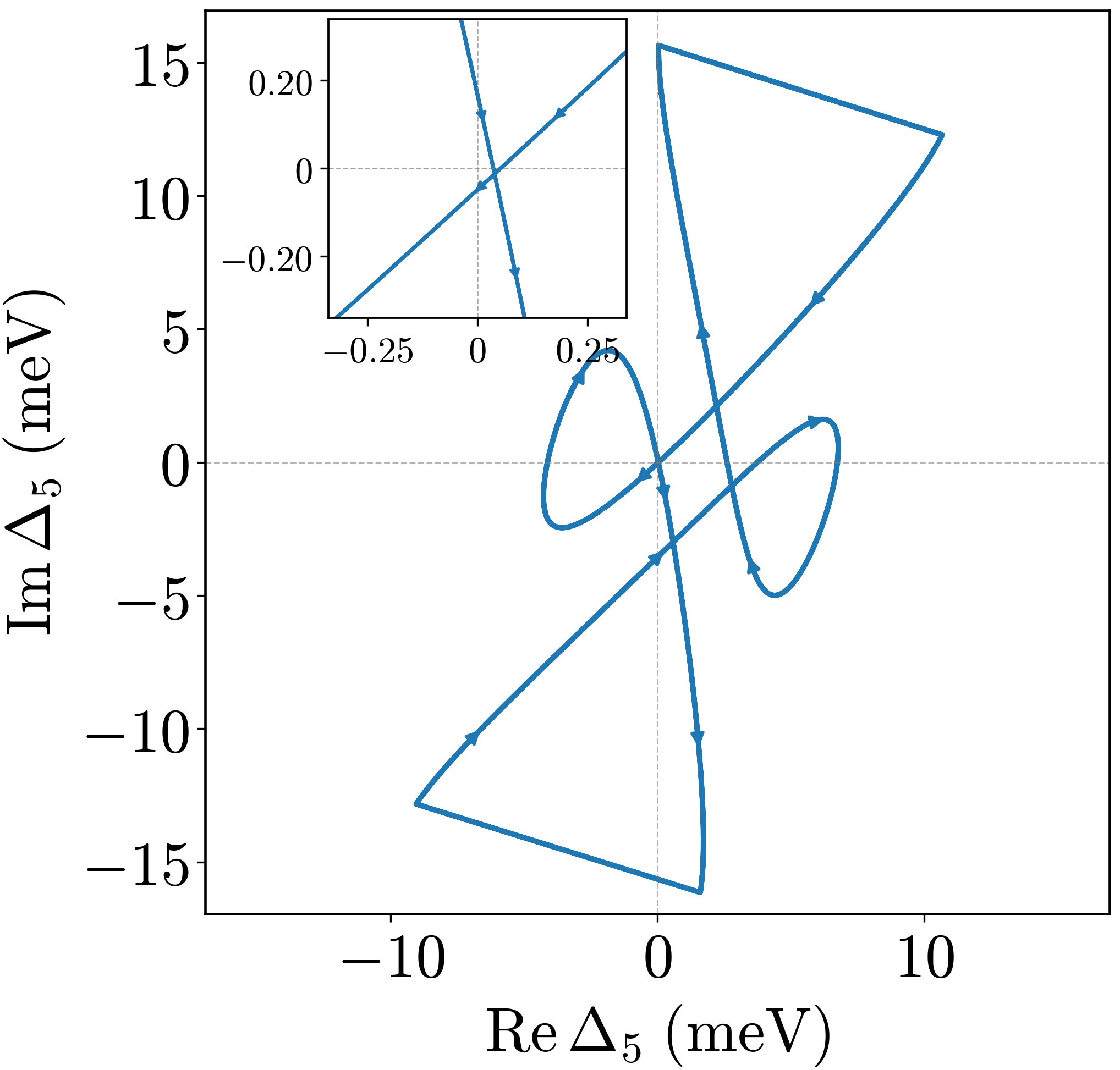} % 2036*1946
        \caption{$T=16$~K, $w_5=-2$.}
        \label{subfig:w5_16}
        \end{subfigure}
    \hspace{0.6cm}
    \begin{subfigure}{0.28\textwidth}
        \centering
        \includegraphics[width=\textwidth]{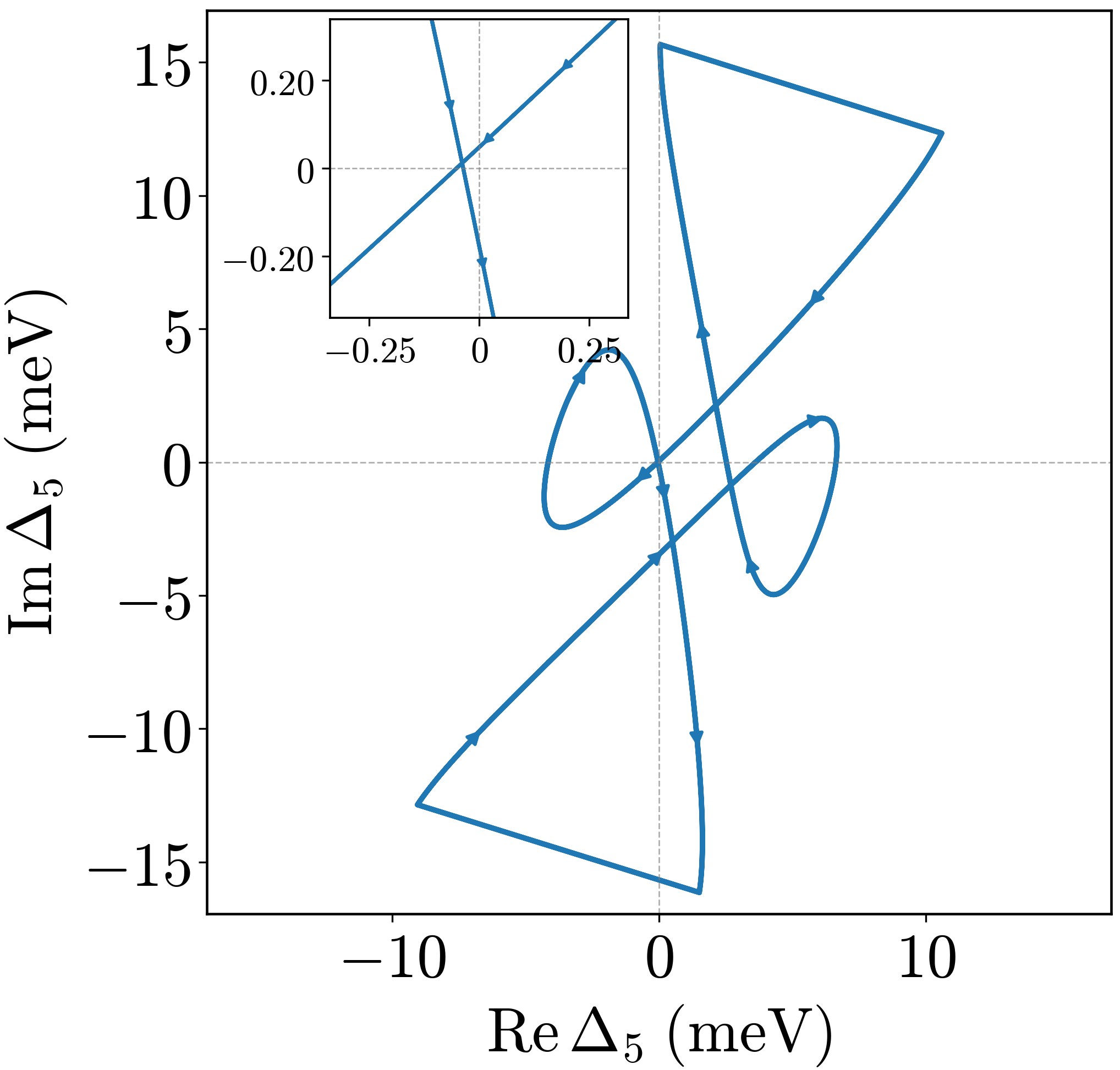} % 2036*1946
        \caption{$T=18$~K, $w_5=2$.}
        \label{subfig:w5_16}
        \end{subfigure}
    \captionsetup{justification=raggedright}
    \caption{Complex winding trajectory of the superconducting pairing projected onto the normal-state Fermi surface of band 5 at (a) $T=16$ and (b) $18$~K,
    the geometric loop shown in each panel is traversed twice in the same direction over the full Fermi surface.
    The insets show the vicinity of the origin.
    Across the transition, the trajectory passes through the origin, and the winding changes from $w_5=-2$ to $w_5=+2$.}
    \label{fig:w5}
\end{figure}
%--------------------------------------------

Although the pairing amplitudes in the self-consistent calculations are not very small compared with the normal-state energy scales, the states considered here can be adiabatically deformed
toward the weak-pairing limit without closing the BdG gap or changing the Chern number.
The Fermi-surface winding therefore provides a useful topological characterization of the two superconducting sectors.

%---------------------------------------------------------------------------------------------------------------------------

\subsection{Fermi-surface evolution and pairing competition}

%---------------------------------------------------------------------------
\begin{figure}[h]
\centering
\begin{subfigure}{0.235\textwidth}
    \centering
    \includegraphics[width=\textwidth]{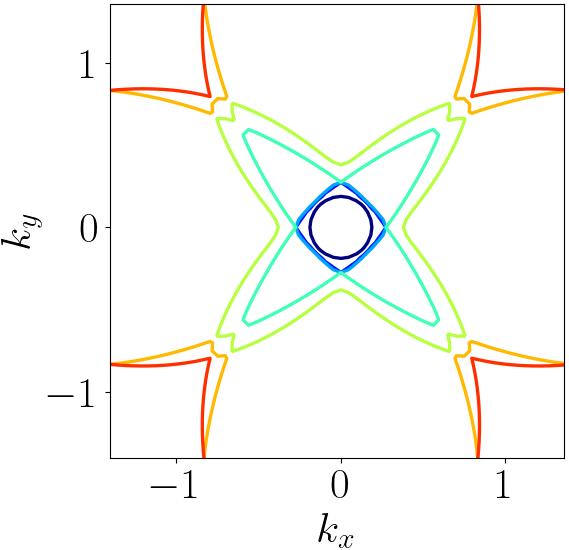} % 569*550
    \caption{$\mu=-0.15t$.}
    \label{subfig:FS_0.15}
\end{subfigure}
\hspace{0.06cm}
\begin{subfigure}{0.235\textwidth}
    \centering
    \includegraphics[width=\textwidth]{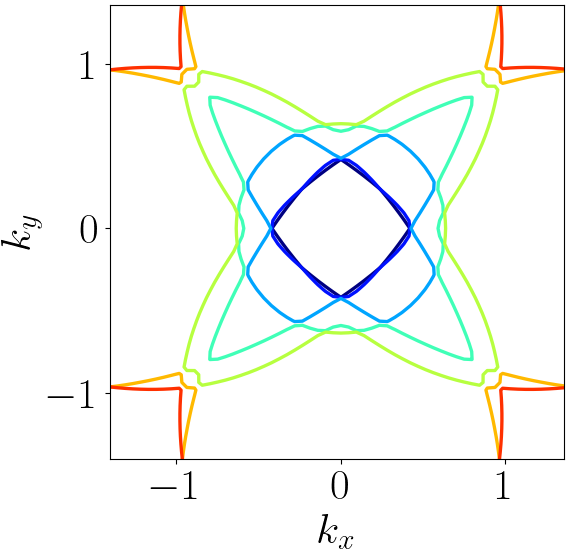} % 569*550
    \caption{$\mu=-0.5t$.}
    \label{subfig:FS_0.5}
\end{subfigure}
\hspace{0.06cm}
\begin{subfigure}{0.235\textwidth}
    \centering
    \includegraphics[width=\textwidth]{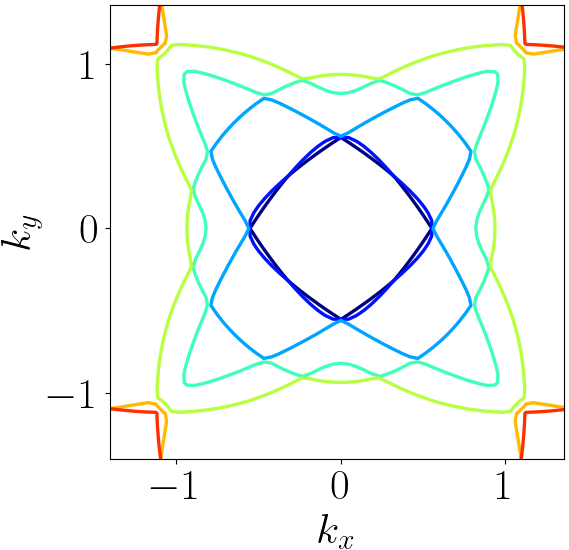} % 569*550
    \caption{$\mu=-0.8t$.}
    \label{subfig:FS_0.8}
\end{subfigure}
\hspace{0.06cm}
\begin{subfigure}{0.235\textwidth}
    \centering
    \includegraphics[width=\textwidth]{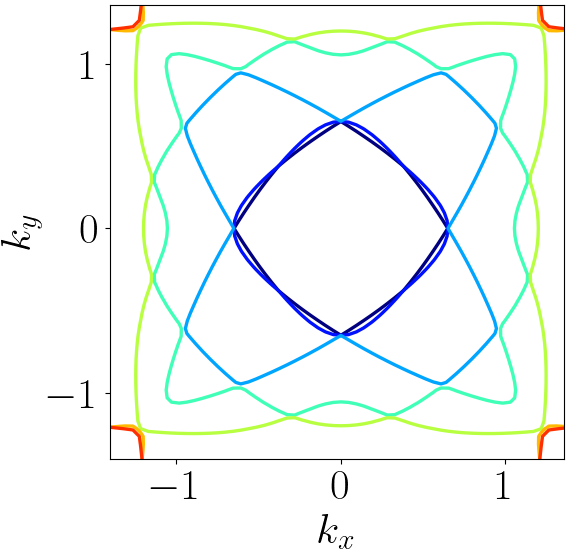} % 569*550
    \caption{$\mu=-1.0t$.}
    \label{subfig:FS_1.0}
\end{subfigure}
\hfill
\begin{subfigure}{0.235\textwidth}
    \centering
    \includegraphics[width=\textwidth]{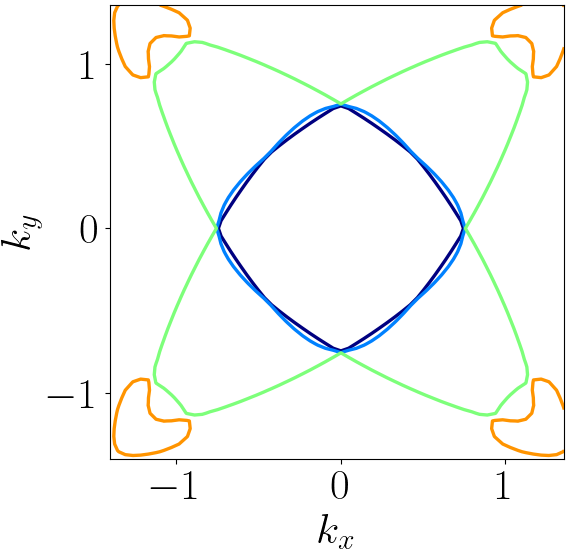} % 569*550
    \caption{$\mu=-1.2t$.}
    \label{fig:FS_1.2}
\end{subfigure}
\hspace{0.06cm}
\begin{subfigure}{0.75\textwidth}
    \centering
    \includegraphics[width=\textwidth]{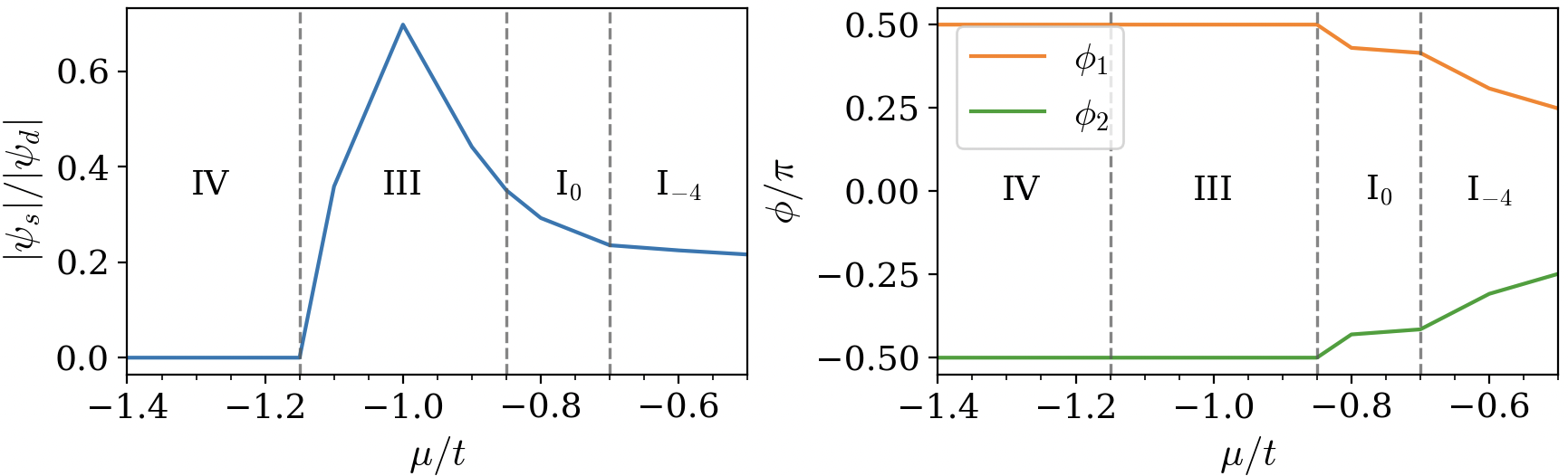} % 1731*526
    \caption{}
    \label{subfig:psi_mu}
\end{subfigure}
\captionsetup{justification=raggedright}
\caption{Chemical-potential evolution of the normal-state Fermi surface and superconducting pairing structure at $\theta=53.13^{\circ}$ and $g_0=0.05$~eV.
Panels (a)--(e) show the normal-state Fermi surfaces at the values of $\mu$ specified in the respective subfigures.
With decreasing $\mu$, the $\Gamma$-centered Fermi-surface structure favorable to $s$-wave pairing evolves from a relatively small loop or annular structure into a larger, slightly distorted loop, and eventually disappears.
Panel (f) shows the corresponding evolution of the relative pairing strength $|\psi_s|/|\psi_d|$ and the relative phases $\phi_{1,2}=\phi_{d1,2}-\phi_s$.
As the $s$-wave-favored Fermi-surface structure expands, the system evolves from the topologically nontrivial $s+d_1e^{i\phi_1}+d_2e^{i\phi_2}$ state ($C=-4$, Region I$_{-4}$) to its topologically trivial counterpart ($C=0$, Region I$_0$), and then to the trivial $s+i(d_1-d_2)$ state ($C=0$, Region III).
Upon further decreasing $\mu$, both this Fermi-surface feature and $|\psi_s|$ vanish, leaving the $d_1-d_2$ state (Region IV).
Vertical dashed lines in panel (f) mark the boundaries between the different pairing regimes.}
\label{fig:FSmu}
\end{figure}
%---------------------------------------------------------------------------
\begin{comment}
\begin{subfigure}{0.195\textwidth}
    \centering
    \includegraphics[width=\textwidth]{Ak_mu-0.15.png} % 1409*1200
    \caption{$\mu=-0.15t$.}
    \label{fig:Ak_0.15}
\end{subfigure}
\end{comment}
%--------------------------------------------------------------------------

To elucidate how the chemical potential controls the competition between the $s$- and $d$-wave components, we examine the normal-state Fermi surfaces at $\theta=53.13^{\circ}$ and $g_0=0.05$~eV for several representative values of $\mu$, as shown in Figs.~\ref{fig:FSmu}~(a)--(e).
Owing to the $C_4$ symmetry of the square lattice, $\theta=53.13^{\circ}$ is symmetry equivalent to $\theta=36.87^\circ$ considered in Fig.~1~(c) in the main text.
As $\mu$ decreases, the $\Gamma$-centered Fermi-surface structure favorable to $s$-wave pairing progressively expands.
Depending on the specific parameters, this structure may appear as a single closed Fermi-surface loop [Figs.~\ref{fig:FSmu}~(a) and \ref{fig:FS}~(b)],
an annular structure composed of several Fermi-surface sheets [Figs.~\ref{fig:FSmu}~(b,c)], or a larger and slightly distorted loop [Fig.~\ref{fig:FSmu}~(d)].

The corresponding evolution of the relative pairing strength $|\psi_s|/|\psi_d|$ and relative phases $\phi_{1,2}=\phi_{d1,2}-\phi_s$ is shown in Fig.~\ref{fig:FSmu}~(f).
For the relatively small $s$-wave-favored Fermi-surface structures in Figs.~\ref{fig:FSmu}~(a,b), the system resides in the topologically nontrivial $s+d_1e^{i\phi_1}+d_2e^{i\phi_2}$ state with $C=-4$ (Region I$_{-4}$).
As this $\Gamma$-centered structure expands [Fig.~\ref{fig:FSmu}~(c)], the enhanced $s$-wave weight drives the system into the topologically trivial sector of the same frustrated three-component state (Region I$_0$).
In both sectors, $\phi_1-\phi_2\neq0,\pi$, indicating broken time-reversal symmetry.
As the Fermi loop becomes larger and more distorted [Fig.~\ref{fig:FSmu}~(d)], the $s$-wave contribution is further enhanced, driving the system into the topologically trivial $s+i(d_1-d_2)$ state (Region III).
Upon further decreasing $\mu$, this $s$-wave-favored feature disappears [Fig.~\ref{fig:FSmu}~(e)], accompanied by the vanishing of $|\psi_s|$, and the system ultimately enters the $d_1-d_2$ state (Region IV).
% This evolution provides a microscopic Fermi-surface origin for the chemical-potential-driven sequence of superconducting phases discussed in the main text.
Therefore, the chemical-potential-driven Fermi-surface evolution provides a microscopic mechanism for tuning the relative $s$-wave weight and the associated superconducting phases discussed in the main text.

%%%%%%%%%%%%%%%%%%%%%%%%%%%%%%%%%%%%%%%%%%%%%%%%%

\section{Experimental relevance of the parameter regime}

To clarify the experimental relevance of the $s+d_1e^{i\phi_1}+d_2e^{i\phi_2}$ state, we first express the model parameters in physical units. % and determine the carrier concentration from the self-consistent superconducting states.
In this work, we employ a multi-band lattice model for twisted bilayer cuprate superconductors.
The intralayer nearest-neighbor and next-nearest-neighbor hopping amplitudes are taken as $t=0.153$~eV and $t^{\prime}=-0.45t=-68.85$~meV, respectively.
These values are consistent with the parametrization used by Ref.~\cite{Can2021SM,can2021SM,Bille2001SM} in twisted double-layer cuprates like Bi$_2$Sr$_2$CaCu$_2$O$_{8+\delta}$ (Bi2212), providing an effective description of the low-energy electronic structure associated with the CuO$_2$ planes.
% This band parametrization follows the lattice model employed by Ref.~\cite{Can2021} to describe the Bi$_2$Sr$_2$CaCu$_2$O$_{8+\delta}$ (Bi2212) Fermi surface.

The interaction parameters in our model should be regarded as phenomenological parameters of the mean-field theory. 
Consistent with previous mean-field studies of twisted cuprates \cite{Can2021SM,can2021SM}, we set the onsite repulsion to $U=0$ and take a nearest-neighbor attractive interaction $V=-0.146$~eV.
This choice does not imply that the onsite Coulomb repulsion in real cuprates is absent; rather, the nearest-neighbor attraction represents an effective interaction in the renormalized pairing channel.
At the mean-field level, this interaction supports order parameters with both $d_{x^2-y^2}$-wave and extended-$s$-wave symmetry, generated by nearest-neighbor bond pairing.

For the interlayer tunneling, we use an exponentially decaying form \cite{Can2021SM,can2021SM}, $g_{ij}=g_0\exp\left[-\left(\sqrt{r_{ij}^2+d^2}-d \right)/\rho \right]$,
where $r_{ij}$ is the in-plane separation between sites $i$ and $j$ on different layers, $d$ is the interlayer spacing, and $\rho$ characterizes the decay length of the interlayer tunneling associated with the spatial extent of the Cu~$4s$ orbitals.
Using the tunneling parametrization for cuprates, we take $d \simeq 12~{\rm \AA}$ \cite{Bansil2005SM}, $\rho \simeq 4$~a.u.$=2.11~{\rm \AA}$ \cite{Can2021SM}, % Bi-based
and the in-plane lattice constant entering the effective tunneling model as $a\simeq 5.4~{\rm \AA}$ \cite{Slezak2008SM}.
These parameters set the physical length scales used in constructing the twisted bilayer Hamiltonian.
In the numerical calculations, all lengths are measured in units of $a$, so that these values correspond to $d=2.22a$ and $\rho=0.39a$.

The interlayer tunneling strength $g_0$, however, is more difficult to determine uniquely for a real twisted interface because it is sensitive to the complicated microscopic crystal structure, interface separation, and electronic reconstruction.
Estimates of $g_0$ in high-$T_c$ cuprates, obtained from different experimental fits and theoretical modeling, vary over a broad range, from a few meV to about $100$~meV \cite{Can2021SM,Tummuru2022SM}.
Motivated by this uncertainty, we scan $g_0$ from $10$ to $80$~meV, covering nearly one order of magnitude.
% and is comparable to that used in previous theoretical phase diagrams of topological superconductivity in twisted double-layer cuprates \cite{Can2021}. % has also been adopted
Therefore, our scan is not intended to assign a single fixed value of $g_0$ to a specific material, but rather to examine whether the $s+d_1e^{i\phi_1}+d_2e^{i\phi_2}$ state can be stabilized within experimentally reasonable interlayer-tunneling energy scales.

To make a more direct comparison with experimental doping levels, we compute the particle number from the self-consistent paired BdG state rather than infer it from the normal state or from the chemical potential alone.
The electron density $n_e$, defined as the average number of electrons per Cu site including spin degeneracy, is evaluated as
\begin{equation}
n_e=\frac{1}{N}\sum_{il\sigma}\langle c^\dagger_{il\sigma}c_{il\sigma}\rangle,
\end{equation}
where $N=2(m^2+n^2)$ denotes the total number of Cu sites in the moir\'e unit cell.
The corresponding signed carrier doping relative to half filling is then defined as $p=1-n$.
With this convention, $p>0$ corresponds to hole doping, while $p<0$ corresponds to electron doping.
This definition provides a more transparent connection to experimental carrier concentrations than quoting $\mu/t$ alone, because in the presence of $t^\prime$, interlayer tunneling, and superconducting pairing, the chemical potential does not translate in a simple way into an obvious doping level. 
% a given chemical potential does not uniquely specify an obvious doping level.
% We note, however, that our calculations are performed at fixed chemical potential. A strictly fixed-doping calculation would require adjusting \(\mu\) self-consistently for each set of parameters and temperature.

In Fig.~1(a) in the main text, we choose $\theta=43.6^\circ$ and $\mu=-0.9t=-0.138$~eV.
The topologically nontrivial $s+d_1e^{i\phi_1}+d_2e^{i\phi_2}$ state appears for $g_0=30$ -- $52$~meV at $T\lesssim 0.4T_c$.
Within this region, the paired-state hole doping is approximately $p=0.016$ -- $0.110$, corresponding to the underdoped hole-doped regime.
The width of this window in $g_0$ is about $22$~meV, indicating that the phase is not restricted to a single fine-tuned parameter point.
In Fig.~1(b) in the main text, we take $\theta=53.13^\circ$ and $\mu=-0.5t=-0.077$~eV.
This three-component state is found for $g_0=43$ -- $60$~meV at $T\lesssim 0.35T_c$, and the corresponding doping is $p=-0.108$ -- $-0.052$, namely on the electron-doped side.
This result shows that this state is not confined to hole doping in the model, but can also extend into the electron-doped regime.
Since electron-doped cuprates are also experimentally established material systems \cite{Tokura1989SM,Armitage2010SM}, this part of the phase diagram suggests a possible route to realizing the $s+d_1e^{i\phi_1}+d_2e^{i\phi_2}$ state beyond hole-doped cuprate devices.

In Fig.~1(c) in the main text, we fix $T=0$~K and $g_0=50$~meV, and scan the chemical potential to determine the stability range of the $s+d_1e^{i\phi_1}+d_2e^{i\phi_2}$ pairing states at different twist angles.
For example, at $\theta=36.87^\circ$, this frustrated three-component state extends from $\mu=-0.5t=-0.077$~eV to $\mu=-0.85t=-0.13$~eV, corresponding to $p=-0.086$ -- $0.082$.
Thus, this phase crosses the electron-doped regime, the vicinity of half filling, and the low-hole-doping regime.
For $\theta=43.6^\circ$, the topologically nontrivial $s+d_1e^{i\phi_1}+d_2e^{i\phi_2}$ state extends from $\mu=-0.5t=-0.077$~eV to $\mu=-1.0t=-0.153$~eV, corresponding to $p=-0.08$ -- $0.181$. % Moving closer to $45^\circ$,
The latter range covers electron doping, the vicinity of half filling, underdoping, optimal doping, and slight overdoping.
Together with the broader stability window found closer to $45^\circ$, these results demonstrate that the $s+d_1e^{i\phi_1}+d_2e^{i\phi_2}$ state can persist over a wide carrier-doping range.

In addition to these regions, we also present several results at smaller $g_0$ in Sec.~V, demonstrating that the frustrated three-component phase (Region~I) is not limited to $30$~meV$<g_0<60$~meV. % self-consistent
For example, at $\theta=53.13^\circ$, $g_0=20$~meV, and $\mu=-1.34t=-0.205$~eV, Region~I persists from $T=0$ to $0.18T_c$, with the hole doping changing from $p=0.2124$ to $p=0.2118$, corresponding to the overdoped regime. % only slightly
At the same twist angle, when $g_0=24$~meV and $\mu=-1.2t=-0.184$~eV, it persists up to $0.16T_c$ with $p=0.141$ -- $0.140$, placing it on the underdoped side close to optimal doping.
Similarly, at $\theta=43.6^\circ$, $g_0=24$~meV, and $\mu=-1.25t=-0.191$~eV, Region~I persists up to $0.31T_c$ with $p=0.177$ -- $0.175$, corresponding to a slightly overdoped regime;
and when $g_0=28$~meV and $\mu=-1.15t=-0.176$~eV, it persists up to about $0.35T_c$ with $p=0.136$ -- $0.135$, corresponding to an underdoped regime close to optimal doping.
These examples cover underdoped, near-optimal, and overdoped hole concentrations, demonstrating that the $s+d_1e^{i\phi_1}+d_2e^{i\phi_2}$ state can remain stable over finite temperature and doping intervals even at relatively weak interlayer tunneling.
In these fixed-$\mu$ temperature scans, the particle number varies only weakly with $T$, with typical changes smaller than $0.002$.
It is therefore reasonable to characterize these phases by the low-temperature or $T=0$ value of $p$.
A strictly fixed-density calculation would require adjusting $\mu$ at each temperature, but the small variation of $p$ indicates that this distinction does not affect the doping classification of the phases discussed here.

Overall, the $s+d_1e^{i\phi_1}+d_2e^{i\phi_2}$ pairing regime identified above spans underdoped, near-optimally doped, and overdoped hole concentrations, and also extend to the electron-doped side in parts of the phase diagram. 
Using the commonly adopted empirical reference $p\simeq 0.16$ for optimal hole doping, these results place the three-component phase within carrier-density regimes that are directly relevant to cuprate superconductors. 
The transition temperatures quoted here are obtained from the self-consistent mean-field calculation, and we therefore focus mainly on the relative temperature range over which Region~I remains stable, together with the corresponding physical energy scales and doping levels.
Taken together, these results show that the $s+d_1e^{i\phi_1}+d_2e^{i\phi_2}$ phase is not restricted to a fine-tuned point in arbitrary units, but appears within finite windows of $g_0$, $\mu$, $\theta$, and $p$, whose energy and carrier-density scales are comparable to those relevant for cuprate experiments.

%%%%%%%%%%%%%%%%%%%%%%%%%%%%%%%%%%%%%%%%%%%%%%%%%

\section{Nematic Kerr Response}

\subsection{Symmetry constraints on the optical conductivity tensor}

We first derive the symmetry constraints on the optical conductivity tensor. % Within linear response theory, 
The optical conductivity is given by 
\begin{equation}
\sigma_{ij}(\boldsymbol{q},\omega)=\frac{e^2}{i\hbar \omega}\left[\pi_{ij}(\boldsymbol{q},\omega)-\pi_{ij}(\boldsymbol{q},0)\right],
\end{equation}
where $\omega$ is the frequency of the incident light, and
\begin{equation}
\pi_{ij}(\boldsymbol{q},\omega)=\frac{1}{i} \int_0^\infty dt e^{i\omega t} \left\langle \left[ \hat{J}_i^\dagger(\boldsymbol{q},t), \hat{J}_j(\boldsymbol{q},0) \right] \right\rangle
\end{equation}
is the retarded current-current correlation function.
Here $\hat{J}_i(\boldsymbol{q},t)$ denotes the $i$'th component of the total current operator and transforms as $\mathcal{T} \hat{J}_i(\boldsymbol{q},t)\mathcal{T}^{-1}=-\hat{J}_i(-\boldsymbol{q},-t)$ under time reversal ($\mathcal{T}$),
satisfying $\hat{J}_i^\dagger(\boldsymbol{q},t)=\hat{J}_i(-\boldsymbol{q},t)$.

To derive the corresponding constraint on the conductivity tensor, we define $O_{ij}(\boldsymbol{q},t)=\left\langle \left[ \hat{J}_i^\dagger(\boldsymbol{q},t), \hat{J}_j(\boldsymbol{q},0) \right] \right\rangle$.
For a time-reversal-invariant equilibrium state,
$\langle \hat{O}\rangle=\langle \mathcal{T}\hat{O}\mathcal{T}^{-1} \rangle^*$.
Therefore,
\begin{equation}
O_{ij}(\boldsymbol{q},t) = \left\langle \mathcal{T} \left[ \hat{J}_i(\boldsymbol{-q},t), \hat{J}_j(\boldsymbol{q},0) \right] \mathcal{T}^{-1} \right\rangle^* = \left\langle \left[\hat{J}_i(\boldsymbol{q},-t), \hat{J}_j(-\boldsymbol{q},0) \right] \right\rangle^* = \left\langle \left[\hat{J}_i(\boldsymbol{q},0), \hat{J}_j(-\boldsymbol{q},t) \right] \right\rangle^*.
\end{equation}
Since $\langle[A,B]\rangle^*=\langle[B^\dagger,A^\dagger]\rangle$, we have
\begin{equation}
O_{ij}(\boldsymbol{q},t) = \left\langle \left[\hat{J}_j^{\dagger}(-\boldsymbol{q},t), \hat{J}_i^{\dagger}(\boldsymbol{q},0) \right] \right\rangle = \left\langle \left[\hat{J}_j^{\dagger}(-\boldsymbol{q},t), \hat{J}_i(-\boldsymbol{q},0) \right] \right\rangle \equiv O_{ji}(-\boldsymbol{q},t),
\end{equation}
and it follows that
\begin{align}
\pi_{ij}(\boldsymbol{q},\omega)&=\frac{1}{i} \int_0^\infty dt e^{i\omega t} O_{ij}(\boldsymbol{q},t) = \pi_{ij}(\boldsymbol{q},\omega)=\frac{1}{i} \int_0^\infty dt e^{i\omega t} O_{ji}(\boldsymbol{-q},t) \equiv \pi_{ji}(-\boldsymbol{q},\omega), \notag\\
\sigma_{ij}(\boldsymbol{q},\omega)&=\frac{e^2}{i\hbar \omega}\left[\pi_{ji}(\boldsymbol{-q},\omega)-\pi_{ji}(-\boldsymbol{q},0)\right] \equiv \sigma_{ji}(-\boldsymbol{q},\omega).
\end{align}
In the optical limit ($\boldsymbol{q}=0$), time-reversal symmetry therefore requires $\sigma_{ij}(\omega)=\sigma_{ji}(\omega)$.

We next consider the interlayer two-fold rotations $l_{x,y,x\pm y}$.
For $l_x$, $\hat{J}_x \to \hat{J}_x$, $\hat{J}_y \to -\hat{J}_y$, and for $l_y$, $\hat{J}_x \to -\hat{J}_x$, $\hat{J}_y \to \hat{J}_y$.
If either of these symmetries is preserved, the off-diagonal conductivity components change sign, $\sigma_{xy,yx}=-\sigma_{xy,yx}$, whereas the diagonal components remain invariant, $\sigma_{xx,yy}=\sigma_{xx,yy}$.
On the other hand, for $l_{x+y}$, $\hat{J}_x \to \hat{J}_y$, $\hat{J}_y \to \hat{J}_x$, and for $l_{x-y}$, $\hat{J}_x \to -\hat{J}_y$, $\hat{J}_y \to -\hat{J}_x$.
Thus, the preservation of $l_{x\pm y}$ enforces $\sigma_{xy,yx}=\sigma_{yx,xy}$ and $\sigma_{xx,yy}=\sigma_{yy,xx}$.

\begin{comment}
Combining $l_{x,y,x\pm y}$ with $\mathcal{T}$,
if the ground state preserves $\mathcal{T}l_{x,y}$, the off-diagonal components transform as $\sigma_{xy} \to -\sigma_{yx}$, while the diagonal components remain unchanged, $\sigma_{xx,yy}\to\sigma_{xx,yy}$.
Therefore, $\mathcal{T}l_{x,y}$ enforces $\sigma_{xy}+\sigma_{yx}=0$, while imposing no constraint on $\sigma_{xx}$ and $\sigma_{yy}$.
In contrast, if the ground state preserves $\mathcal{T}l_{x\pm y}$, the diagonal components are interchanged, $\sigma_{xx} \leftrightarrow \sigma_{yy}$, whereas the off-diagonal components remain individually invariant.
Hence, $\mathcal{T}l_{x\pm y}$ enforces $\sigma_{yy}-\sigma_{xx}=0$, while leaving $\sigma_{xy}$ and $\sigma_{yx}$ unconstrained.
\end{comment}

%-----------------------------------------------------------------------------------------------------------------

\subsection{Derivation of the Kerr rotation} % without $C_4$ symmetry

In this section we derive the polar Kerr rotation of an optically thin twisted bilayer without imposing $C_4$ rotational symmetry.
For normal incidence and treat the bilayer as a two-dimensional conducting sheet in vacuum, the tangential electric field ($\boldsymbol{E}$) is continuous across the interface, % and the sheet current is given by $\boldsymbol{J}=\mathbf{\Sigma}\boldsymbol{E}$, where
and the physical sheet conductivity ($\mathbf{\Sigma}$) can be written as
\begin{equation}
\mathbf{\Sigma}=\frac{e^2}{\hbar}\boldsymbol{\sigma}=\frac{e^2}{\hbar}
\begin{pmatrix}
\sigma_{xx} & \sigma_{xy}\\
\sigma_{yx} & \sigma_{yy}
\end{pmatrix}.
\end{equation}
Here $\boldsymbol{\sigma}$ denotes the dimensionless in-plane optical conductivity tensor, obtained by attaching the fine-structure constant $\alpha$, i.e., by setting $e^2/\hbar=\alpha$ in natural units.
No symmetry constraint ($\sigma_{xx}=\sigma_{yy}$ or $\sigma_{xy}=-\sigma_{yx}$) is imposed.

The incident light propagates along the surface normal, and its linear polarization is fixed along the laboratory $x$-direction.
Both the incident ($\boldsymbol{E_i}$) and reflected ($\boldsymbol{E_r}$) electric fields are expressed in the same laboratory basis, with
\begin{equation}
\boldsymbol{E_r}=\mathbf{r} \boldsymbol{E_i}=
\begin{pmatrix}
r_{xx} & r_{xy}\\
r_{yx} & r_{yy}
\end{pmatrix}
\boldsymbol{E_i},
\end{equation}
where $\mathbf{r}$ is the Jones matrix.
For the original sample orientation, $\boldsymbol{E_i}=E_0(1,0)^T$, and $\boldsymbol{E_r}=E_0(r_{xx},r_{yx})^T$,
therefore, the relevant complex polarization ratio is $\rho^{(0)}=r_{yx}/r_{xx}$.
After physically rotating the sample by $90^\circ$ while keeping the light propagation direction and the laboratory axes fixed,
we have $\boldsymbol{E_r}^{(90)}=E_0(r_{yy},-r_{xy})^T$, and $\rho^{(90)}=-r_{xy}/r_{yy}$.
The rotation angle of the major axis of the reflected polarization ellipse is obtained from the Stokes parameters,
\begin{equation}
\theta_K=\text{Re} \arctan \rho.
\label{eq:theta_rho}
\end{equation}

We now express the Jones matrix in terms of the optical conductivity tensor. For a suspended thin sheet,
\begin{equation}
\mathbf{r}=-\left(\mathbf{I}+C\boldsymbol{\sigma} \right)^{-1}C\boldsymbol{\sigma}=-\frac{C}{L}
\begin{pmatrix}
\sigma_{xx}+CD & \sigma_{xy}\\
\sigma_{yx} & \sigma_{yy}+CD
\end{pmatrix}.
\label{eq:r}
\end{equation}
where $C=2\pi \alpha$, $D=\det \boldsymbol{\sigma}=\sigma_{xx}\sigma_{yy}-\sigma_{xy}\sigma_{yx}$ and $L=1+C(\sigma_{xx}+\sigma_{yy})+C^2D$.

% This form shows that the determinant term originates from the full matrix inversion of the conductivity tensor and does not rely on $C_4$ symmetry.

Substituting Eq.~\eqref{eq:r} into Eq.~\eqref{eq:theta_rho},  the general polarization-resolved Kerr rotations are given by
\begin{align}
\theta_K^{(0)}&=\text{Re} \arctan \left[ \frac{\sigma_{yx}} {\sigma_{xx}+2 \pi \alpha (\sigma_{xx} \sigma_{yy} - \sigma_{xy} \sigma_{yx})} \right], \notag\\
\theta_K^{(90)}&=\text{Re} \arctan \left[\frac{-\sigma_{xy}}{\sigma_{yy}+2\pi\alpha (\sigma_{yy}\sigma_{xx}-\sigma_{yx}\sigma_{xy})} \right].
\end{align}
in which the superscripts $(0)$ and $(90)$ denotes the original and rotated nematic orientation of the sample.
The difference between $\theta_K^{(0)}$ and $\theta_K^{(90)}$ captures the nematic anisotropy of the reflected polarization response.

%%%%%%%%%%%%%%%%%%%%%%%%%%%%%%%%%%%%%%%%%%%%%%%%%

\subsection{Nematic Kerr signals at $\theta=53.13^{\circ}$}

Fig.~\ref{fig:Kerr12} summarizes the nematic optical response for the $\phi_1+\phi_2=0$ mode at twist angle $\theta=53.13^{\circ}$.
Panel~(a) displays the diagonal anisotropy $\sigma_{xx}-\sigma_{yy}$, while panel~(b) presents the associated polarization-resolved Kerr rotations $\theta_K^{(0)}$ and $\theta_K^{(90)}$.
The two nematic orientations of the sample yield distinguishable Kerr responses, with $\theta_K \sim \mathcal{O}(1)$~mrad and an orientational contrast $|\theta_K^{(90)}-\theta_K^{(0)}| \sim \mathcal{O}(0.1)$~mrad.
Although these values are smaller than those obtained for the $\phi_1+\phi_2=\pi$ mode discussed in the main text, they remain well within experimental reach,
as modern Kerr measurements can achieve resolutions at the sub-$\mu$rad and even tens-of-nanoradian level \cite{Schemm2015SM,Saykin2023SM,Farhang2026SM}.
Therefore, the predicted Kerr rotations and their nematic splitting should be readily accessible experimentally.

%-----------------------------------------------
\begin{figure}[htb]
    \centering
    \begin{subfigure}{0.23\textwidth}
        \centering
        \includegraphics[width=\textwidth]{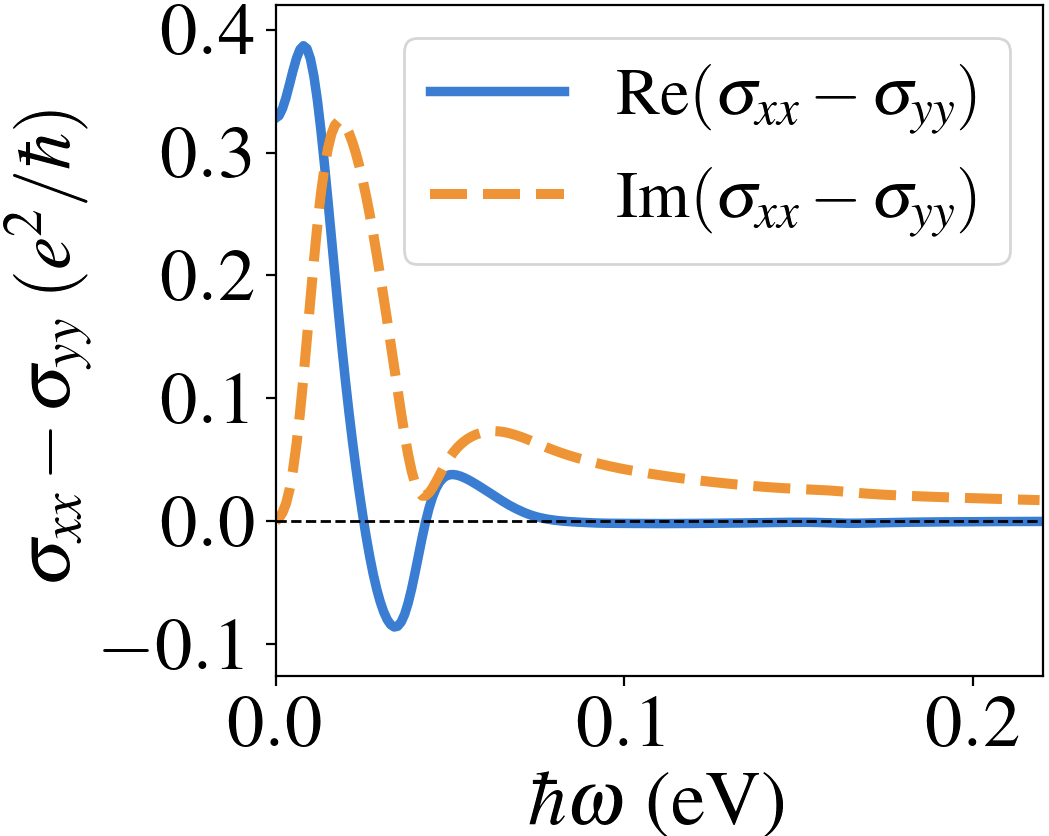} % 1049*836
        \caption{}
        \label{subfig:xx-yy}
        \end{subfigure}
    \hspace{0.6cm}
    \begin{subfigure}{0.23\textwidth}
        \centering
        \includegraphics[width=\textwidth]{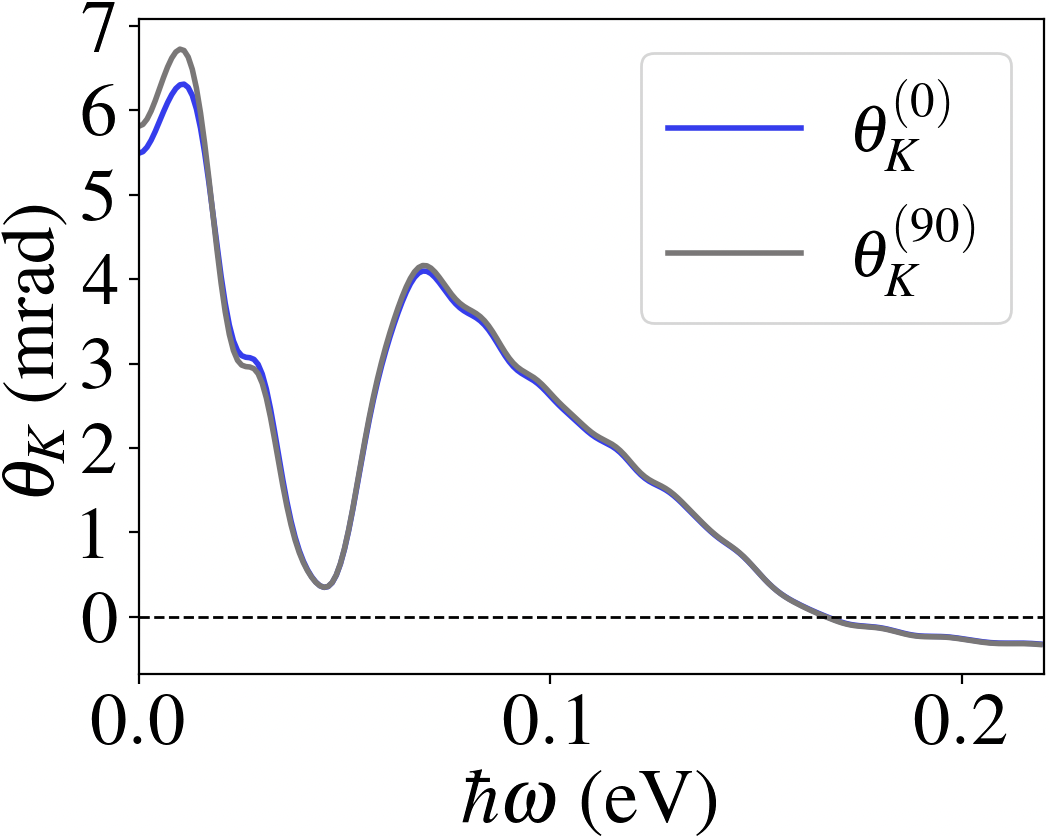} % 1049*836
        \caption{}
        \label{subfig:Kerr}
    \end{subfigure}
    \captionsetup{justification=raggedright}
    \caption{Nematic (a) optical conductivity and (b) Kerr rotation of the $s+d_1e^{i\phi_1}+d_2e^{i\phi_2}$ state at $\theta=53.13^\circ$.
The parameters are chosen as $t=0.153$~eV, $t^{\prime}=-0.45t$, $\mu=-1.0t$, $g_0=0.032$~eV, $V=0.146$~eV, $U=0$, and $T=0$~K.}
    \label{fig:Kerr12}
\end{figure}
%-----------------------------------------------

%%%%%%%%%%%%%%%%%%%%%%%%%%%%%%%%%%%%%%%%%%%%%%%%%

\subsection{Optical penetration to the twisted interface}

The photon-energy window considered here, $0<\hbar\omega\lesssim0.2$~eV, corresponds to $0<f\lesssim48.4$~THz and $\lambda_0\gtrsim6.2~\mu$m,
where $f=\omega/(2\pi)$ is the frequency and $\lambda_0=2\pi c/\omega=hc/(\hbar\omega)$ is the vacuum wavelength.
For light normally incident along the $c$-axis, the electric field lies in the $ab$-plane, and the relevant complex refractive index is $\tilde{n}_{ab}=n_{ab}+i\kappa_{ab}=\sqrt{\tilde{\epsilon}_{ab}}$, in which $\widetilde\epsilon_{ab}=\epsilon_{\infty,ab}+i\widetilde\sigma_{ab}/(\epsilon_0\omega)$.
The intensity therefore decays as $I(z)=I(0)e^{-\alpha(\omega)z}$, where $\alpha(\omega)=2\omega\kappa_{ab}(\omega)/c$ is the absorption coefficient.
The corresponding intensity penetration depth is
\begin{equation}
\delta_I(\omega)=\frac{1}{\alpha(\omega)}=\frac{c}{2\omega\kappa_{ab}(\omega)}=\frac{\lambda_0}{4\pi\kappa_{ab}(\omega)}.
\end{equation}

At representative mid-infrared energies within this photon-energy window,
experiments report optical penetration depths of about $200$~nm in La$_{1.675}$Eu$_{0.2}$Sr$_{0.125}$CuO$_4$ (LESCO) at $\hbar\omega=80$~meV ($f\simeq 19.3$~THz) \cite{Fausti2011SM}
and about $1~\mu$m in YBa$_2$Cu$_3$O$_{6.48}$ (YBCO) at $\hbar\omega\simeq83$~meV ($f\simeq20$~THz)~\cite{Fava2024SM}.
For comparison, experimentally realized incident-side cuprate flakes are about $18$--$80$~nm thick~\cite{Zhao2023SM,Zhu2021SM}, smaller than both representative penetration depths above. % in the frequency window considered here.
Therefore, mid-infrared light within the calculated window can readily reach the twisted interface.

%%%%%%%%%%%%%%%%%%%%%%%%%%%%%%%%%%%%%%%%%%%%%%%%%

\end{widetext}

%%%%%%%%%%%%%%%%%%%%%%%%%%%%%%%%%%%%%%%%%%%%%%%%%


\begin{thebibliography}{10}
\bibitem{s12d12}
Although the intralayer $s$-wave pairing allows both symmetric ($s_1+s_2$) and antisymmetric ($s_1-s_2$) combinations,
the leading interlayer Josephson coupling $-(s_1s_2^*+s_1^*s_2)$ energetically favors $\phi_{s1}=\phi_{s2}$ and higher-order couplings are much weaker (detailed in Sec~IV~C in SM).
Accordingly, the results of self-consistent mean-field calculations show that $s_1-s_2$ is negligibly small. 
Therefore, we retain only $s=s_1+s_2$ and $d_{1,2}$ as a reasonable simplification. % for the critical temperatures and GL free energy analyses. % the layer-resolved $d_{x^2-y^2}$-wave channels,
\bibitem{GLc}
In the superconducting phase, one has $\alpha_{s,d}<0$ and $\beta_{s,d}>0$.
$\gamma_{dd}$, $\gamma_{sd}$ describe the phase-independent couplings among the three components, and the $\gamma_{dd}$ term reduces the $SO(2)$ rotational symmetry in the $(\psi_1,\psi_2)$ space to $C_4$.
\end{thebibliography}
\end{document}